\documentclass[fleqn,usenatbib]{mnras}

\usepackage{newtxtext,newtxmath}

\usepackage[T1]{fontenc}
\usepackage{ae,aecompl}

\pdfoutput=1

\usepackage{graphicx}	
\usepackage{amsmath}	
\usepackage{amssymb}	
\usepackage{xcolor}
\usepackage{soul}
\usepackage{silence}
\definecolor{scc}{rgb}{0.0, 0.26, 0.15}
\definecolor{mbc}{rgb}{0.26, 0.15, 0.0}
\definecolor{grc}{rgb}{0.0, 0.15, 0.3}

\title[Dusty filaments in the Galactic Plane]{The Hi-GAL catalogue of dusty filamentary structures in the Galactic Plane}

\author[E. Schisano et al.]{
Eugenio Schisano,$^{1}$\thanks{E-mail: eugenio.schisano@iaps.inaf.it} 
S. Molinari,$^{1}$
D. Elia,$^{1}$
M. Benedettini,$^{1}$
L. Olmi,$^{2}$
S. Pezzuto,$^{1}$
\newauthor
A. Traficante,$^{1}$
M. Brescia, $^{3}$
S. Cavuoti,$^{3,4}$
A.~M. di Giorgio,$^{1}$
S.~J. Liu,$^{1}$
\newauthor
T.~J.~T. Moore,$^{5}$
A. Noriega-Crespo,$^{6}$
G. Riccio,$^{3}$
A. Baldeschi,$^{1}$
U. Becciani,$^{7}$
\newauthor
N. Peretto,$^{8}$
M. Merello,$^{9}$
F. Vitello,$^{7}$
A. Zavagno,$^{10}$
M.~T. Beltr{\'a}n,$^{2}$
L. Cambr{\'e}sy,$^{11}$
\newauthor
D.~J. Eden,$^{5}$
G. Li~Causi,$^{1,12}$
M. Molinaro,$^{13}$
P. Palmeirim,$^{14}$
E. Sciacca,$^{7}$
\newauthor
L. Testi,$^{2,15}$
G. Umana,$^{7}$
and A.~P. Whitworth$^{8}$
\\
\\
$^{1}$INAF-IAPS, Via Fosso del Cavaliere 100, Rome, Italy \\
$^{2}$INAF, Osservatorio Astrofisico di Arcetri, Largo E. Fermi 5, I-50125 Firenze, Italy \\
$^{3}$INAF - Astronomical Observatory of Capodimonte, via Moiariello 16, I-80131 Napoli, Italy \\
$^{4}$Department of Physics 'E.Pancini', University Federico II, via Cinthia 6, I-80126 Napoli, Italy \\
$^{5}$Astrophysics Research Institute, Liverpool John Moores University, Liverpool Science Park Ic2, 146 Brownlow Hill, Liverpool, L3 5RF, UK\\
$^{6}$Space Telescope Science Institute, 3700 San Martin Dr., Baltimore, MD, 21218, USA \\
$^{7}$INAF - Astrophysical Observatory of Catania, Via Santa Sofia 78, I-95123 Catania, Italy \\
$^{8}$School of Physics and Astronomy, Cardiff University, Cardiff CF24 3AA, UK \\
$^{9}$Departamento de Astronom{\'i}a, Universidad de Chile, Casilla 36-D, Santiago, Chile\\
$^{10}$Aix Marseille Univ., CNRS, LAM, Laboratoire d'Astrophysique de Marseille, F-13388 Marseille, France\\
$^{11}$Universit\'e de Strasbourg, CNRS, Observatoire Astronomique de Strasbourg, UMR 7550, 67000, Strasbourg, France\\
$^{12}$INAF - Osservatorio Astronomico di Roma, Via Frascati 33, 00040, Monte Porzio Catone (RM), Italy\\
$^{13}$INAF - Osservatorio Astronomico di Trieste, via G.B. Tiepolo 11, I-34131 Trieste, Italy  \\
$^{14}$Instituto de Astrof{\'i}sica e Ci{\^e}ncias do Espa\c{c}o, Universidade do Porto, CAUP, Rua das Estrelas, PT4150-762 Porto, Portugal \\
$^{15}$European Southern Observatory, Karl Schwarzschild str. 2, D-85748 Garching, Germany \\
}

\date{Accepted 2019 December 3. Received  2019 November 19; in original form 2018 May 11}

\pubyear{2019}

\begin{document}
\label{firstpage}
\pagerange{\pageref{firstpage}--\pageref{lastpage}}
\maketitle

\begin{abstract}

The recent data collected by {\it Herschel} have confirmed that interstellar structures with filamentary shape are ubiquitously present in the Milky Way. Filaments are thought to be formed by several physical mechanisms acting from the large Galactic scales down to the sub-pc fractions of molecular clouds, and they might represent a possible link between star formation and the large-scale structure of the Galaxy. In order to study this potential link, a statistically significant sample of filaments spread throughout the Galaxy is required. In this work we present the first catalogue of $32,059$ candidate filaments automatically identified in the Hi-GAL survey of the entire Galactic Plane. For these objects we determined  morphological (length, $l^{a}$, and geometrical shape) and physical (average column density, $N_{\rm H_{2}}$, and average temperature, $T$) properties. We identified filaments with a wide range of properties:  2$'$\,$\leq l^{a}\leq$\, 100$'$, $10^{20} \leq N_{\rm H_{2}} \leq 10^{23}$\,cm$^{-2}$ and $10 \leq T\leq$ 35\,K. We discuss their association with the Hi-GAL compact sources, finding that the most tenuous (and stable) structures do not host any major condensation and we also assign a distance to $\sim 18,400$ filaments for which we determine mass, physical size, stability conditions and Galactic distribution. When compared to the spiral arms structure, we find no significant  difference between the physical properties of on-arm and inter-arm filaments. We compared our sample with previous studies, finding that our Hi-GAL filament catalogue represents a significant  extension in terms of Galactic coverage and sensitivity. This catalogue represents an unique and important tool for future studies devoted to understanding the filament life-cycle.
\end{abstract}




\section{Introduction}

Observations of the Galaxy reveal that matter in the interstellar medium (ISM) is mostly distributed in structures with a filamentary shape, resembling the appearance of Earth clouds. These structures are identified through different tracers in all Galactic environments.  They were initially observed in the diffuse ISM by the far-IR all-sky IRAS survey \citep{Low1984} and called Galactic cirri. Observations in HI \citep{McClureGriffiths2006} and CO \citep{Ungerechts1987, Bally1987, Goldsmith2008} revealed that also molecular clouds are formed by complex networks of hairlike filaments. A closer inspection of the denser regions of molecular clouds shows that they have pronounced elongated shapes, with signs of internal fragmentation \citep{Schneider1979,  Motte1998, Lada2007}.
More recently, the high sensitivity and spatial resolution of the {\it Herschel} Space Observatory \citep{Pilbratt2010} allowed to study the emission from the cold (10-50\,K) dust component of the ISM and revealed, with plenty of detail, the ubiquitous presence of filamentary features \citep{Molinari2010, Andre2010}. Filaments are present in all {\it Herschel} observations; they appear in any cloud mapped by  Gould Belt \citep{Andre2010} and HOBYS \citep{Motte2010} surveys, regardless of the cloud distance, mass or star-formation content \citep{Arzoumanian2011, Hill2011,  Hennemann2012, Peretto2012, Schneider2012, Palmeirim2013,  Konyves2015}, and in any images of the {\it Herschel} Infrared Galactic Plane Survey \citep[Hi-GAL,][]{Molinari2010,Schisano2014}.

The large {\it Herschel} dataset reveals the wide range of sizes, densities and morphologies that filaments can have. Their size ranges from almost 100 parsec long \citep{Wang2015} down to sub-parsec substructures \citep{Schisano2014, Arzoumanian2019}. They vary from diffuse, almost translucent features with column densities $N_{\rm H_{2}}\sim10^{20}$\,cm$^{-2}$ up to dense, optically thick objects with $N_{\rm H_{2}}\sim10^{23}$\,cm$^{-2}$. Moreover, their shapes can vary from isolated, well defined  and approximately linear structures to twisted and irregular complexes composed of groups of filaments, often nesting within each other.

The exact origin of filaments is still unclear, although they are thought to be connected to turbulence present in the ISM \citep{Padoan2001}. In fact, filamentary structures (and shell-like features) are formed after the passage of a shock wave and/or at the interface between two colliding flows \citep{Koyama2000, VazquezSemaden2007}. On the other hand, the observed variety of shapes may conceal different  physical mechanisms leading to their formation. Supersonic turbulence, gravity, cloud-cloud collision, fragmentation of expanding shells, magnetic fields, shadowing forming cometary clouds and galactic shear have been proved to form filamentary morphologies \citep{Nagai1998, Hartmann2007, Heitsch2008, Molinari2014}. Simulations show that filaments form at all scale: they are present as substructures of molecular clouds \citep{Padoan2007, Hennebelle2008, VazquezSemadeni2011, Federrath2013, Gomez2014}, but also as major structures of the Galaxy \citep{Dobbs2006, Smith2014}. Indeed,  at large scales, the ISM is shaped by Galactic rotation and large-scale turbulence, and filaments are found to form between spiral arms (inter-arm space) \citep{Smith2014, DuarteCabral2016} or in gravitational wells of the main spiral arms \citep{Dobbs2013}. These features have been recently observed, with long filamentary clouds found both associated with the spiral arms, and defined as  Galactic ``bones'' \citep{Goodman2014, Zucker2015}, or located in the vast inter-arm space \citep{Ragan2014}. However, filaments are also observed at the smaller scales of molecular clouds: both inactive and active star-forming clouds appear highly filamentary \citep{Andre2010}. Furthermore, the youngest star-forming cores are mostly observed to be spatially correlated to filaments \citep{Molinari2010,Andre2010}. These evidences together suggest that filaments are pre-existing and set up the conditions for star formation \citep{Andre2014}; the formation of stars is therefore derived from the fragmentation processes in these cylindrical geometries \citep{Inutsuka1992,Larson2005}.

All these results inspire a connection between the processes acting at the largest Galactic scale with the formation of stars, passing through the shaping of local (sub)structures within  molecular clouds. This potential link can be explored through a systematic study of the formation, evolution and destruction of filaments, task carried on with the detailed study of individual clouds \citep{Arzoumanian2011, Hacar2011, Kirk2013, Ragan2014,Salji2015, Wang2014, Benedettini2015} and the statistical analysis of large samples of filamentary structures in portion of the Galactic Plane \citep{Schisano2014, Li2016, Wang2016}.  In this context, this work aims to provide the first catalogue of candidate filaments in the entire Galactic Plane. We have therefore used the data from the {\it Herschel} Hi-GAL survey,  re-processing the entire dataset in order to produce mosaics and to compute column density maps (Section \ref{Sect:HiGALdata}). We identify features in these data with an automatic extraction algorithm (Section \ref{Sect:IdentifyFilaments}). We select all the features resembling filamentary shapes, measure general physical properties for each of these objects and build the catalogue (Section \ref{Sect:HiGALcatalog}). Then, we discuss the global properties of the filamentary features in the catalogue, their spatial distribution, their association with compact clumps and the implications in terms of the Galactic structure (Section \ref{sect:global}). 
We compare our catalogue with the other catalogues available in literature: the ATLASGAL filamentary catalogue \citep{Li2016} and the IRDC catalogue by \citealt{Peretto2009} (Section \ref{sect:comparison}). Finally we summarize our results and draw some conclusions (Section \ref{sect:conclusions})

\section{\emph{Herschel} / Hi-GAL data}
\label{Sect:HiGALdata}

\begin{figure*} 
\includegraphics[width=18cm]{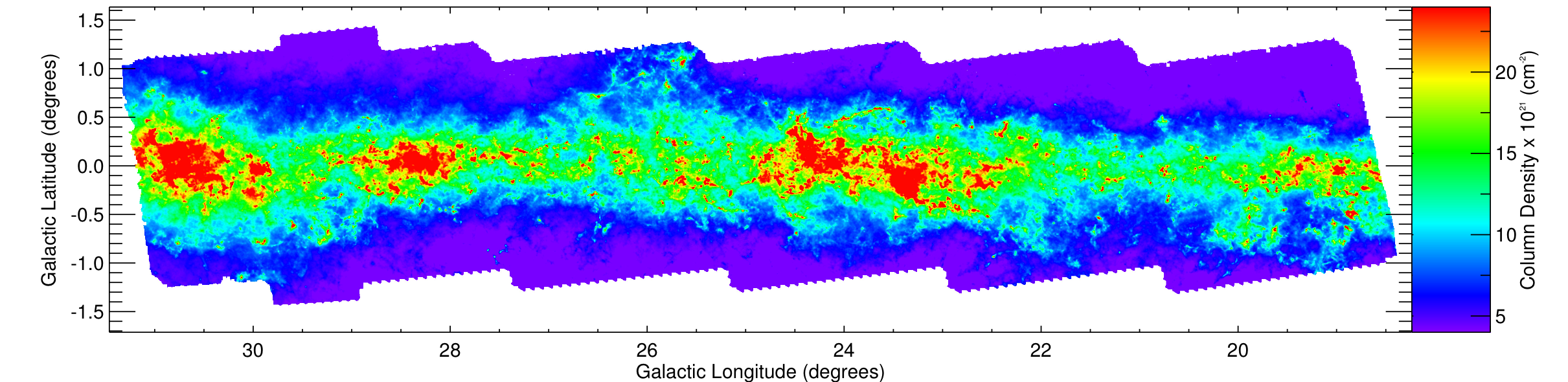}
\includegraphics[width=18cm]{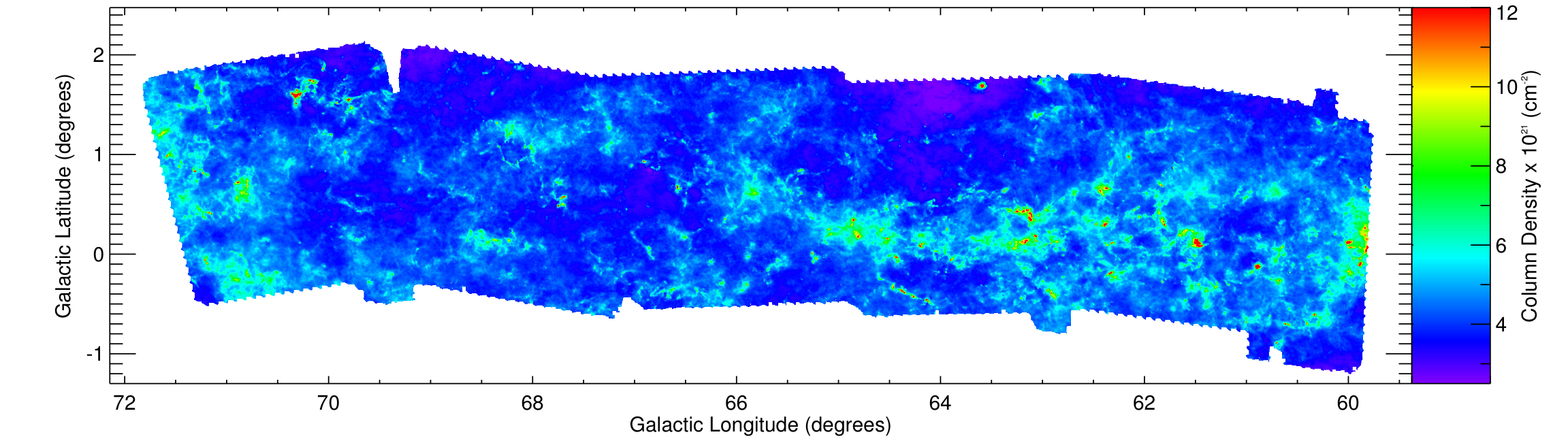}
\caption{Two examples of column-density maps computed from the Hi-GAL mosaics covering  the Galactic longitude ranges $l$\,=\,19$^{\circ}-30^{\circ}$ and $l$\,=\,$60^{\circ}-70^{\circ}$, respectively. }
\label{Fig:ColumnDensity}
\end{figure*}

\subsection{The Hi-GAL photometric mosaics}

The Hi-GAL project \citep{Molinari2010} is a photometric survey designed to map the entire Galactic plane (GP) with the {\it Herschel} Space Observatory \citep{Pilbratt2010} in the wavelength range from 70 to 500\,$\mu$m through the two instruments PACS \citep{Poglitsch2010} and SPIRE \citep{Griffin2010}. The GP is fully covered with 166 individual maps, called ``tiles'', each one covering a region of the sky of 2.2$^{\rm o}\,\times\,$2.2$^{\rm o}$, scanned along two orthogonal directions, and overlapping with its neighbours by $\sim$20\,arcmin. The first Hi-GAL public data release DR1 is derived from 65 tiles covering the inner Milky Way in the longitude range $-70^{\rm o}$\,$\geq$\,{\it l}\,$\geq\,68^{\rm o}$ \citep{Molinari2016}. These tiles were processed with the ROMAGAL pipeline \citep{Traficante2011} and photometrically calibrated with the help of IRAS/{\it Planck} data. The remaining $101$ tiles, related to the fainter outer Galaxy, will be delivered in the next Hi-GAL release (Molinari et al. in prep.). 

The main goal of this work is to identify filament-like features that extend potentially over large portions of the sky. In literature there are cases of giant filamentary clouds with  sizes greater than 1$^{\circ}$ reaching extension up to $\sim$5$^{\circ}$ \citep{Li2013,Ragan2014}. This implies that some filaments can potentially extend  beyond the borders of a single 2\fdg 2$ \times2\fdg 2$ tile. Therefore, we decided to reprocess the Hi-GAL raw data, in order to build mosaics larger than a single tile and to avoid dealing with the splitting of filamentary structures over contiguous Hi-GAL tiles.  We adopted the UNIMAP map maker \citep{Piazzo2015}  to reprocess the entire dataset. UNIMAP has been already used to produce high-quality individual Hi-GAL tiles in the outer Galaxy (Molinari et al., in prep.). Here we processed together the raw datasets of adjacent tiles in a single computation run of the map maker to obtain maps larger than a single tile.  This approach has two main advantages: first, it automatically delivers in a single run a larger element to build a mosaic, secondly, it directly combines the data in the overlapping region between two adjacent tiles. The overlapping region has a portion that was scanned along only one direction during the observation of a single tile. Therefore, the map derived from the individual dataset presents beam distortions and a lower signal-to-noise along its border. The simple mosaicking of the single tiles retains distortions and low-quality artefacts, while they are not present when UNIMAP processed together the observations of neighbour tiles. 
The details of the mosaics and their computation are reported in Appendix~\ref{App:mosaics}. The entire GP is covered with the footprints of $37$ mosaics, each one spanning $\sim$10$^{\circ}$ in Galactic longitude. We chose the mosaic footprints in order to have an overlap of $\sim$2$^{\circ}$ to properly recover any extended structure lying over two adjacent mosaics. 

\subsection{Column density and temperature maps from Hi-GAL dataset}
\label{Sect:ColumnDensity}

The high sensitivity of {\it Herschel} observations allows us to trace the distribution of material, even in structures with a low density. In particular, Hi-GAL observations guarantee detection of material down to column densities of $\sim0.7\times10^{20}$\,cm$^{-2}$, derived from the brightness sensitivities predicted for the observing strategy \citep{Molinari2016}, with the assumption of the dust emission model described below and an average dust temperature of $17$\,K. This indicates that these data are the natural dataset to identify a complete Galaxy-wide census of filamentary structures.

We computed $N_{\rm H_{2}}$ column-density and temperature maps from the photometrically calibrated Hi-GAL mosaics following the approach described in \citet{Elia2013}. In short, we convolved  the {\it Herschel} data to the $500$-$\mu$m resolution ($\sim36\arcsec$) and re-binned on that map grid. Afterwards we performed a pixel-by-pixel fitting of the single-temperature grey body function given by: 

\begin{equation}
F_{\nu} = N({\rm H}_{2})\, \mu\, m_{\rm H}\, \Delta\theta_{500}^{2}\, \kappa_{0}\, \left( {\frac{\nu}{\nu_{0}}} \right)^{\beta} \,B_{\nu}(T)
\label{Eq:CDEstimate}
\end{equation}

\noindent 
where {\it F}$_{\nu}$ is the pixel intensity, $\mu$ is the mean molecular weight assumed equal to $2.8$ for the classical cosmic abundance ratio, $\Delta\theta_{500}$ is the angular pixel size in the $500$\,$\mu$m map, while $B_{\nu}(T) $ is the Planck function at temperature $T$. We adopted the dust opacity law from the prescription of \citet{Hildebrand1983} as in other works dealing with {\it Herschel} data \citep{Schneider2013, Elia2013, Konyves2015, Benedettini2015}: $\kappa_{0}\,=\,0.1\,{\rm cm}^{2}\,{\rm g}^{-1}$ at $\nu_{0}\,=\,1000\,$GHz, which takes into account a gas-to-dust ratio by mass of $100$, and a fixed value for the spectral index $\beta\,=\,2$. We included in the fit the {\it Herschel} intensities in the wavelength range from $160$ to $500$\,$\mu$m.

\begin{figure}
\includegraphics[width=0.45\textwidth]{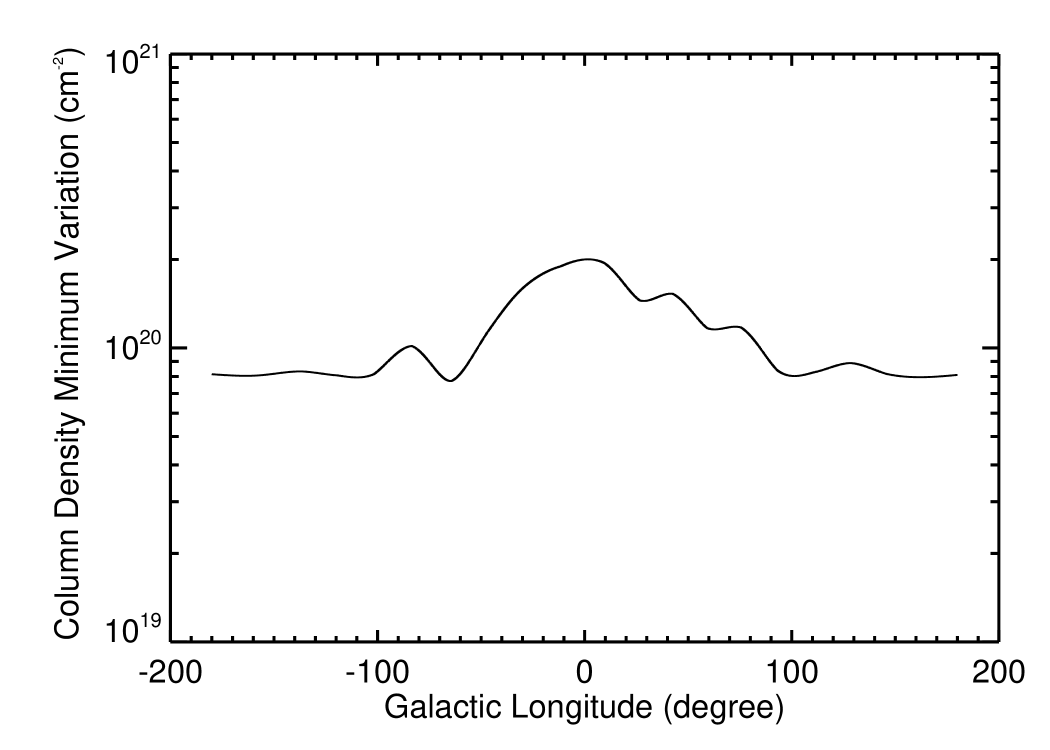}
\caption{Minimum variations in column density, $\Delta N^{\rm min}_{\rm H_{2}}$, detectable in Hi-GAL maps as a function of Galactic longitude. The $\Delta N^{\rm min}_{\rm H_{2}}$ is derived from the lowest measurable brightness, ascribed to the emission of cirrus at constant temperature equal to $17$\,K.}
\label{Fig:MinimumNh2variation}
\end{figure}

Figure~\ref{Fig:ColumnDensity} shows two examples of the column-density maps derived for two different regions of the GP.  We assumed a $20$ per cent uncertainty on the intensity at each band in the fit to take into account for any systematic error in the calibration of the mosaics. This translates into a systematic uncertainty of the order of $\sim$9 per cent on the fitted parameters $N_{\rm H_{2}}$ and $T$. However, we point out that this value refers to an overall uncertainty on the absolute $N_{\rm H_{2}}$ due to systematic errors. The random pixel-by-pixel fluctuations measured in the column-density maps are instead smaller. For the aims of our work, we evaluated the minimum increment in $N_{\rm H_{2}}$, $\Delta N^{min}_{\rm H_{2}}$, that a structure has to show to be significant and detectable in the Hi-GAL data. We estimated $\Delta N^{min}_{\rm H_{2}}$ as a function of Galactic longitude from the photometric maps as follows. First, we identified in each map the regions with the faintest emission, measuring the brightness in each band, $I_{\lambda}$, and the corresponding standard deviation, $\sigma_{I_\lambda}$. The measurements are estimates of the cirrus brightness and its associated noise, that are the intrinsic photometric limits of the Hi-GAL dataset instead of the $Herschel$ instrumental sensitivities \citep{Molinari2016}.  We define the minimum significant column-density variation $\Delta N^{min}_{\rm H_{2}}$:

\begin{equation}
\Delta N^{min}_{\rm H_{2}} = N_{\rm H_{2}}^{(+\sigma)} -  N_{\rm H_{2}}^{(-\sigma)}
\label{Eq:MinColumn}
\end{equation}

\noindent 
where $N_{\rm H_{2}}^{(+\sigma)}$ and $N_{\rm H_{2}}^{(-\sigma)}$ are the column densities, averaged in all the bands, derived from $I_{\lambda} + \sigma_{I_{\lambda}}$ and $I_{\lambda} - \sigma_{I_{\lambda}}$, respectively, and  a uniform temperature for the cirrus of $T\sim17$\,K. Fig.\,\ref{Fig:MinimumNh2variation} shows the resulting $\Delta N^{\rm min}_{\rm H_{2}}$ as a function of Galactic longitude indicating  the effective limit under which a detected structure should not be considered significant.  The amplitude is found to increase from $0.8\times10^{20}$\,cm$^{-2}$ in the outskirts of the Galaxy, up to $\sim2\times10^{20}$\,cm$^{-2}$ towards the Galactic centre, while there are small increases at longitudes where large cloud complexes cover large portions of the Hi-GAL data, such as Cygnus ($l\approx80^{\circ}$), W3-W5 ($l\approx110^{\circ}$) and Carina ($l\approx280^{\circ}$). 

\subsubsection{Effect of dust opacity on N$_{\rm H_{2}}$ and T}

The column-density and temperature maps presented in Sect.\ref{Sect:ColumnDensity} are computed under the assumption that the dust properties are the same everywhere in the Galaxy. However, there are several indications that these properties may vary throughout the Galaxy \citep{Cambresy2001,Paradis2011}.  The {\it Planck} collaboration found that, while the emission spectrum in the far infrared/submillimetre regime ($\lambda\,\leq\,850\,\mu$m) is well fitted by a single grey-body function with a spectral index $\beta$ \citep{Planck2011XXV}, the value of $\beta$ depends on the fraction of molecular gas \cite{Planck2014XIV}. {\it Planck} results points towards a median value of $\beta = 1.88$ in the GP, slightly shallower than the value adopted in this work, but ranging from $1.75$ in the atomic medium up to $1.98$ in molecular gas \citep{Planck2014XIV}. We evaluated how a different spectral index affects our results by recomputing the column-density maps assuming $\beta$ equal to $1.8$. The adoption of a shallower value for $\beta$ has the net effect of decreasing and increasing the resulting column density and temperature, respectively. We found that the average ratio of $N_{\rm H_{2}}^{\beta = 1.8}$ over  $N_{\rm H_{2}}^{\beta = 2.0}$  is equal to $0.81\,\pm\,0.01$ so, on average, the column density decreases systematically by $\sim$20 per cent. The temperature variations are smaller, with an increment of about $0.9$-$1.2$\,K that corresponds to $5$ and $7$ per cent of the average temperature over the maps. Therefore, we conclude that different assumptions on the dust opacity exponent affect marginally the temperature estimates of filaments reported here, but they can alter their column density. These measurements are more appropriate for dense filaments, mostly made by molecular gas for which the $\beta$ assumed here matches with {\it Planck} measurements. Contrariwise, our  column densities are possibly overestimated in the case of tenuous structures, where the material is mostly dominated by gas in atomic phase and a shallower $\beta$ should be applied.
 
\section{Identification of filamentary features}
\label{Sect:IdentifyFilaments}

This section describes the approach used to identify filamentary structures in the {\it Herschel} column-density maps. We start by defining ``filamentary feature'' in the most generic 
The description starts from a general definition for ``filamentary feature'', discusses the algorithm (Sect.\,\ref{Sect:Algorithm}) and the choice of extraction parameters tailored to identify any region corresponding to our definition (Sect.\,\ref{Sect:FeatureExtraction}). In Sect.\,\ref{Sect:Segmentation} we introduce a further decomposition into different substructures that are listed in the final catalogue of filamentary features.

\subsection{Methods for filament detection}
\label{Sect:Algorithm}

To build a catalogue of filamentary features it is necessary to translate the qualitative description of ``filamentary appearance'', often cited in the literature when describing the ISM \citep{Low1984, Schlegel1998}, into an unbiased and quantitative definition for ``filaments'', i.e. a structures in the images. Then, it is possible to characterize filaments with a set of measurable parameters to select them from other features, allowing an automatic identification and extraction of filament-like features. In the recent literature, there are various  definitions of filaments \citep{Arzoumanian2011, Hill2011, Hennemann2012, Andre2014, Schisano2014} and methods for their detection \citep{Sousbie2011, Menshchikov2013, Schisano2014, Salji2015, Koch2015}, some of which have been already applied to {\it Herschel} maps. 

In this work we choose to call a ``filament'' {\it any extended, two dimensional, cylindrical-like feature that is elongated and shows a higher brightness contrast with respect to its surroundings}. Our definition is extremely general and includes several types of features, all with ``filamentary'' morphology, present on an image, including the physical interstellar structures  discussed in the recent star formation studies \citep{Arzoumanian2011,Andre2014,Arzoumanian2019}. The features so defined are easily identified with the help of the image Hessian Matrix, $H(x,y)$, its eigenvalues and/or their linear combination. These tools are adopted in some algorithms \citep{Schisano2014, Salji2015, Planck2016XXXII} among which we selected the one described by \citet{Schisano2014} that has been already tested on and applied to {\it Herschel} GP data. We refer to \citet{Schisano2014} for the  description of the algorithm, its detection and reliability performances determined through simulated filaments. In sort, the algorithm relies on the Hessian Matrix $H(x,y)$ of the intensity map, $I(x,y)$ (in our case the N$_{\rm H_{2}}(x,y)$ map), to enhance elongated regions with respect any other emission. In fact, the second derivative of $I(x,y)$ present in $H(x,y)$ performs a spatial filtering, damping the large-scale and slowly varying emission of the background and amplifying the contrast of any small-scale feature, where the emission changes rapidly. The detailed description of the effect of the second derivative transformation on {\it Herschel} intensity map is discussed in \cite{Molinari2016}.  In that case, the second derivative was implemented in the CuTEx photometry package \citep{Molinari2011}, but it was computed only along specific directions, i.e. the $x$-axis and $y$-axis of the image, to identify compact sources with a circular shape. Instead,  it is necessary to probe of any angular direction in the case of filaments, due to their geometry and orientation in the plane of the sky. To address this, the filament extraction algorithm diagonalizes $H(x,y)$ and compute the eigenvectors and eigenvalues, $\lambda_{a}$ and $\lambda_{b}$ (with $\lambda_{a}\,\leq\,\lambda_{b}$) \citep{Schisano2014}. The diagonalization of $H(x,y)$ is equivalent to the rotation of axes towards the $x'y'$ directions  where $I(x,y)$ has the maximum and minimum variations, that are measured by $\lambda_{a}$ and $\lambda_{b}$, respectively. This property is useful to select regions where the local emission has a cylindrical``ridge'' shape that corresponds to positions where $\lambda_{a}\,\ll\,\lambda_{b}\,\leq\,0$ \citep{Schisano2014}. The enhancement of these features done by $H(x,y)$ allows to detect and extract even tenuous filaments with a low contrast \citep{Schisano2014}.  

\begin{figure*} 
\includegraphics[width=15cm]{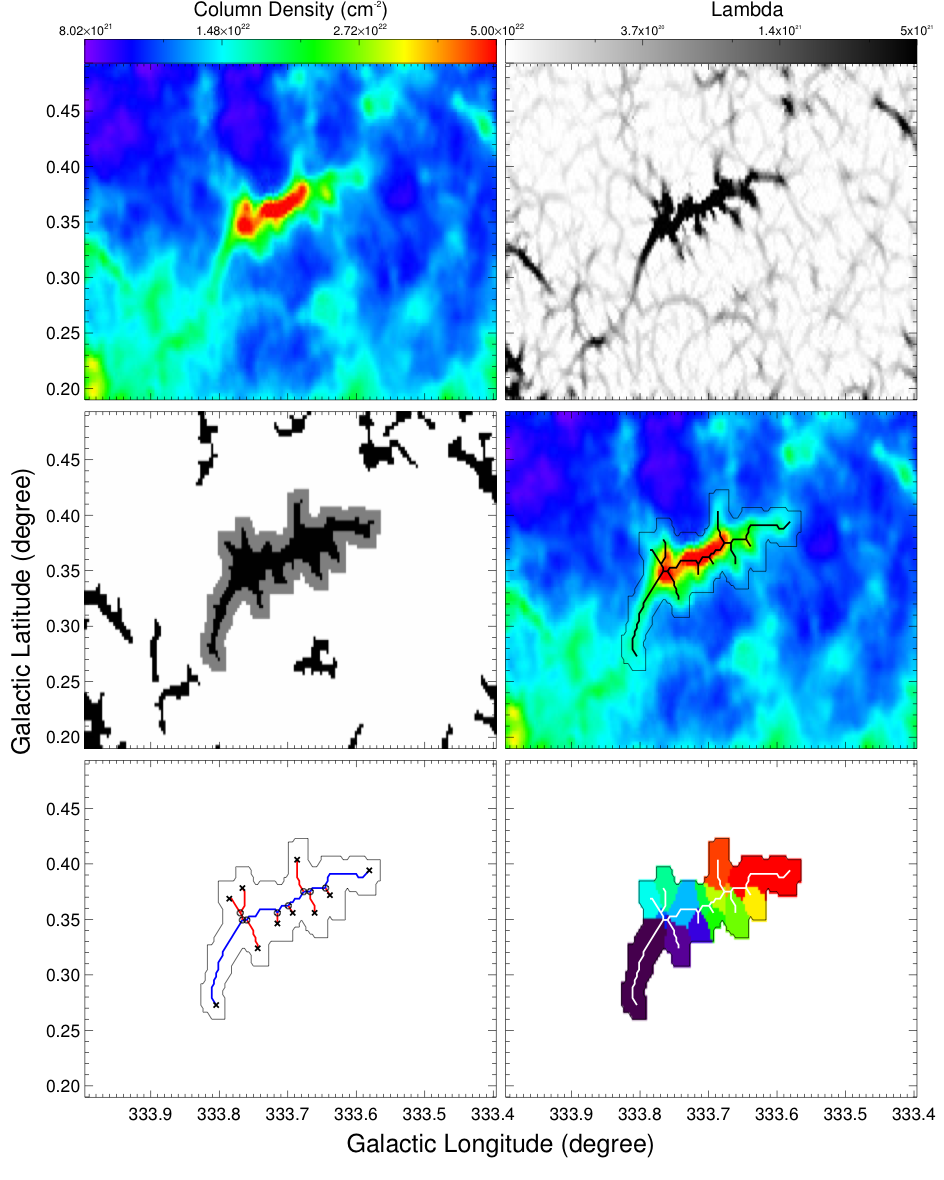}
\caption{Example of the filament extraction and the definitions adopted in this work. The upper left panel  shows the column-density map of a region centred at ($l$,$b$) = ($333.7^{\circ}$, $0.35^{\circ}$), showing an extended, elongated filamentary feature. The upper right panel shows the  eigenvalue $\lambda_{a}$ map  where the filamentary morphologies are enhanced by the Hessian Matrix transformation. The middle left panel shows the initial masks (in black) identified by a $3$-$\sigma$ thresholding of the $\lambda_{a}$ map (see Sect.~\ref{Sect:FeatureExtraction}). The grey region around the central structure identifies the extended mask derived by the dilation of the initial mask, including all the emission ascribed to the filament. The middle right panel shows the contour of the extended mask on the column density map and its central skeleton built up by multiple segments, called $1$-D branches. The 1-D branches trace the departure of the initial mask from the linear cylindrical shape, each of them mapping a peculiar asymmetry of the candidate object.
The lower left panel shows once again the $1$-D branches and the associated singular points: vertices, indicated with crosses, i.e., the ending position of the skeleton, and nodes, indicated by circles, i.e., the connecting position between two or more branches on the skeleton, respectively.  A $1$-D branch is any segment of the skeleton between two singular points, indicated with blue and red colour, respectively. The spine is defined as the longest path on the skeleton connecting two different vertices and is shown in blue. The lower right panel shows the segmentation of the extended mask into multiple subregions defining the $2$-D branches. The splitting is done by associating each pixel of the extended mask with the closest $1$-D branch.
}

\label{Fig:ObjDefinitions}
\end{figure*}

Figure\,\ref{Fig:ObjDefinitions} shows an example of the ability of the algorithm to identify and extract filaments in the simple case of an isolated and extended feature. The filamentary feature is recognizable by eye in the column density map, as shown in the upper left panel of Figure\,\ref{Fig:ObjDefinitions}, but it is enhanced in the eigenvalue map $\lambda_{a}$ (upper right panel of Figure \ref{Fig:ObjDefinitions}). In fact, as discussed above, any bright background emission is strongly attenuated in the eigenvalue map $\lambda_{a}$ (inverse colour scale).  Moreover, the intensity on this map depends on the intensity, the contrast and how strong is the downward concavity of the feature, in other words on the amplitude of the variation from one pixel to its neighbours in the $I(x,y)$. The (absolute) intensity of $\lambda_{a}$ is stronger where the $I(x,y)$ variations are higher, i.e. elongated and high contrasted features. On the other hand, the intensity of $\lambda_{a}$ quickly drops where there are more modest $I(x,y)$ variation, corresponding to low-contrast, faint and/or less connected structures. In theory, the selection of cylinder-like features requires the analysis of both $\lambda_{a}$ and $\lambda_{b}$: the latter traces the cylinder main axis and $\lambda_{a}$ the orthogonal direction, usually with a stronger concavity  \citep{Schisano2014}. However, real physical filaments host compact sources \citep{Molinari2010, Konyves2015} whose presence alters and increases the concavity along the filament main axis. This fact strongly affects $\lambda_{b}$, limiting its use, since by selecting only the pixels where  $\lambda_{a}\,\ll\,\lambda_{b}$ we would exclude any source lying within the structure. To avoid this, we use only the $\lambda_{a}$ map, hence our initial thresholding of the eigenvalue map does not include only pixels belonging to filamentary-like features, but it will require further criteria to remove possible contaminants. We discuss the adopted criteria later in the article.\\
The thresholding of $\lambda_{a}$ defines a binary mask composed by separated regions that we call  {\it candidate region}. We refer to any group of pixels identified by the thresholding that belong to a distinct region as the {\it ``initial mask''}. Examples of initial masks relative to a $3$-$\sigma$ thresholding of the $\lambda_{a}$ map are shown in black in the middle left panel of Figure\,\ref{Fig:ObjDefinitions}. We stress that, with this approach,  we do not trace only cylindrical shapes, but include also roundish, clump-like, features, although the idea behind the use of $H(x,y)$ is to enhance mostly the contrast of the filamentary morphologies. This means that real physical filamentary structures would be only a subsample of the entire list of candidate regions and should be selected through a further process (see Sect.~\ref{Sect:HiGALcatalog}). 

The algorithm requires only two parameters to run: a threshold value and a dilation parameter. The threshold value defines the cut-off level to be applied to $\lambda_{a}$ to identify the initial masks. Its choice fixes the total number of candidate regions identified and the shape of their initial masks. The dilation parameter determines the borders of each region ascribed to the filament and beyond which we estimate the local background emission. The initial mask borders are not suitable, since they only refer to the central portion of the feature extending up, at most, to the inflection point of the intensity profile of the filament since, by construction, $\lambda_{a}$ selects regions where the emission $I(x,y)$ has only a downward concavity. 
The dilation allows us to extend this mask further until it encompasses the entire area of the filament, including the wings of its profile. We refer to this final region as the {\it extended mask}'',  shown in grey in the middle left  panel of Figure\,\ref{Fig:ObjDefinitions} for a choice of the dilation parameter.  We further discuss how we select the values for these parameters in Appendix~\ref{App:Parameters}.

\subsection{Feature substructures: branches, spine and singular points}
\label{Sect:Segmentation}

We introduce here some definitions referring to substructures that are listed in our final catalogue. We start by considering that the classic physical model for a filament approximates it as a  $1$-D feature where the width along the radial direction, $R$, is much smaller than its length, $L$. In this simplified model, the filament is fully defined by quantities measured only in its central inner region \citep{Ostriker1964,Fiege2000}. This fact explains why algorithms such as DisPerSE \citep{Sousbie2011} and FilFinder \citep{Koch2015} trace filaments as linear segments, i.e., the main axis of the structure, usually referred to as the {\it spine} in the literature.  However, our definition and algorithm consider the feature as a two-dimensional portion of the map (see Sect.\,\ref{Sect:Algorithm}). To include in our catalogue quantities measured on the filament central region, required for comparison with $1$-D models, we adopted a $1$-D representation for each  region. Before introducing  such a representation, we make some important remarks on the typical candidate regions.

The shapes of the candidate regions are generally not regular. Even in the simplest cases, such as that presented in Fig.\,\ref{Fig:ObjDefinitions}, the candidates may show a main structure with several elongated appendages. This means that each candidate region is likely to trace a large cloud with several substructures, as in many filaments observed in nearby regions \citep{Arzoumanian2011,Hacar2011,Palmeirim2013}. It is not uncommon that filaments are contiguously connected to the extended portion of a cloud. The so called  {\it hub-filament} configuration, where multiple filaments orientated along different directions nest on a dense and spherical feature, is recurrent in the Galaxy and associated with high-mass star and cluster formation \citep{Myers2009,Schneider2012}. These cases can be potentially identified as regions with irregular shape by our algorithm, so it is possible that a single entry in our catalogue is associated with multiple physical filaments. We take into account this possibility in our scheme by tracing all the asymmetries of the region in our $1$-D representation.

We built the $1$-D representation as a group of segments that we called {\it $1$-D branches} or simply {\it branches} \citep{Schisano2014}. We use the ``skeleton'' of the binary mask to this aim. The ``skeleton''  is the smallest group of pixels that still allows us to trace the topology of the candidate region  \citep{Gonzalez2006}. Basically, it preserves the region extension, main connectivity and general shape, without losing any information about all its asymmetries. We trace the ``skeleton'' with a thinning algorithm that computes the medial axis transform of the initial mask. This operation identifies all the positions that have more than one pixel on the region boundary as the closest one; in other words, they are the axis of the region. We then connected the pixels of the ``skeleton'' into segments with a minimum spanning tree (MST) algorithm. An example of a region skeleton and of individual branches is shown in the middle right panel of Figure\,\ref{Fig:ObjDefinitions}.  Each segment of the skeleton has two extreme pixels that we divide into {\it nodes}, if they nest in another segment, or {\it vertices}, if they are an ending point without any adjacent pixel. Finally, we need to define a {\it main axis}, or simply {\it filament spine}, from this group of segments. We identify this as the set of branches creating the longest possible path that connects two distinct vertices. We mark these branches in order to measure an upper limit for the entire filament length (see Sect.\ref{Sect:Length}).
As said above, it is possible that the asymmetries traced by the branches correspond to single filaments. To measure average properties of these substructures, we split the extended mask into subregions, named {\it 2-D branches}, each one associated with a single $1$-D branch. We define this splitting by assigning each pixel of the filament to the closest $1$-D branch. This criterion segments the candidate region into multiple subregions as shown in the bottom right panel of Fig.\,\ref{Fig:ObjDefinitions}, where each  $2$-D branches resulting from the segmentation of the extended mask are drawn with a different colour.

\section{The Hi-GAL candidate filament catalogue}
\label{Sect:HiGALcatalog}

This section describes the Hi-GAL catalogue of candidate filamentary features. We introduce the criteria applied to the list of candidate regions to remove spurious detections and to select the candidate filaments to be included in the final catalogue (Sect.~\ref{Sect:SelCriteria}). We then present the quantities we determined for each object in the catalogue: the  quality control values, such as contrast and relevance (see Sect.~\ref{Sect:Contrast}), and the measurements of length (see Sect.~\ref{Sect:Length}), column density and temperature (see Sect.~\ref{Sect:2C2Tmodel}). The complete description of the tables and their columns in the Hi-GAL filament catalogue is in Appendix~\ref{App:catalog}. 

\begin{figure*} 
\includegraphics[width=18cm]{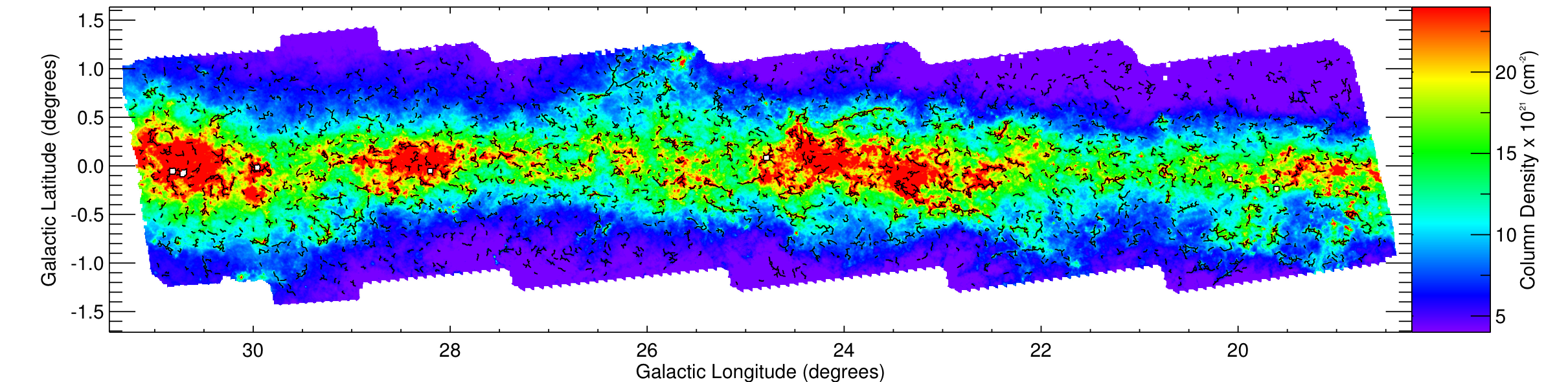}
\includegraphics[width=18cm]{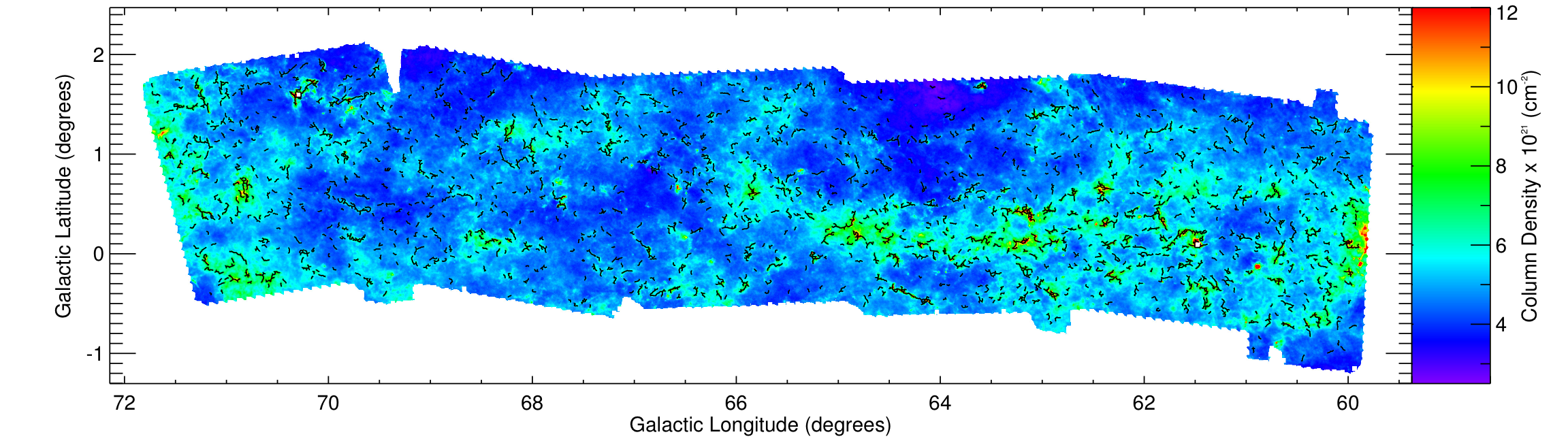}
\caption{Example of candidate filaments identified in the two mosaics covering Galactic longitudes $l$\,=\,19$^{\circ}\,-\,30^{\circ}$ and $l$\,=\,$60^{\circ}\,-\,70^{\circ}$ shown in Figure\,\ref{Fig:ColumnDensity}. For sake of clarity, the figure shows only the branches as black lines. }
\label{Fig:OutputDetection}
\end{figure*}

\subsection{The candidate filaments}
\label{Sect:SelCriteria}

We applied the algorithm for filament detection to all Hi-GAL column-density mosaics, adopting the extraction parameters described in Appendix~\ref{App:Parameters}. We removed any candidate region whose area was smaller than $15$ pixels (see discussion in Sect.\,\ref{Sect:FeatureExtraction}). We also filtered out any region with a main spine (see Sect.\,\ref{Sect:Segmentation}) shorter than $2$ arcmin, corresponding to $\sim4$ times the spatial resolution of the column density map.
Even after this cleaning, we identified a large number of candidate regions ( $\sim$10,000) in each mosaic, confirming that ISM appears highly filamentary. However, not all these regions should be classified as candidate filaments. In order to produce a reliable catalogue of filaments in the Galaxy, we have introduced further selection criteria based on the shape of these objects.   

\begin{figure} 
\includegraphics[width=0.45\textwidth]{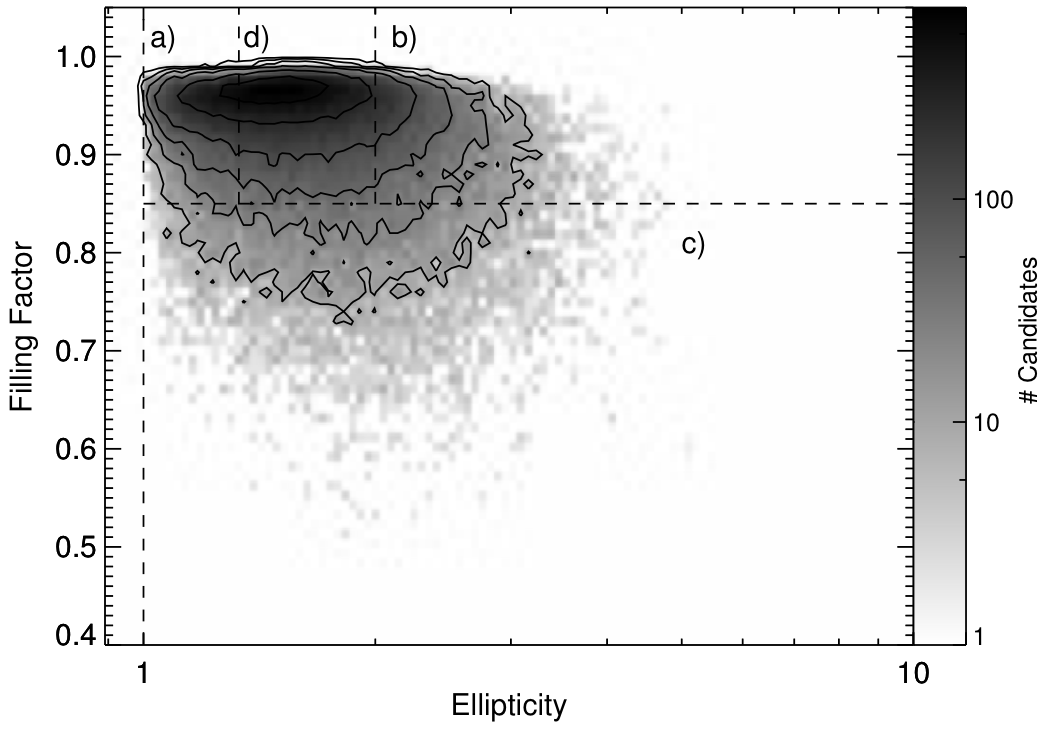}
\caption{Density plot of the distribution of ellipticity and filling factor for the candidate regions initially extracted from the Hi-GAL mosaics in bins of $\Delta$$f$ 0.01 and $\Delta$$\log_{10}{e}$ 0.01 dex . The contour lines correspond to  regions where there are more than 10, 25, 50, 100, 250, 500 objects per bins, respectively.  The dashed lines divide the plane into four sections, each associated with a peculiar morphology (see text for details).}
\label{Fig:EllvsFill}
\end{figure}

The masks obtained after the thresholding of $\lambda_{a}$  show a large variety of shapes. Indeed, \citet{Wang2015} and \citet{Li2016} noticed already that filaments may present different shapes, and attempted to classify them by visual inspection. Such approach is unfeasible in our case where are involved a much larger number of regions than the one ($\sim100$) of these early works. Thus, we adopted a simpler classification scheme based on two measurable quantities derived from the ellipse fitting of all the initial masks: the ratio between the lengths of the major and minor axes, or {\it ellipticity} $e$, and the ratio between the area of the initial mask and that of the fitting ellipse, or {\it filling factor} $f$. Fig.\,\ref{Fig:EllvsFill} shows the distribution of these parameters for all the candidate regions. The $e$ and $f$ parameters are used to divide our sample into four different morphological types that include all possible features already identified in previous works: ({\it a}\,) extended and approximately round clumps; ({\it b}\,) approximately linear regions with few asymmetries; ({\it c}\,)  curved or twisted regions with few asymmetries (like arcs or edges of bubbles); ({\it d}\,) pronged regions with several branches. 
We removed from our sample all candidates of type {\it a}, the remaining objects are all features showing an elongated, filamentary-like shape. The type {\it a} structures are features resembling a filled ellipse with low ellipticity, similar to those observed in clump-like structures \citep{Molinari2016},  selected by $f\,\geq\,0.85$ and $e\,\leq\,1.3$. These cut-off values were chosen from the modal value of the axis ratio of {\it Herschel} compact sources, equal to $\sim$1.3 \citep{Molinari2016}, and noticing type {\it a} features must have a high $f$, i.e. they are similar to the fitted ellipse.

\begin{figure*} 
\includegraphics[width=16cm]{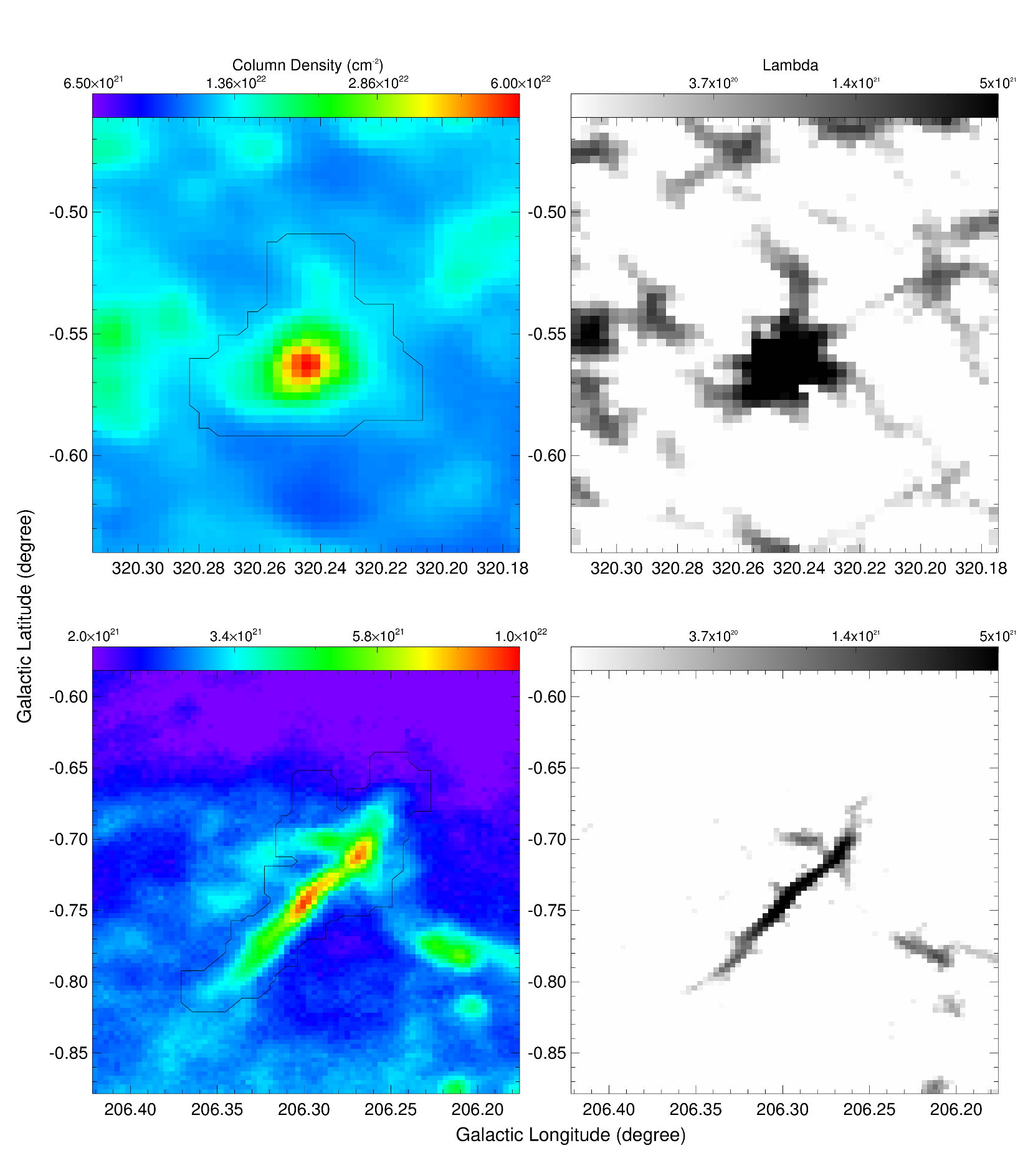}
\caption{{\it Top Panels}: Example of a type (a) candidate, as identified in the column-density map (left panel) from the $\lambda_{a}$ eigenvalue map (right panel ). Having $e=1.2$ and $f=0.97$, this feature is excluded from the filament catalogue.  {\it Bottom Panels}: Example of a type (b) candidate. The feature shown has $e=2.07$ and a $f=0.89$.  }
\label{Fig:ExamplePanel1}
\end{figure*}

\begin{figure*} 
\includegraphics[width=16cm]{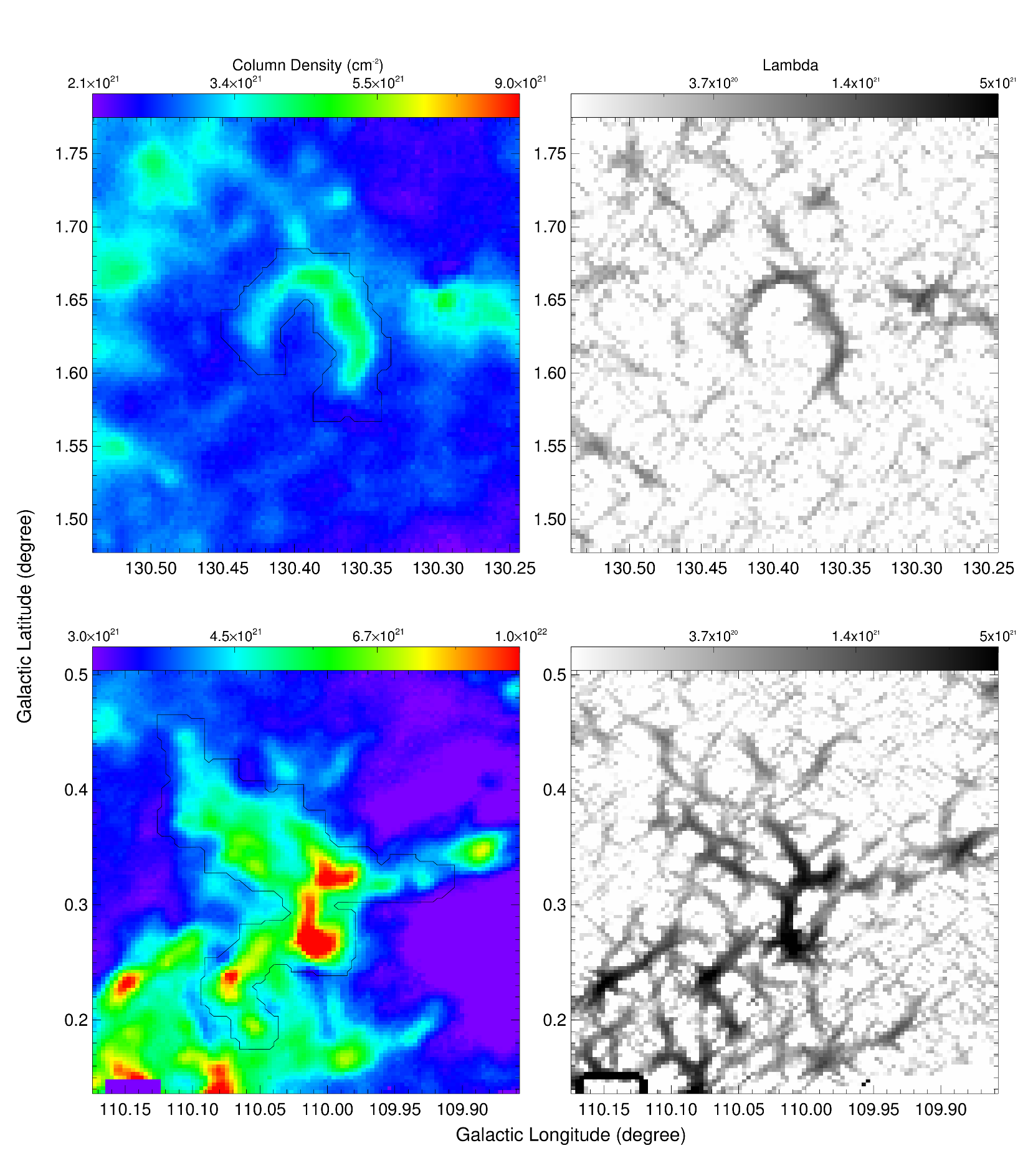}
\caption{{\it Top Panels}: Example of a type (c) candidate, as identified in the column-density map (left panel) from the $\lambda_{a}$ eigenvalue map (right panel). The feature shown is a strongly curved arc for which we measured $e=1.35$ and $f=0.47$ .  {\it Bottom Panels}: Example of a type (d) candidate. The entire feature drawn from the thresholding of the $\lambda_{a}$ map is composed of multiple individual features. The entire complex has $e=1.3$ and a $f=0.87$. }
\label{Fig:ExamplePanel2}
\end{figure*}

Any region left after removing all features of type {\it a} is named  {\it candidate filament} and it matches our generic definition introduced in  Sect.\,\ref{Sect:Algorithm}. Type  {\it b} regions are the most elongated candidates, having a high similarity between the initial mask and the fitted ellipse, so we identified them by selecting the objects with $f\,\geq\,0.85$ and $e\,\geq\,2$. The remaining two morphological types do not clearly separate from $e$ and $f$ values. We note anyway that type {\it c} features do not generally resemble their fitted ellipse due to their curved shapes, while type {\it d} are generally very extended and have a low ellipticity. So we attempted to classify all the features with $f\leq,0.85$ as type {\it c} and the ones with $f\geq\,0.85$ and $1.3\,\leq\,e\,\leq\,2$ as type {\it d}.  We stress that the separation of the candidate filaments in types {\it b}, {\it c} and {\it d} is merely qualitative.  Nevertheless, such a classification allows us to select subsamples of structures sharing a common morphology. For example, type {\it b} structures include all linear and highly elongated features, ideal for follow-up studies on the physics of filaments. Examples of candidates representative of the various types are shown in Fig.~\ref{Fig:ExamplePanel1} and Fig.~\ref{Fig:ExamplePanel2}.

Using these criteria, we identify a total of $39,768$  candidate filaments in all mosaics. This sample, however, contains duplicates: the structures falling in the overlapping area between mosaics. We identified these duplicates by matching the relative masks. In general matched masks across two mosaics do not show the same exact coverage since there are differences in the two mosaics ascribed to flux calibration, column density distributions and local threshold values. We chose to keep in our final list the matched objects with the larger area, removing from the sample $5306$ duplicates. Finally, we also removed any feature that lies on a mosaic borders for a large fraction of its area. These features have a high probability to be artefacts introduced by the derivative (and then $\lambda_{a}$) due to the lacking of measurement outside the edge of the map.

After applying these filters, we ended up with a final catalogue of $32,059$ candidate filaments across the entire GP fulfilling the selection criteria on $\lambda_{a}$ in terms of thresholds, length, area coverage, elongation and morphology, as summarized by the following: 

\begin{itemize}
\item the candidate filaments must have an approximate cylindrical intensity  profile, with a high curvature along at least one direction ($\lambda_{a}\,\geq\,3\,\times\sigma^{\rm local}_{\lambda_{a}}$).
\item they must have a length, measured along their major axis, longer than $\sim$2 arcmin.
\item they must have the bulk of the emission (the central region represented by the initial mask) extending over an area larger than 15 pixels.
\item they must have an estimated ellipticity $e \geq 1.3$ or a filling factor $f \leq 0.85$.
\end{itemize}

\noindent For each of these candidates we estimated the morphological and physical parameters from the {\it Herschel} data, as discussed in the following sections. Associated with this catalogue we also identified $140,525$ branches and $172,584$ singular points, whose positions and physical properties are listed in separated tables. The subregions identified from the segmentation do not always refer to a separate set of filamentary substructures. They require further data at higher angular resolution to confirm their real nature. Nevertheless, we still decided to list in a separate table all the features that can be traced in {\it Herschel} images.

\subsection{Contrast and Relevance}
\label{Sect:Contrast}

The GP emission observed by {\it Herschel} is highly structured, variable and complex \citep{Molinari2010}. The variations of the background inhibit the definition of a parameter to characterize the reliability of a source, as discussed extensively by \cite{Molinari2016}. Sources that appear to be reliable upon visual inspection show very different values of any  parameter that is typically adopted as quality flags for the detection (see their Fig.\,17). This problem is made more complex by the wide range of sizes of the observed sources: criteria that are calibrated for point-like objects generally fail for extended ones. Filamentary structures show similar, and more enhanced, issues due to their large extension. Nevertheless, we tried to define quantities that can be used as a first guess  for the ``quality'' of the extracted feature. Hence, we characterize our candidate  filaments by defining two parameters: the {\it contrast}, $C$, and the {\it relevance}, $R$, that we discuss below.
The filament contrast $C$ is adopted as an estimate of how much more intense the structure appears, on average, compared to the surrounding emission. The relevance $R$ estimates the S/N ratio for extended, and irregular, features.

\begin{figure} 
\includegraphics[width=0.45\textwidth]{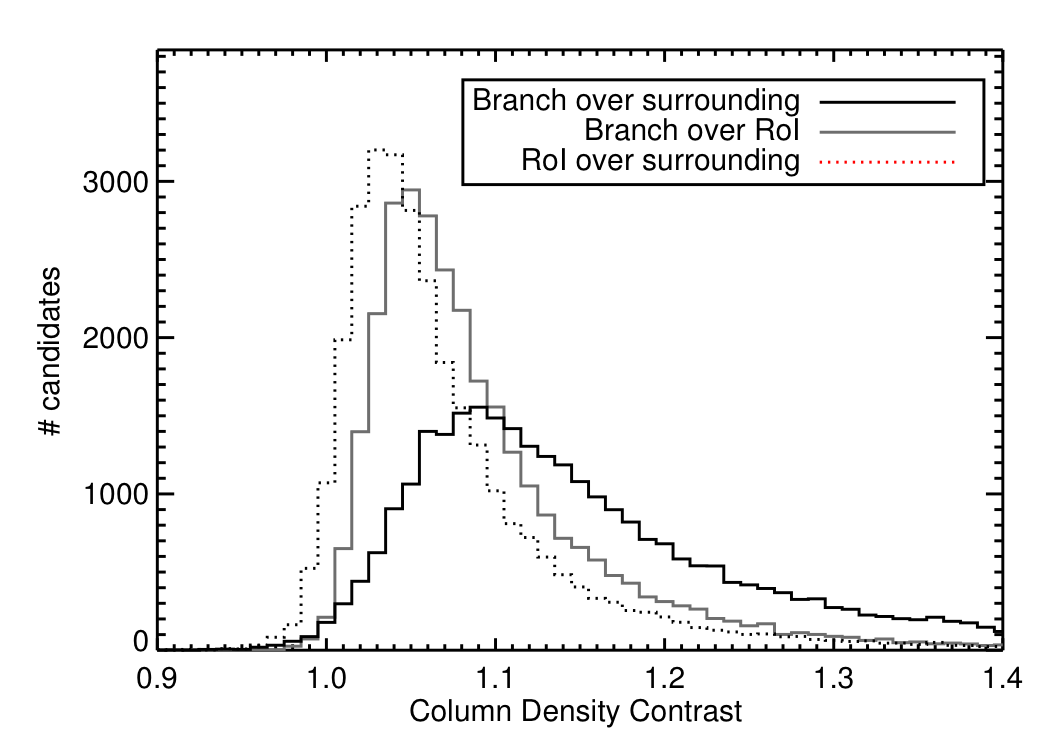}
\caption{Average column-density contrast distributions between different features included in the catalogue. The thin black line indicates the contrast of the central portion of the region, i.e., the branches, while the dotted line indicates the contrast of the entire candidate region both with respect to the local surroundings, estimated in a 2-pixels-wide ring around the extended mask. The red line instead shows the distribution of contrasts of the branches with respect to the region itself. 
}
 \label{Fig:ContrastDistribution}
\end{figure}

We define the contrast $C$ of a candidate as:
\begin{equation}
C^{\rm feature}_{\rm surr} = { \frac{\overline N_{{\rm H}_{2}}^{\rm feature}}{\overline N_{{\rm H}_{2}}^{\rm surr}} }
\end{equation}

\noindent where ${\overline N_{\rm H_{2}}^{\rm \,feature}}$,  ${\overline N_{\rm H_{2}}^{\rm \,surr}}$ are the  average column densities of the filament and of the local surroundings, respectively, the latter defined in a $2$-pixels-wide region around the candidate perimeter.  In the top panel of Fig.\,\ref{Fig:ContrastDistribution} we show the histograms of the contrast of the entire region of interest (RoI), i.e. the extended mask, $C^{\rm roi}_{\rm surr}$ and of the central branch, $C^{\rm br}_{\rm surr}$, with respect to the local surroundings. For completeness, we also show the contrast of the branches with respect to the RoI, $C^{\rm br}_{\rm roi}$ (red line). 

The contrast, as defined above, is a measurement of how much the column density varies from the surrounding background to the filament itself. Filamentary structures are denser in their centres so, while the intensity averaged over the entire feature has only a marginal increment with respect to the background, as shown by a median value for  $C^{\rm roi}_{\rm surr}\sim1.04$ (equivalent to a $4$ per cent increment), the branches are effectively brigther than the rest of the filamentary region, with $C^{\rm br}_{\rm roi}$ distribution peaking at $\sim1.06$ (or $6$ per cent increment). The combined effect of these $N_{\rm H_{2}}$ variations is shown by the $C^{\rm br}_{\rm surr}$ distribution that peaks around a value of $\sim1.1$, but dropping quickly for smaller values. This means that the average column density in the central regions of the majority of our candidates is systematically $10$ per cent higher than the local background. We point out that the observed increment represents a lower estimate of the real contrast. In fact, we averaged the column density over all the branches, including the fainter substructures, so the estimated ${\overline N_{\rm H_{2}}^{\rm \,br}}$ is lower than the effective contrast of the centre of the filament. 
The measured contrasts map how the emission increases for line of sights separated by few pixels. In other words, even small values of $C$ trace sharp variations of the intensity as expected by structures that are prominent upon visual inspection. Indeed, we checked some features randomly and confirmed that they effectively stand out from the surrounding emission if we stretch the intensity scale. However, the visual inspection does not suggest that $C^{\rm roi}_{\rm surr}$ parameter can select the more robust features. We checked some low-contrast structures and found that they are often faint, but sufficiently enhanced in our opinion to be considered real features. Therefore, we are in the same case found for Hi-GAL compact sources \cite{Molinari2016} where contrast alone is not sufficient to characterize the reliability. 
We complement the indication from the contrast parameter $C$ with an additional quantity, that we call {\it relevance} $R$, defined as 

\begin{equation}
R  = { \frac{{\overline N_{\rm H_{2}}^{\rm \,br}} - {\overline N_{\rm H_{2}}^{\rm \,surr}}} {\sigma_{\rm H_{2}}^{\rm surr}} },
\end{equation}

\noindent where we require that $\sigma_{\rm H_{2}}^{\rm surr}$ measures the column density fluctuations locally around the feature. The value of $\sigma_{\rm H_{2}}^{\rm surr}$ is challenging to be measured for the extended features. We tried different methods to estimate these fluctuations $\sigma_{\rm H_{2}}^{\rm surr}$. A first guess is derived from the $\Delta N^{\rm min}_{\rm H_{2}}$ defined in Sect.\,\ref{Sect:ColumnDensity}. This quantity should be considered just as a lower limit since it is estimated in a portion of the map that can be quite distant from the feature and quantifies the relevance of the feature with respect to the stochastic random ``noise'' produced by the cirrus emission. However, this definition ignores any other variations of the local background that limit the detectability of the source. To take into account of these intrinsic limits, we measured the standard deviation of the column density determined in a $2$-pixels-wide margin around the extended mask perimeter. This appears to be a reasonable estimate for isolated and small features, but fails in the case of objects that extend over several arcmin and/or are located on a background that monotonically varies. In fact, a constant gradient in the background would produce a large standard deviation over the $2$-pixel-wide border even if any fluctuations (whose amplitude we aim to measure) would be absent. To overcome this issue, we first subtracted a linear fit from the values over the $2$-pixel-wide border, representing the underlying background large scale spatial gradient, and then computed the standard deviation of the residual background in the filament mask, $\sigma_{\rm H_{2}}^{\rm back}$. For the reasoning described above  $\sigma_{\rm H_{2}}^{\rm back}$ can be assumed a proper estimate for the column density fluctuations around the feature. 

\begin{figure} 
\includegraphics[width=0.45\textwidth]{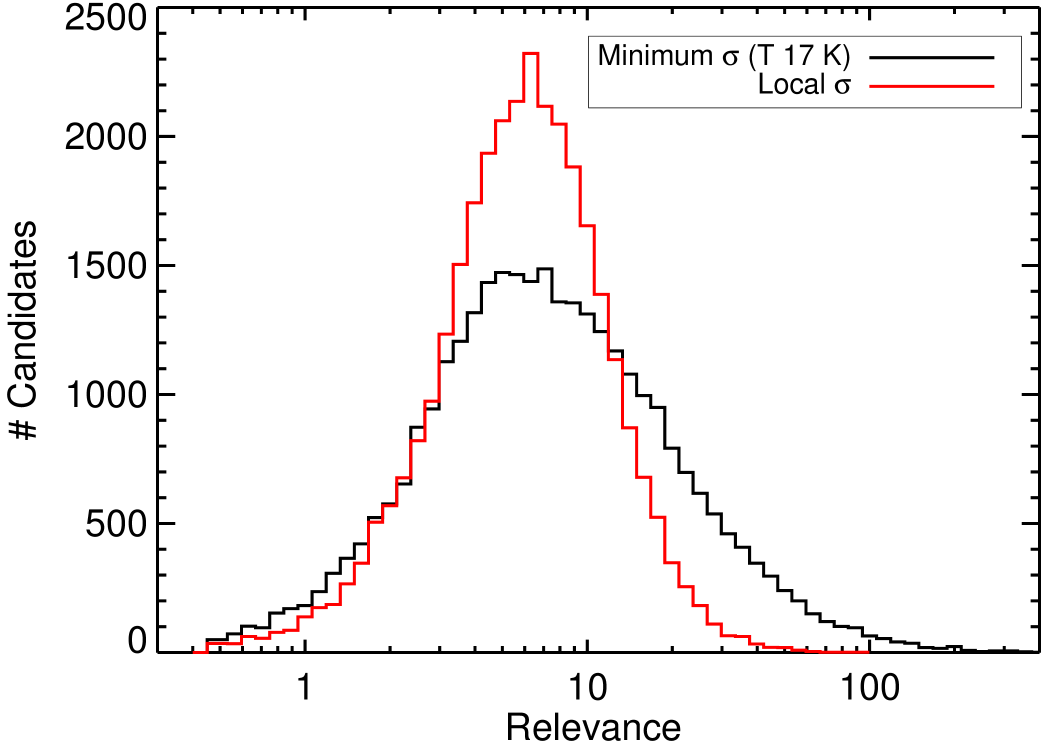}
\caption{Distribution of the relevance $R$ determined assuming that the column density fluctuations are dominated by the cirrus emission at the Galactic longitude of the candidate, $\Delta N^{min}_{\rm H_{2}}(l)$, (in {\it black}) or from the local variations of the background emission, $\sigma_{\rm H_{2}}^{\rm back}$, (in {\it red}).  
}
\label{Fig:HistogramRelevance}
\end{figure}

We present in Fig.\,\ref{Fig:HistogramRelevance} the distribution of $R$ over our entire sample, estimating $\sigma_{\rm H_{2}}^{\rm surr}$ both as $\Delta N^{min}_{\rm H_{2}}$  and $\sigma_{\rm H_{2}}^{\rm back}$. In the first case, the distribution is quite broad and extends up to values of $\sim200$, in the other case, $R$ values are more limited, with the highest values around $\sim40$. The difference in the higher tail of the distribution reflects the presence of highly structured background emission, whose variations do not depend on the cirrus fluctuations. The two distributions converge towards the lower tail, with both peaking at $\sim5$-$6$. The large majority of the extracted candidate filaments have $R\,\ge\,3$, confirming the results from the visual inspection where the most of our features appear to be sufficiently enhanced than their surroundings to be considered real features.

\begin{figure} 
\includegraphics[width=0.45\textwidth]{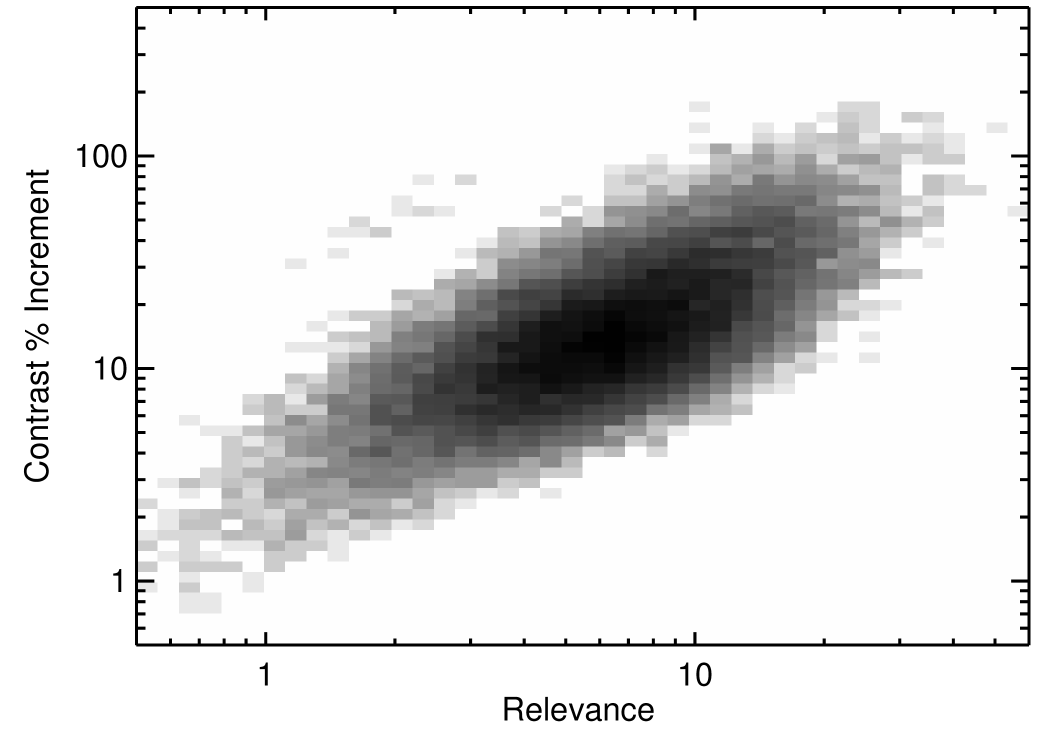}
\includegraphics[width=0.45\textwidth]{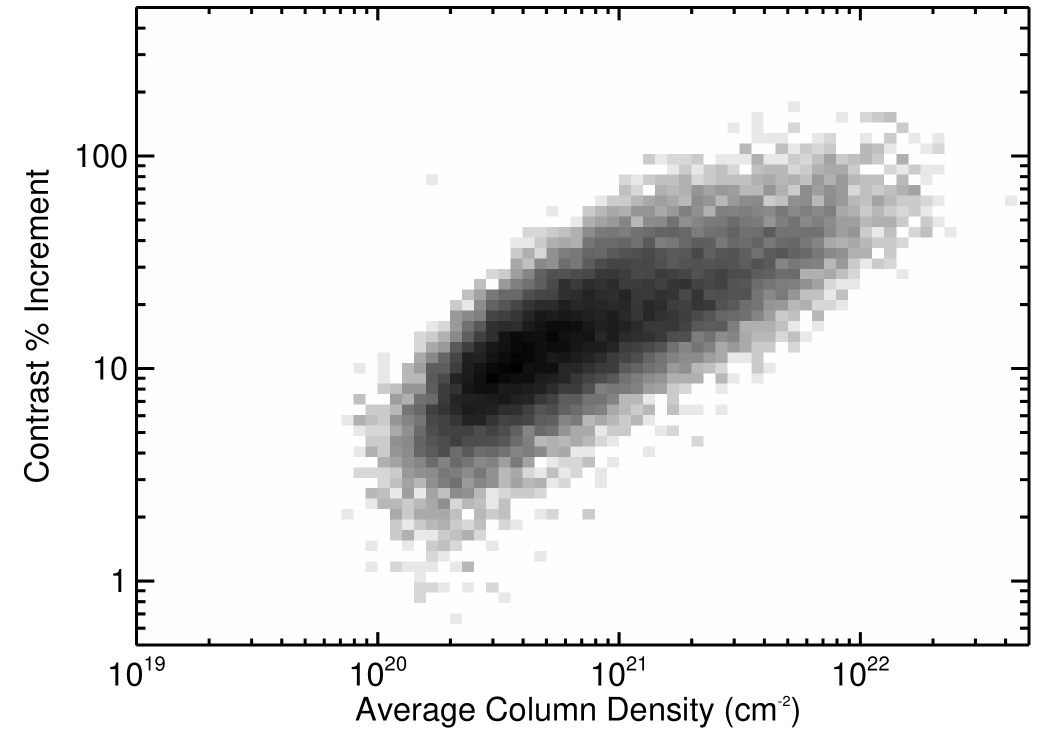}
\caption{{\it Top panel}: Relation between the contrast of candidate regions and their relevance $R$ ({\it see text}) shown as density plot.
 {\it Bottom panel}: Density plot showing the relation between the contrast of the identified candidate filaments  and their average column density. 
}
\label{Fig:ContrastvsRelevance}
\end{figure}

Finally, we discuss the relation between the contrast $C$, relevance $R$ and average column density of the candidates, $ {\overline N_{\rm H_{2}}^{\rm \,roi}}$. Fig.\,\ref{Fig:ContrastvsRelevance} shows density plots illustrating how these parameters relate to each other, where we express $C$ as the column density enhancement, i.e. $C^{\rm \,br}_{\rm \,surr}$ as a percentage. 

Features with high values of $R$ have also a stronger contrast, and typically correspond to higher average column densities. On the contrary, the structures with the smallest contrast enhancement ($\leq4$ per cent) are among the least dense in our sample, with ${\overline N_{\rm H_{2}}^{\rm \,roi}}\,\lesssim\,2$-$3\times$10$^{20}$\,cm$^{-2}$, but their relevance $R$ goes from very low values ($\leq1$, or unreliable features) up to $5$ (a real and evident feature).
We decided not to exclude any objects from the catalogue based on $C$ and $R$, since an arbitrary cut-off would only reflect our personal choice of the features we consider trustworthy. However, we point out that features with low values of both $C^{\rm br}_{\rm surr}$ and $R$ should be considered as unreliable.

\subsection{Lengths of candidates}
\label{Sect:Length}

\begin{figure} 
\includegraphics[width=0.45\textwidth]{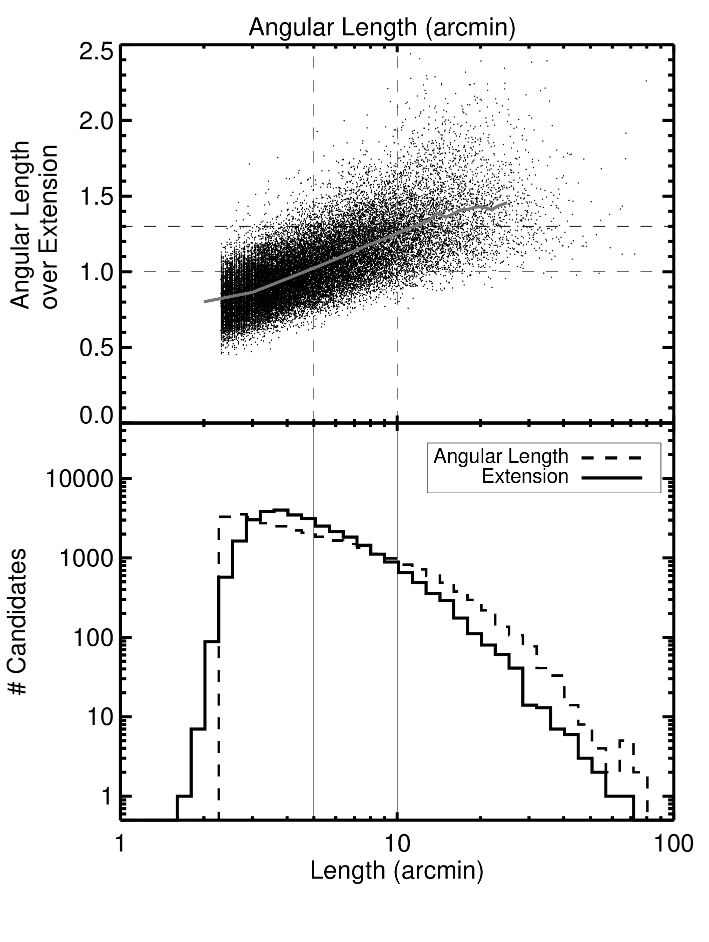}
\caption{{\it Top Panel}: Ratio between $l^{e}$ and $l^{a}$  as a function of $l^{a}$. The red line shows the trend of the median value of this ratio measured in bins of $l^{a}$ 1\,arcmin wide.
{\it Lower Panel}: Distribution of angular sizes of candidates  identified in Hi-GAL mosaics. Two different estimators are adopted: the filament extension, $l^{e}$, defined as twice the major axis of the best-fitted ellipse to the extended region  (dashed line), and the filament angular length, $l^{a}$, defined as sum of the linear distances between consecutive pixels of the candidate spines (full line). 
 }
\label{Fig:AngularLength}
\end{figure}

The angular size of the candidate filaments is measured using two different estimates: a) the length of the major axis of the ellipse fitted to the mask region, also defined as the {\it extension of the filament, l$^{\,e}$}; b) the total  length obtained by adding the distance between consecutive positions along the spine, also defined as the {\it  angular length, l$^{\,a}$}. In the bottom panel of Fig.\,\ref{Fig:AngularLength} we show the distribution of these two quantities for the entire catalogue,  truncated at the lower end by the  selection criteria described in Sect.~\ref{Sect:FeatureExtraction}. Most of the structures in our catalogue are short, with 87 per cent of the candidates having a length of less than 10\,arcmin, and yet there are still more than $\sim$2,200 features with a larger size. The two distributions are in agreement within 3--8 per cent (depending on the length) for candidate filaments with lengths between 5 and 10\,arcmin, but they differ for shorter and longer structures. In the top panel of Fig.\,\ref{Fig:AngularLength} we show the ratio between the two length estimators as a function of the angular length, with the grey line representing its median value estimated in bins of 1\,arcmin. This ratio slowly increases with $l^a$ and we use this dependency to compare the two estimators by splitting our sample into three groups, depending on the filament length: short structures, $l^{a} \leq 5$\,arcmin, intermediate structures, $5\leq l^{a} \leq 10$\,arcmin, and long structures, $l^{a} \geq 10$\,arcmin.  

Short structures have  angular lengths, $l^{a}$, that are on average 10 per cent shorter than their extension $l^{e}$. This is due to the finite thickness of a region affecting the elliptical fitting and the derived estimates which, as expected, become less relevant with larger regions. The two estimates  $l^{a}$ and  $l^{e}$ are consistent for structures with $l^{a} \approx 5$\,arcmin, where the median of the ratio $l^{a}/l^{e} \simeq 1$, but for all the intermediate structures, despite the similarity of their distributions, $l^{a}$ is always larger than $l^{e}$ with discrepancies as high as 30 per cent. On the other hand, for long structures the two estimates are quite different  with $l^{a}$ more than  30 per cent larger than $l^{e}$. This discrepancy can be ascribed to the morphology of the candidate filaments, that are not generally straight. In these cases $l^{e}$ is expected to underestimate the real projected length of the filament, while  $l^{a}$ is expected to give a more realistic estimate. However, this also depends upon the definition of spine (see Sect.~\ref{Sect:Segmentation}). In the case of  large complexes with several branches or strongly pronged structures or, more generally, for candidates where the path connecting the spine points is strongly twisted, then  $l^{a}$ could overestimate the real size. Therefore, we report both estimates as the linear length of the candidate filament, pointing out that they are coincident and equal to the real linear size in the simple case of a straight, linear filament.

\subsection{Column density and temperature: different modelling for filament and background}
\label{Sect:2C2Tmodel}

We estimated the  main physical properties of the filaments from the column density and temperature maps. As previously stated in Section\,\ref{Sect:FeatureExtraction}, the extended mask defines the region associated with each candidate filament. We assume that, in this mask, there are only two physical components (2C, hereafter): 1) the structure classified as filament; 2) a  ``background'' contribution,  including any emission not associated with the filament itself (i.e., the real background and perhaps some foreground emission). Thus, it is fundamental to estimate the background emission in order to measure that associated with the filament alone. The extraction algorithm determines an estimate of the background, taking into account that it may change over the footprint of the filament. This is done starting from each pixel associated with the branches, identifying the direction perpendicular to the branch to which it belongs and interpolating, inwards this direction, the values measured in a 2-pixels-wide ring around the extended mask \citep{Schisano2014}. This procedure is repeated for all the pixels in the branches, providing a background estimate for the large majority of the extended mask. This approach usually leaves only a few pixels of the mask not covered, where we estimated the background through a simple bilinear interpolation of neighbour values.

Initially, we applied this decomposition directly to the column density maps. 
One can also define a {\it two component - one temperature} model (2C1T, hereafter), which is equivalent to assuming that the filament and the background are at the same temperature $T(x,y)$. The temperature is estimated at position $(x,y)$ by a single grey-body fit to the observed fluxes. Hence, the computed column density in each pixel $(x,y)$ can be estimated as :

\begin{equation}
N_{\rm H_{2}}^{\rm meas}(x,y) = N_{\rm H_{2}}^{\rm fil}(x,y) + N_{\rm H_{2}}^{\rm back}(x,y)
\label{Eq:LinearNH2}
\end{equation}

\noindent where $N_{\rm H_{2}}^{\rm fil}(x,y)$, $N_{\rm H_{2}}^{\rm back}(x,y)$ and $N_{\rm H_{2}}^{\rm meas}(x,y) $ are the column densities of the filament, background and the total  value, respectively, at position $(x,y)$ on the map. This simple model is reliable in regions where the observed photometric flux is dominated by the filament component. In these cases, the temperature $T(x,y)$ is  only slightly affected by the presence of any background emission and the uncertainty on the column density of the filament only depends on how well the background contribution is estimated.

On the other hand, this model does not yield to a proper estimate of the physical properties of the candidate filament when there is strong background emission and/or the background temperature differs from that of the filament. The background temperature is of the order of 17.5\,K, i.e., the typical average temperature of the ISM \citep{Boulanger1996}. In this case, the linear decomposition in equation \ref{Eq:LinearNH2} does not hold directly for the column density, and it should instead be applied to the observed fluxes, $F_{\lambda}^{\rm obs}$, for each {\it Herschel} band: 

\begin{equation}
F_{\lambda}^{\rm obs}(x,y) = F_{\lambda}^{\rm fil}(x,y,T_{\rm fil}(x,y)) + F_{\lambda}^{\rm back}(x,y,T_{\rm back}(x,y))
\label{Eq:LinearFlux}
\end{equation}

\noindent Therefore, we used the extended mask of the candidate filaments on the {\it Herschel} maps and, for each object in our catalogue and at each {\it Herschel} waveband, we used the method described above to estimate the contribution of the background, $F_{\lambda}^{\rm back}(x,y)$, and of the filament, $F_{\lambda}^{\rm fil}(x,y)$. Then, for each separated component, we fitted pixel-by-pixel the single-temperature grey-body function described by Eq.\,\ref{Eq:CDEstimate}, obtaining the column density and temperature for the filament, $N_{\rm H_{2}}^{\rm fil}$ and $T^{\rm fil}$, and  for the background,  $N_{\rm H_{2}}^{\rm back}$ and $T^{\rm back}$. This {\it two-component, two-temperatures}  model (2C2T, hereafter) allows us to determine a more realistic estimate for $T^{\rm fil}$ and $N_{\rm H_{2}}^{\rm fil}$. 

\begin{figure}
\includegraphics[width=0.45\textwidth]{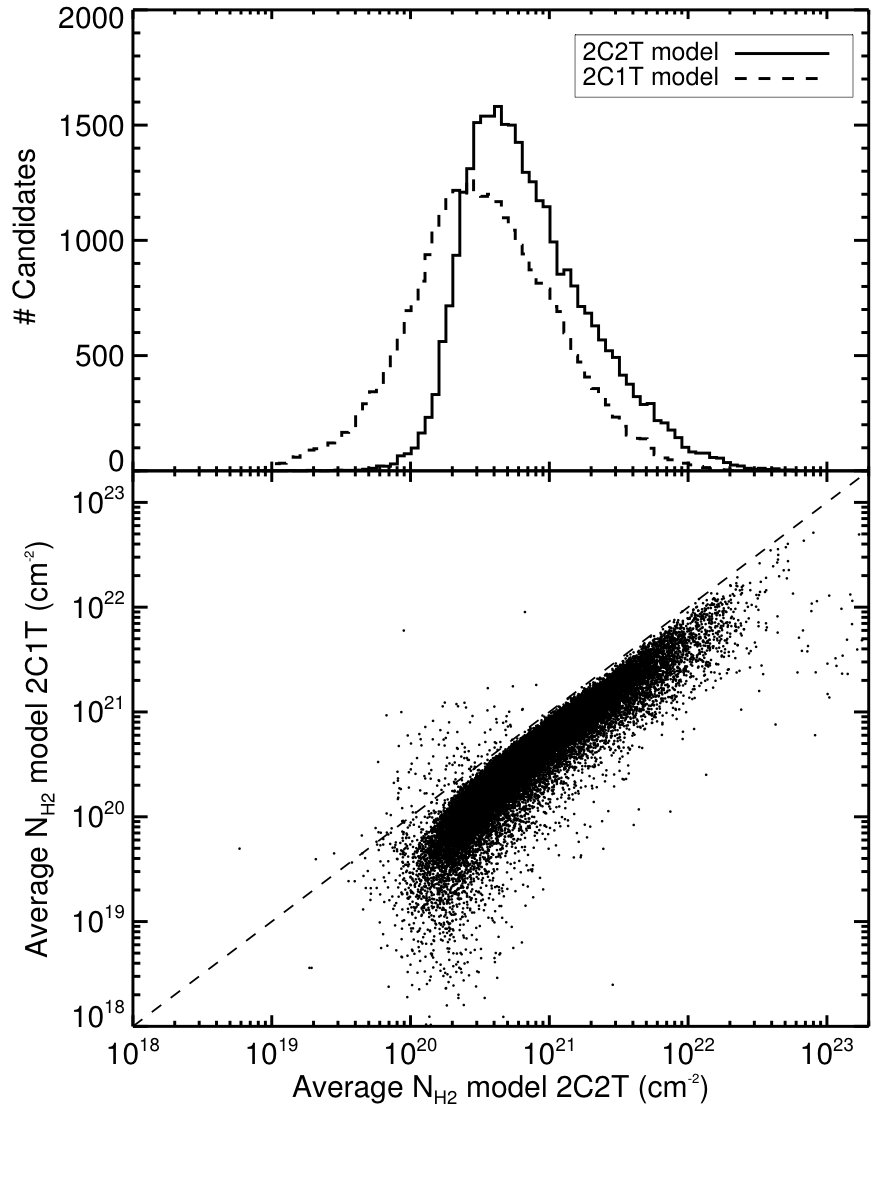}
\caption{ {\it Top Panel}: Distribution of the average column densities measured in the filament candidates with the two approaches described in the text, i.e., the 2C1T $ {\overline N_{\rm H_{2}}^{\rm 1T}}$ (dashed line),  and 2C2T $ {\overline N_{\rm H_{2}}^{\rm 2T}}$ (solid line) modelling  . {\it Bottom Panel}: Relation between the ${\overline N_{\rm H_{2}}^{\rm 1T}}$ and ${\overline N_{\rm H_{2}}^{\rm 2T}}$.  
}
\label{Fig:ComparisonColumnDensityDistribution}
\end{figure}

\begin{figure} 
\includegraphics[width=0.45\textwidth]{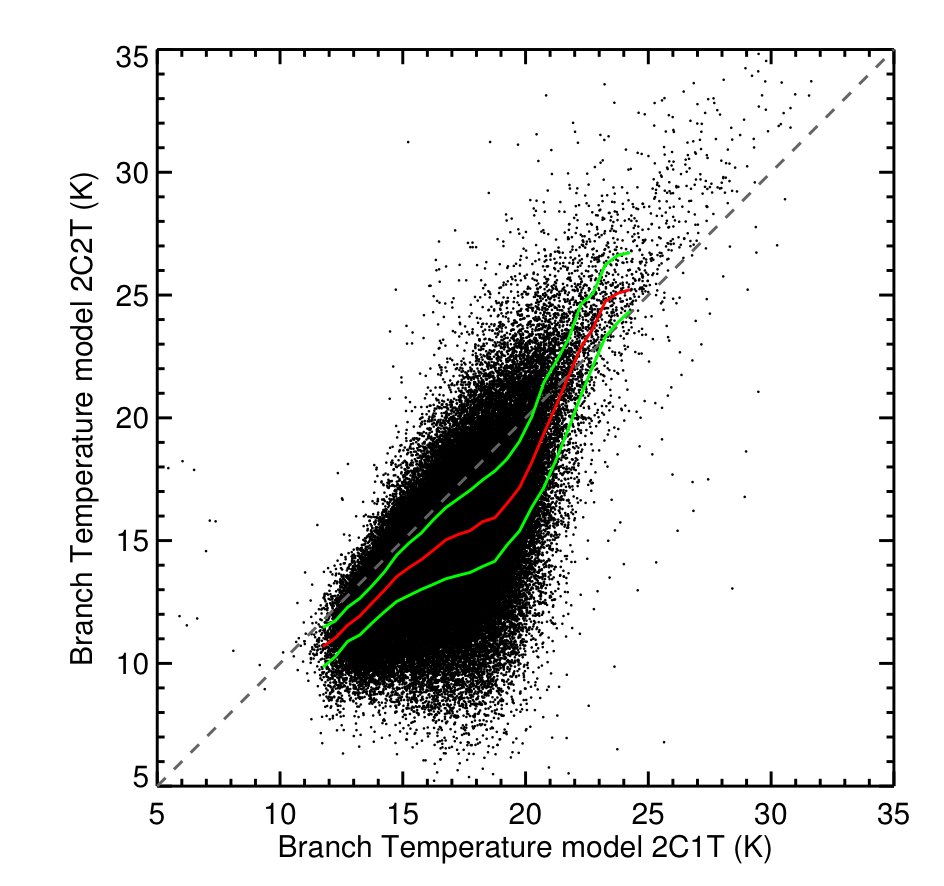}
\caption{Relation between the average temperatures along the candidate branches estimated with the two approaches described in the text. i.e., the 2C1T and 2C2T modelling. The lines show the median (in red) and the quartiles (in green) of the 2C2T branch temperature distribution estimated in bins of 0.5\,K. 
}
\label{Fig:ComparisonBranchTemperatureDistribution}
\end{figure}

We compare here the results of the two models over the entire dataset, then we proceed to discuss their differences in Sect.~\ref{Sect:Differences2C} by analysing the example of the specific filament shown in Fig.\,\ref{Fig:ObjDefinitions}. \\
Fig.\,\ref{Fig:ComparisonColumnDensityDistribution} and Fig.\,\ref{Fig:ComparisonBranchTemperatureDistribution} show the comparison between the average column densities and temperature estimates from the two models. We note that the average column densities derived from the 2C2T models, $ {\overline N_{\rm H_{2}}^{\rm 2T}}$,  and the central temperatures  measured along the branches, $  {\overline T_{\rm branch}^{\rm 2T}}$,  are systematically higher and lower, respectively, than the relative counterparts obtained with the 2C1T model, i.e., $ {\overline N_{\rm H_{2}}^{\rm 1T}}$ and $ {\overline T_{\rm branch}^{\rm 1T}}$. The $ {\overline N_{\rm H_{2}}^{\rm 1T}}$ and $ {\overline N_{\rm H_{2}}^{\rm 2T}}$ values show a good correlation in the range $3\times 10^{20} \leq {\overline N_{\rm H_{2}}} \leq \,10^{22}$\,cm$^{-2}$, but the 2C2T results are  typically 1.94$^{+0.68}_{-0.36}$ (median, first and third quartiles of the distribution of their ratio) times higher than those obtained with the 2C1T model.
Low-density candidates  ($  {\overline N_{\rm H_{2}}}\,\le\,3\times\,10^{20}\,$cm$^{-2}$) show the largest differences between the two estimates: $  {\overline N_{\rm H_{2}}^{\rm 2T}}$ tends to concentrate towards a lower limit of $\sim\,10^{20}\,$cm$^{-2}$, while  $ {\overline N_{\rm H_{2}}^{\rm 1T}}$ continuously decreases toward lower values, finally dropping to values of the order of $\sim\,10^{19}\,$cm$^{-2}$.  
We found that $  {\overline N_{\rm H_{2}}^{\rm 1T}}\,\ge\, {\overline N_{\rm H_{2}}^{\rm 2T}}$ only for a few low-density candidates but, in these cases, the results from 2C2T model are  affected by the large uncertainties introduced at some wavelengths when the flux is separated into the two components (see below). 
The correlation breaks down in the high density regime (${\overline N_{\rm H_{2}}} \ge 10^{22}$\,cm$^{-2}$), where the results from 2C1T tend to cluster towards much lower values than for 2C2T and never reach column densities as high as ${\overline N_{\rm H_{2}}} \ge 10^{23}$\,cm$^{-2}$).  

The relation between the average central temperatures estimated along the branches, $ {\overline T_{\rm branch}}$, from the two models is shown in Fig.\,\ref{Fig:ComparisonBranchTemperatureDistribution}. We overplotted the median and the quartiles (red and green lines, respectively) of the ${\overline T_{\rm branch}^{\rm 2T}}$ estimated over bins of  ${\overline T_{\rm branch}^{\rm 1T}}$ to facilitate the visualization of the plot. We adopted bins ${\overline T_{\rm branch}^{\rm 1T}}$ which are 0.5\,K wide. The average central temperature determined from the 2C2T is generally lower than that estimated with the 2C1T model. This occurs in particular in the range between 12 and 20\,K, where the discrepancy is between 1--3\,K. The largest discrepancy is found at temperature ${\overline T_{\rm branch}^{\rm 1T}}$ of 18--19\,K, where there are even candidates where we measured ${\overline T_{\rm branch}^{\rm 2T}}$ as low as low as 10--12\,K. The temperatures estimates from the two models tend to converge for ${\overline T_{\rm branch}}\,\geq\,20$\,K, where ${\overline T_{\rm branch}^{\rm 2T}}$ only slightly exceeds ${\overline T_{\rm branch}^{\rm 1T}}$. \\

\subsubsection{Differences between the two models}
\label{Sect:Differences2C}

\begin{figure*}
\begin{center}
\includegraphics[width=14cm]{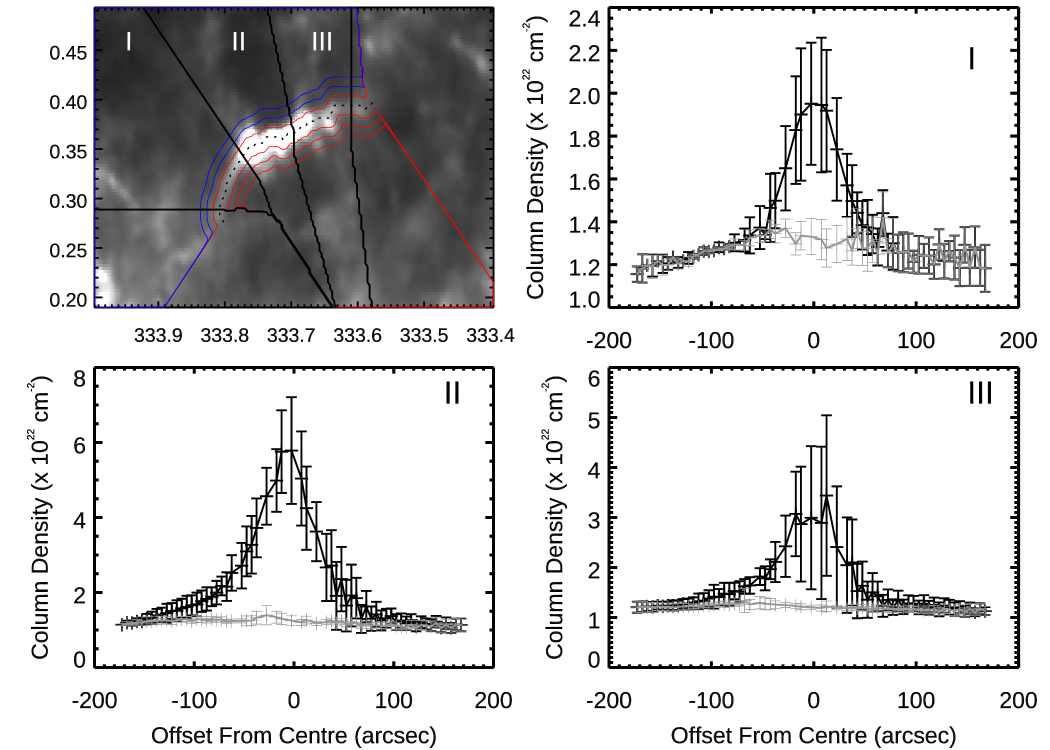}
\caption{Top left panel: Column-density map of the candidate shown in Figure\ref{Fig:ObjDefinitions}, where the region is divided into three sections delimited by thick lines, each including a few branches. The dotted line shows the position of the main spine, the coloured lines trace isodistance curves from it, with positive distances in blue and negative distances in red. Top right, bottom left and bottom right panels: Filament average column density (thick black line) and estimated background (light grey curve) radial profiles with respect to the main spine in the three sections. The column density for the filament and the background components are estimated with the 2C1T model, i.e., assuming that they share the same temperature. The plotted bar at each radial distance defines the amplitude of the $N_{\rm H_{2}}$ variation in the entire section. 
} 
\label{Fig:TwoComponentSingleTSeparation}
\end{center}
\end{figure*}

Fig.\,\ref{Fig:TwoComponentSingleTSeparation} shows the filament presented in Fig.\,\ref{Fig:ObjDefinitions}. We split the filament structure into three different sections corresponding to different groups of 2D branches, as described in Sect.\,\ref{Sect:Segmentation}. The central section, labelled as {\rm II}, represents the densest portion of the candidate filament, while the other two, {\rm I } and {\rm III}, cover low-density regions. The two sections {\rm I} and {\rm III} span a similar range of column densities, but the emission at 160\,$\mu$m in section {\rm III} is weaker than in {\rm I}, and the filamentary shape is barely detectable at this wavelength. 

The average radial column-density profiles, measured in the three sections, plus the corresponding estimated background using the 2C1T model are also shown in Fig.\,\ref{Fig:TwoComponentSingleTSeparation}. The procedure described in Sect.~\ref{Sect:2C2Tmodel} is able to properly separate the two components, filament and background, as shown by the rather regular estimated background on the filament extended mask.  This mask expands up to radial distances where the emission of the two components matches and appears to include the whole emission ascribed to the filament.

\begin{figure*}
\begin{center}
\includegraphics[width=14cm]{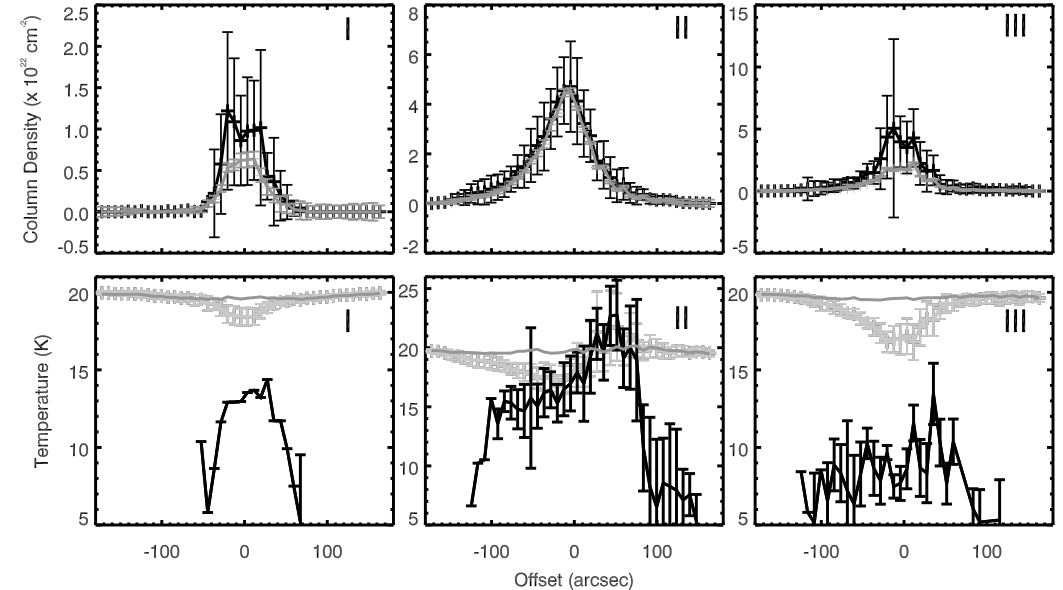}
\caption{Top three panels: Differences between the filament column-density radial profiles obtained from the 2C2T modelling, i.e., with different temperatures for the filament and background components, and 2C1T model. The filament column-density radial profiles from the 2C2T modelling are shown in black, while the same quantities from the 2C1T modelling, already presented in Figure\,\ref{Fig:TwoComponentSingleTSeparation} and from which the background component has been subtracted, are shown in light grey for comparison purposes. Results for the three sections are shown in different panels from left to right. Bottom three panels: Temperature radial profiles determined with the 2C2T modelling, relative to the filament (black) and the background (dark grey) components, compared to the same profile obtained through the 2C1T modelling and shared by both components (light grey). }
\label{Fig:TwoComponentTwoTSeparationNH2}
\end{center}
\end{figure*}

\begin{figure}
\includegraphics[width=0.45\textwidth]{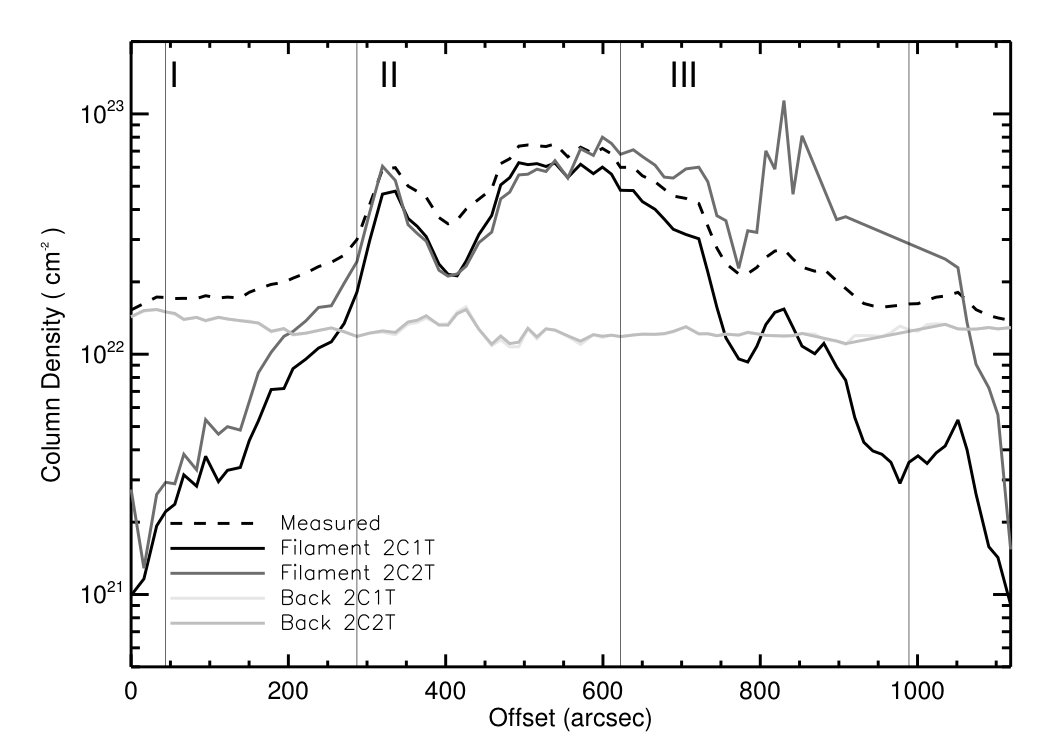}
\includegraphics[width=0.45\textwidth]{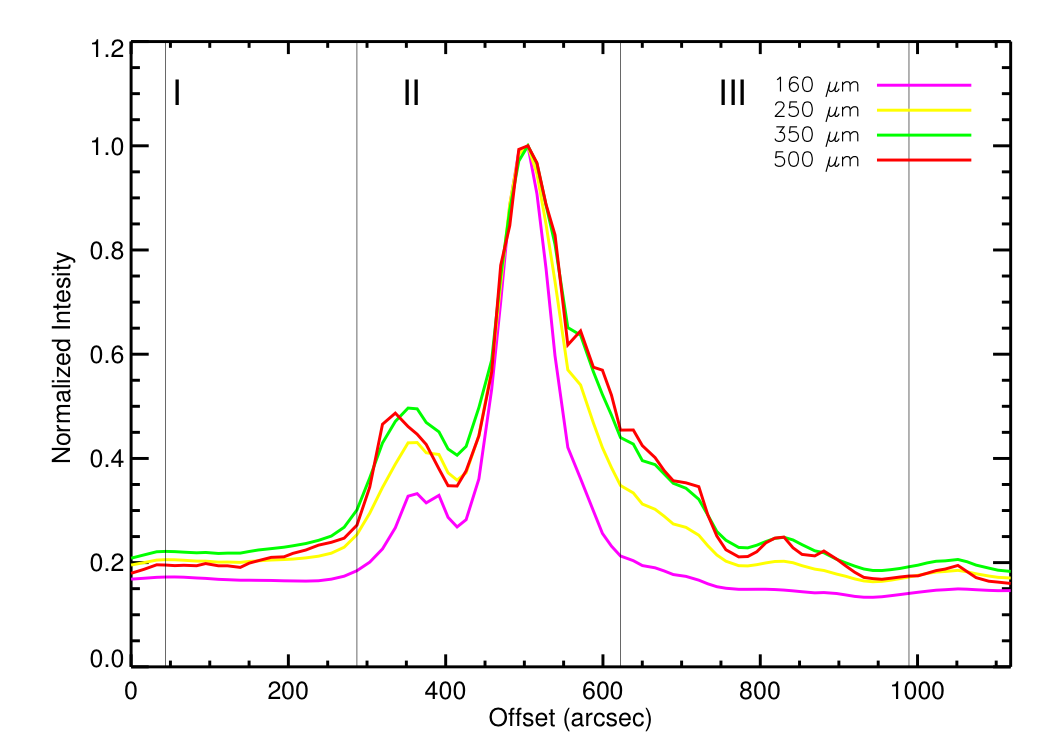}
\caption{{\it Top panel}: Column-density profile measured along the spine for the candidate shown in Figure\,\ref{Fig:TwoComponentSingleTSeparation}. The offset is measured starting from the south-eastern side and follows the spine.  The black and light grey continuous lines are column-density profiles estimated with the 2C1T model of the filament and background component, respectively.  Their sum is equal to the measured column-density profile, reported as a dashed line. The other lines show the same for the 2C2T model, relative to the filament (dark grey) and the background (light grey basically overlapping to the 2C1T case).  {\it Bottom panel: } Measured intensity profiles along the candidate filament spine at 160, 250, 350, 500\,$\mu$m. Section {\it III} shows several features detected only for $\lambda\,\geq\,250\,\mu$m data; correctly identified as density enhancements with the model 2C2T.}
\label{Fig:RidgeProfiles}
\end{figure}

The filament contribution is estimated from Eq.\,\ref{Eq:LinearNH2} and the resulting radial profiles are compared with those obtained from the 2C2T model in the top panels of Fig.\,\ref{Fig:TwoComponentTwoTSeparationNH2}. The two models give the same results in section {\rm II} and consistent results in {\rm I}, but they greatly differ in section {\rm III}. On the contrary, the column density of the background component is found  not to be affected by the specific model. Similar results are also found for the profiles along the main spine, as derived from the two models shown in the top panel of Fig.\,\ref{Fig:RidgeProfiles}. This effect can be explained in terms of different emission and estimated temperature in the three different sections, as shown in the bottom panels of Fig.\,\ref{Fig:TwoComponentTwoTSeparationNH2} and Fig.\,\ref{Fig:RidgeProfiles}.

The temperature profiles obtained with the 2C1T model show a temperature drop from about $\sim$\,20\,K measured at large radial distances, to $\sim\,16-18$\,K in the central region of the three sections. 
The temperature measured on the filament is still surprisingly close to the typical thermal temperature of the cold dust in the diffuse phase of ISM, expected to range between 17.5 and 19.5 \citep{Boulanger1996, Finkbeiner1999,Bernard2010}. Such a value is an unrealistic estimate for the temperature in the dense and shielded environment  of the filamentary molecular clouds, which is expected to be colder \citep{Stepnik2003}. On the other hand, the temperature estimated with the 2C2T model drops to more realistic and lower values: $\sim13-14$\,K in section {\rm I} and to $\sim$\,10\,K in section {\rm III}, values consistent with the measurements in  molecular clouds \citep{Stepnik2003,Pillai2006,Flagey2009,Peretto2010, Battersby2014}.

The central section {\rm II} hosts the {\rm H\,II} region IRAS+16164-4929 that warms up the filament. Therefore, the difference in temperature of the two components (filament and background) is greatly reduced and the two models are consistent, since a single temperature reproduces correctly the observed emission. We registered the largest discrepancies between the two models in section {\rm III}, where the temperature drops to a value, $T \leq 10\,$K, lower than in section {\rm I}. We verified that this low temperature is not due to issues in the separation of the two components by showing the observed intensity profiles along the filament spine in the bottom panel of Fig.~\ref{Fig:RidgeProfiles}. These profiles, normalized to their maximum, have the same shape in  sections {\rm I} and {\rm II}, independent of the wavelength. This is not found in section {\rm III}, where several features, not present at shorter wavelengths, appear at $\lambda \geq 250\,\mu$m.
The features found in section {\rm III} are high-density condensations which can effectively shield the material from the interstellar radiation field allowing the dust to cool down to the measured lower temperature, T\,$\lesssim 10\,K$. When this happens, the filament component dominates the emission at wavelengths longer than $\lambda \geq$ 250\,$\mu$m, whereas it is dimmer than the background components at shorter wavelengths $\leq 160\,\mu$m.

This discussion indicates that, in general, the 2C2T model provides a more realistic estimate of the column density and temperature of the filament, compared to the 2C1T model. 
On the other hand, we point out that the results from the 2C2T model are subject to larger errors since they require a correct estimate of the background level in four {\it Herschel} photometric bands instead of a single map. It may happen that the weakness of the filamentary emission makes such an estimate particularly difficult and uncertain, especially at  160\,$\mu$m. 
In these cases, errors in the background subtraction in some pixels produce profiles with spikes such as those observed in section {\rm III} and shown in the top panel of Fig.\,\ref{Fig:RidgeProfiles}. So, we decided to report in the catalogue the column density and temperature determined by both models. 
Better estimates for the filament component are possible, but they require a dedicated radiative transfer model \citep{Stepnik2003,Steinacker2016} that cannot be easily applied to a large dataset.

\section{Global analysis of the filament catalogue}\label{sect:global}

This section is dedicated to the analysis of the catalogue of candidate filaments. First, we discuss the Galactic distribution of the filaments (Sect.~\ref{Sect:GalacticDistribution}), then we correlate them with the catalogue of compact objects (Sect.~\ref{Sect:AssociationSources}) to determine whether there are differences between structures hosting dense condensations or not (Sect.~\ref{Sect:TenouosDenseFilaments}). More relevantly, we assign distances to the filaments hosting clumps (Sect.~\ref{Sect:Distances}), allowing us to determine physical properties of the filaments, like length (Sect.~\ref{Sect:LengthFilaments}), mass and linear density (Sect.~\ref{Sect:MassMlin}) that we discuss in relation to the Galactic structure.

\subsection{Galactic distribution}
\label{Sect:GalacticDistribution}

\begin{figure} 
\includegraphics[width=0.45\textwidth]{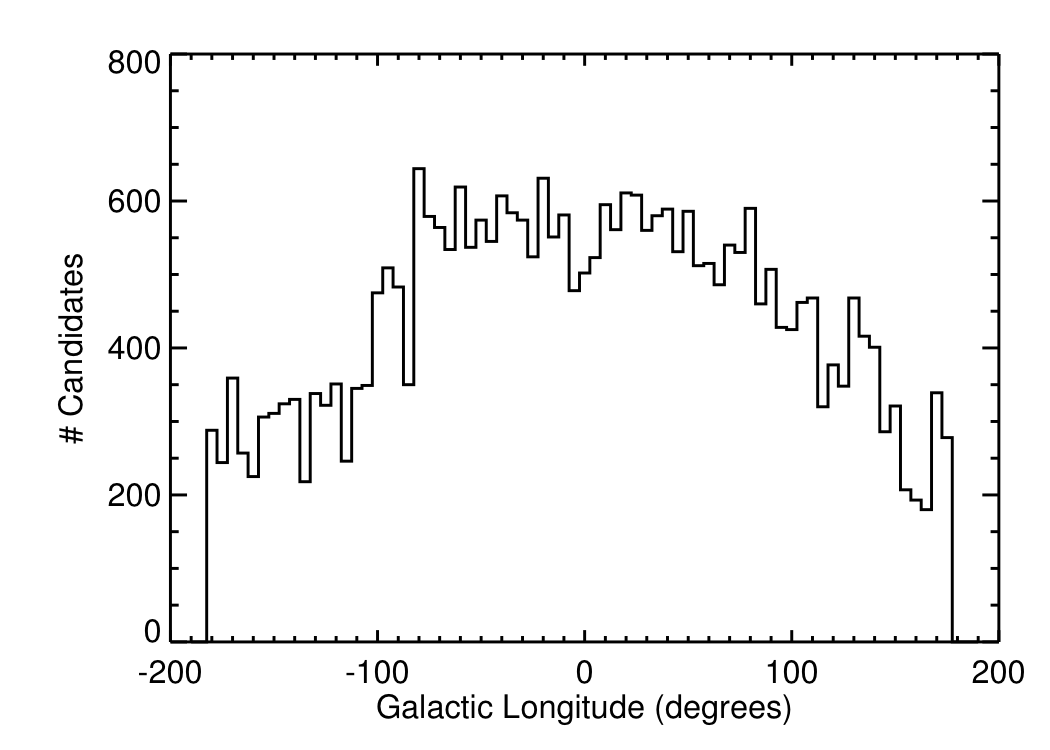}
\includegraphics[width=0.45\textwidth]{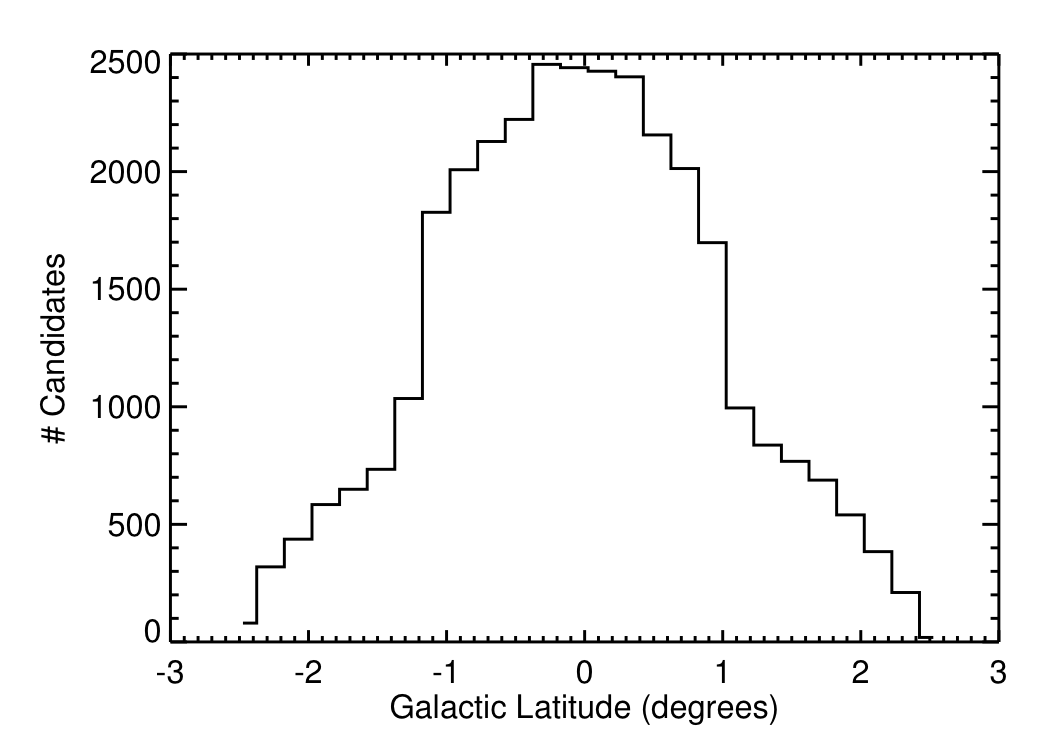}
\caption{Spatial distribution of the objects in the Hi-GAL filament catalogue as a function of Galactic longitude (top) and latitude (bottom), respectively. }
\label{Fig:GalacticDistribution}
\end{figure}

Fig.\,\ref{Fig:GalacticDistribution} shows the distribution of candidate filaments as a function of Galactic longitude (top) and latitude (bottom), respectively. The identified structures are distributed smoothly as a function of longitude, with higher density toward the inner Galaxy ($|\,l\,| \leq 70^{\circ}$), when compared with the outer Galaxy ($|\,l\,| \geq 70^{\circ}$). Also, the structures are almost uniformly distributed in the range of Galactic latitude $-0\fdg 4 \leq b \leq 0\fdg 4$, while their concentration decreases  outside this range, especially for $|\,b\,| \geq1$. Note, however, that this plot can be misleading, since the Galactic latitude is not uniformly sampled by the Hi-GAL observations, which are designed to follow the Galactic warp \citep{Molinari2010} in the outer Galaxy. On the other hand, regions with low Galactic latitude dominate the statistics, thus partially mitigating this bias.  

\begin{figure*} 
\includegraphics[width=\textwidth]{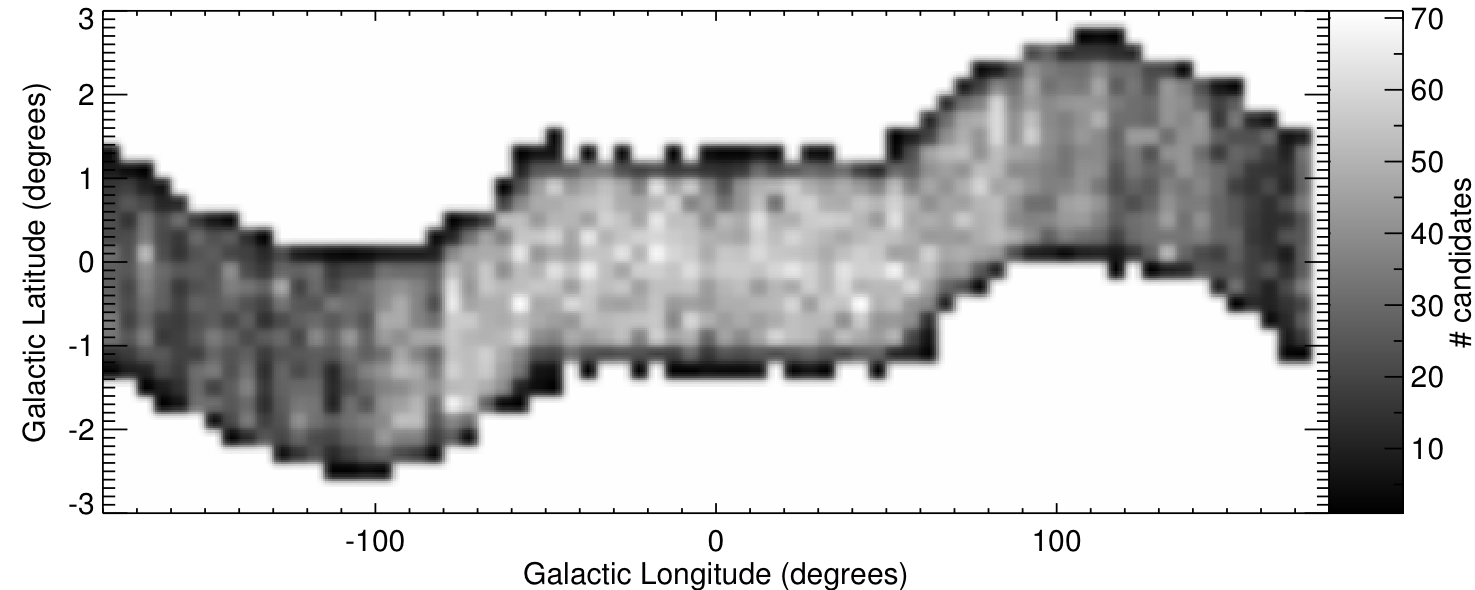}
\caption{Number density distribution of Hi-GAL filaments across the entire Galactic plane, estimated in bins of 5$^{\circ}\times\,0.2^{\circ}$. The twisted shape with the variation with Galactic latitude follows the coverage of Hi-GAL observations designed to trace the warp of the Galactic plane.  }
\label{Fig:DistributionFilaments}
\end{figure*}

Fig.\,\ref{Fig:DistributionFilaments}  shows the distribution of the candidate filaments as a function of {\it l} and {\it b},  in bins of 5$^{\circ} \times \,0.\!\!^{\circ}2$. This number density varies with the Galactic longitude from $\sim$\,60 to $\sim$25 moving outward from the inner Galaxy, but it is rather uniform with the Galactic latitude. We note that the total number of candidates directly depends on the selected threshold value which, in turn, depends on the local surroundings at each Galactic location (Sect.\,\ref{Sect:FeatureExtraction}). 
This local adaptive approach implies that the absolute threshold value decreases in less crowded regions where there are fewer fluctuations of $\lambda_{a}$, resulting in more detections that include more faint structures with less contrast. This is typically the case in the outer Galaxy, where confusion is generally lower, making it possible to identify a larger number of faint structures, which we include in our current catalogue.  

Fig.\,\ref{Fig:GalacticDistribution} shows differences in the distribution between $l \geq 0^{\circ}$ and $l\leq 0^{\circ}$. For positive $l$, covering the first and second Galactic quadrants, there is a steady decline in the number of  filaments moving toward the outer Galaxy, while for negative $l$, i.e., in the third and fourth  quadrants,  this decreasing decline is steeper, with a sharp transition in the range $-100^{\circ} \leq l \leq -80^{\circ}$. We interpret this as an effect of the Galaxy asymmetry with respect to $l = 0^{\circ}$ produced by the presence of the  spiral arms and, in particular,  of  the Local arm \citep{Xu2013} connecting the Sagittarius and Outer arms and crossing the Perseus arm. The material belonging to this arm dominates the observed features in the first quadrant in the region $55^{\circ} \leq l \leq\,72^{\circ}$, when compared to the more distant Perseus arm (distances $d\sim\,$6--8\,kpc). At larger longitudes ($l \geq 80^{\circ}$ and for the whole  second quadrant), the Perseus arm becomes the nearest major structure with $d \leq 4$\,kpc. A similar distribution is not found in the third and fourth quadrants, for $l \geq -100^{\circ}$, i.e., at the location of the Vela Molecular Ridge \citep{May1988, Vazquez2008}, where the line of sight crosses a wide inter-arm space between the Perseus and Carina-Sagittarius spiral arms up to the location  of the Carina arm tangent point  ($l \approx -70^{\circ}$). In this longitude range, the major Galactic structure is the Perseus arm, located in this case at distances of $\geq$8--9\,kpc, which may explain the measured abrupt change in the number of detections. 

\begin{figure} 
\includegraphics[width=0.45\textwidth]{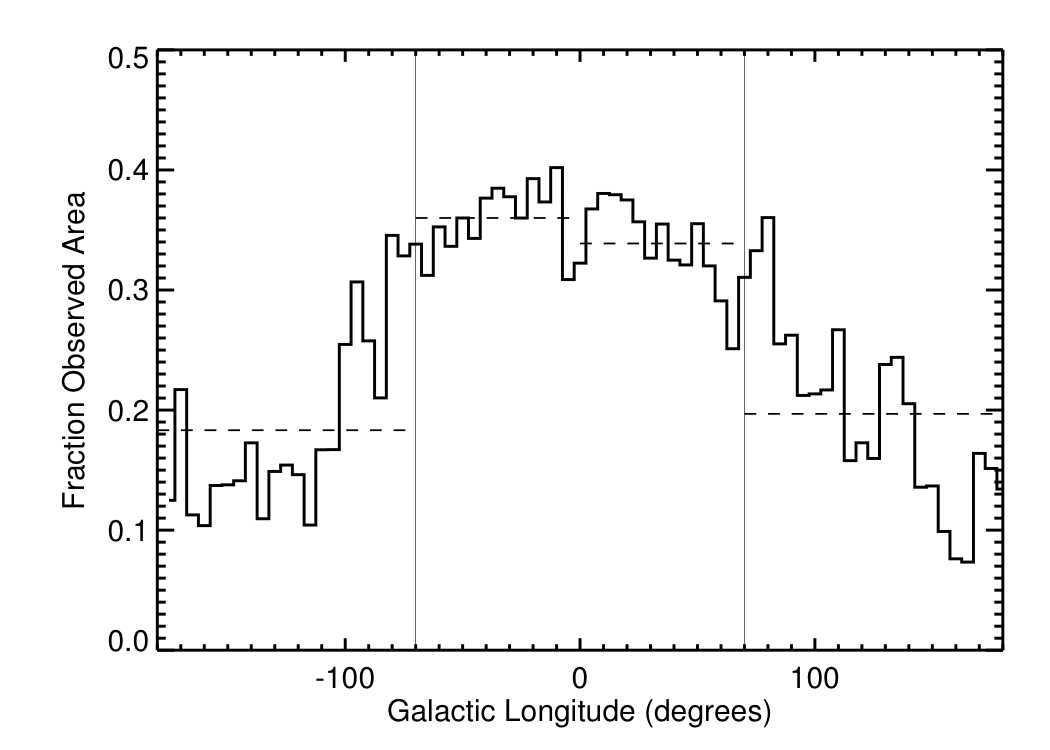}
\caption{Fraction of the observed Galactic plane belonging to candidate filaments as a function of the Galactic longitude. Vertical lines separate the inner ($|l|\,\leq\,80^{\circ}$) from the outer Galaxy ($-280^{\circ}\,\leq l\,\leq\,80^{\circ}$). Horizontal lines represent the median values in the inner (estimated separately for $l\,\geq\,10^{\circ}$ and $l\,\leq\,-10^{\circ}$) and  outer Galaxy, respectively.  }
\label{Fig:FractionArea}
\end{figure}

The distribution of candidate filaments allows to determinate how common these features are in the ISM, and thus to parametrize its ``degree of filamentarity''. The diffuse ISM is often described to be ``filamentary'', since it shows   abundant and recurrent filamentary morphologies \citep{Low1984, Schlegel1998, Miville-Deschenes2010}. A parameter called ``filamentarity'' has  already been introduced to describe the number of 1-D filaments (distribution of galaxies along linear features) forming in cosmological dark-matter simulations \citep{Barrow1985,Shandarin1998} and it has been proposed to discriminate among  cosmological models when applied to surveys of galaxies at large redshifts \citep{Dave1997}.  
Likewise, an  estimate of a similar parameter in the case of ISM observations may allow a comparison with large-scale Galactic simulations. To investigate this, here we use a simplified approach where we estimate the fraction of the observed area of the Galactic plane associated with our sample of candidate filaments. 
This fraction is plotted in Fig.~\ref{Fig:FractionArea} as a function of the Galactic longitude and one can see that it varies by a factor of two, changing from $\sim\,$34--36 per cent in the inner Galaxy ($|\,l\,| \leq 80^{\circ}$), to $\sim\,$18--19 per cent in the outer Galaxy. A larger fraction was  indeed expected in the inner Galaxy, due to a more likely overlap of different components along the line of sight, which may both increase the total number density of physically coherent filaments and creates apparent structures in the 2D maps due to projection artefacts. On the other hand, the effective area fraction in the outer Galaxy is influenced by the peaks at $\sim-95^{\circ}$ and $\sim 80^{\circ}$, caused by the presence of the Local arm/Vela spur and Cygnus star-forming regions, respectively. These two complexes  are close to the Sun $d\,\lesssim\,1-2$\,kpc and extend over a few degrees on the Galactic plane. $Herschel$  was able to easily resolve the  substructures of these two regions, so we found a large number of detections.
The average fraction of the area in the outer Galaxy covered by filaments  drops to $\sim$\,12--14 per cent when we exclude these two nearby regions, less than half the fraction found in the inner Galaxy.

These numbers suggest that the degree of ``filamentarity'' of our Galaxy, defined as the fraction covered by filaments, is  $\sim 15-40$ per cent. Therefore, despite filamentary regions appears to be ubiquitous, there is still a considerable fraction of the emission associated to diffuse and non-filamentary features.
 
\subsection{Association with compact sources}
\label{Sect:AssociationSources}

Filaments are currently considered the places where star formation  preferentially occurs \citep{Andre2014,Schisano2014}. The large catalogue presented here allows to study statistically the relation between filamentary morphologies and star formation by relating filament properties to the ones of the hosted star-forming objects, i.e., compact sources in early evolutionary phases. We present here the association between these two types of structures, discussing the related statistic and adopting it to assign distances to filaments. We  defer the analysis of the relation between filament and clump properties to a future work.

Several studies have been dedicated to find and characterize young and compact (point-like or poorly resolved) sources in extended portions of the Galactic plane \citep{Elia2013, Elia2017,  Contreras2013, Lumsden2013, Traficante2015, Gutermuth2015, Molinari2016, Urquhart2014, Urquhart2018}: their results are suitable for the cross-matching with our filament catalogue.  Here, we choose to compare with the full Hi-GAL compact-source catalogue, which is currently the largest available catalogue of FIR/submm sources. This catalogue covers the entire Galactic plane extending the work over the inner Galaxy ($-71^{\circ} \leq l \leq 67^{\circ}$), done by \cite{Molinari2016} and  \cite{Elia2017}, dedicated to the photometric detection and physical characterization of compact sources respectively. The full Hi-GAL compact-source catalogue contains a total of 150,223 sources, including the 100,922 objects already presented in \citet{Elia2017}.  The detection, photometry and physical characterization of these sources is described in detail in Molinari et al. (in preparation) and Elia et al. (in preparation). The objects listed in this catalogue are detected in at least three consecutive {\it Herschel} bands, ensuring  a robust reliability. This do not exclude that some of these objects could be portion of an underlying filament whose emission is split up into multiple pieces. We do not take into account this possibility, postponing its analysis to the future work focused on a statistical comparison of filaments and compact-source  properties.

The match between filaments and sources is done by associating to each candidate filament all the sources whose centroids fall within the filament boundaries, traced by the extended mask contour (see Sect.\,\ref{Sect:Algorithm}). As a result, we identified $78,815$ compact objects located in the area ascribed to filament candidates. This means that slightly more than half ($\sim\,52$) per cent of the total) of the Hi-GAL source population is angularly correlated to filamentary structures. If the distribution of compact sources would be completely unrelated from the filaments one, the associated sources would be about $\sim20-35\,$\% of the entire sample, since it would only depends on  the fraction of the observed area ascribed to filaments, see Fig.\ref{Fig:FractionArea}. The measured fraction instead suggests that there is a link between these two type of structures.
Not all filaments are associated with compact sources: in fact, $10,660$ regions (i.e., $\sim\,33\,$per cent of the total) have no associations, compared to $21,399$ objects ($\sim\,67$ per cent) containing at least one compact source. The distribution of the number of associations, represented by the grey line in the top right panel of Fig.~\ref{Fig:CumulativeAssociation}, shows a large spread, reaching values as high as $\sim$80 associations (not shown in the figure). The average number of compact sources per candidate filament is $\sim$4.1. It is very common to find features associated with only one or two sources: they are $10,714$ cases, i.e. 50 per cent of the sample of objects hosting sources, that represent a substantial fraction of the $16,966$ filaments hosting $\leq\,5$ sources number. Filaments with multiple sources ($\geq\,6)$ are rare with only $4,433$ cases.

Projection effects influence the results from the angular association described above. The associated sources include objects located at different heliocentric distances, that are aligned along the line of sight of a filamentary cloud. In order to mitigate this effect, we refined the association between filaments and sources, using the radial velocity measurements (RV) and the associated kinematic distance estimates available for several compact sources, see Sect.\,\ref{Sect:Distances}.
For each filament, we first determined the median RV, $\overline{v_{s}}$, of all the initially associated sources, and selected the subgroup with RVs within one median absolute deviation from $\overline{v_{s}}$. We skimmed the sources based on their RVs instead of the distances, since they are  independent from the assumed Galactic rotation curve. We also favoured the median absolute deviation than the standard deviation since it is more resilient against outlying values. The resulting subgroups are composed by the sources confined in a narrow velocity interval around the median.  However, sources with compatible RVs might still lie at two different locations in the Galaxy, since the lines of sight inside the Solar circle  are affected by the near/far ambiguity in the kinematic distances (KDA) \citep{Russeil2011}. In each filament where this may happen, we verified which distance solution between the near and far has been adopted for the majority of the sources, and we selected the corresponding subgroup. In short, the robust association is composed by all the sources that have a compatible RVs and a similar distance choice.  
The criteria described above cannot be applied to candidates hosting two or fewer sources, where we were forced to retain the results from the angular association.

The resulting distribution from the robust association is shown as a black line in the top  panel of Fig.~\ref{Fig:CumulativeAssociation}. The association fraction decreases with respect to the case of simple angular matching: in total, there are $61,176$ compact sources  associated with filaments, equal to $\sim48$ per cent of the $128,326$ Hi-GAL sources with a RV estimate, see Sect.~\ref{Sect:Distances}.  The number of filaments associated to at least one compact source, the average number of associations,
and the number of filaments with multiple ($\geq 6$) sources drop to $18,389$ ($\sim57$ per cent of the entire sample), $\sim3.2$, and $2,6181$ filaments respectively. The drop is mostly due to the fact we are referring to a smaller sample than the entire Hi-GAL catalogue, but the results are still consistent with the ones from the angular association.

These two association criteria are the two extreme cases that can be considered. On the one hand, the simple angular association is a very loose criterion strongly influenced by line-of-sight projections. On the other, the criteria for the robust association are the most restrictive possible with the currently available data. The outcome of the robust association is influenced by several effects such as the existence or not of a RV estimate, the tracers adopted for RV measurement, how RV is assigned to a compact source, how the KDA is solved, etc., see Sect.\,\ref{Sect:Distances}. 
All these possibilities indicate that the robust association can miss some compact sources; therefore, the reported estimates for the fraction of filaments with sources and the average number of associations should be considered as lower and upper limits. 

\subsubsection{Are filaments chains of sources?}

\begin{figure} 
\includegraphics[width=0.48\textwidth]{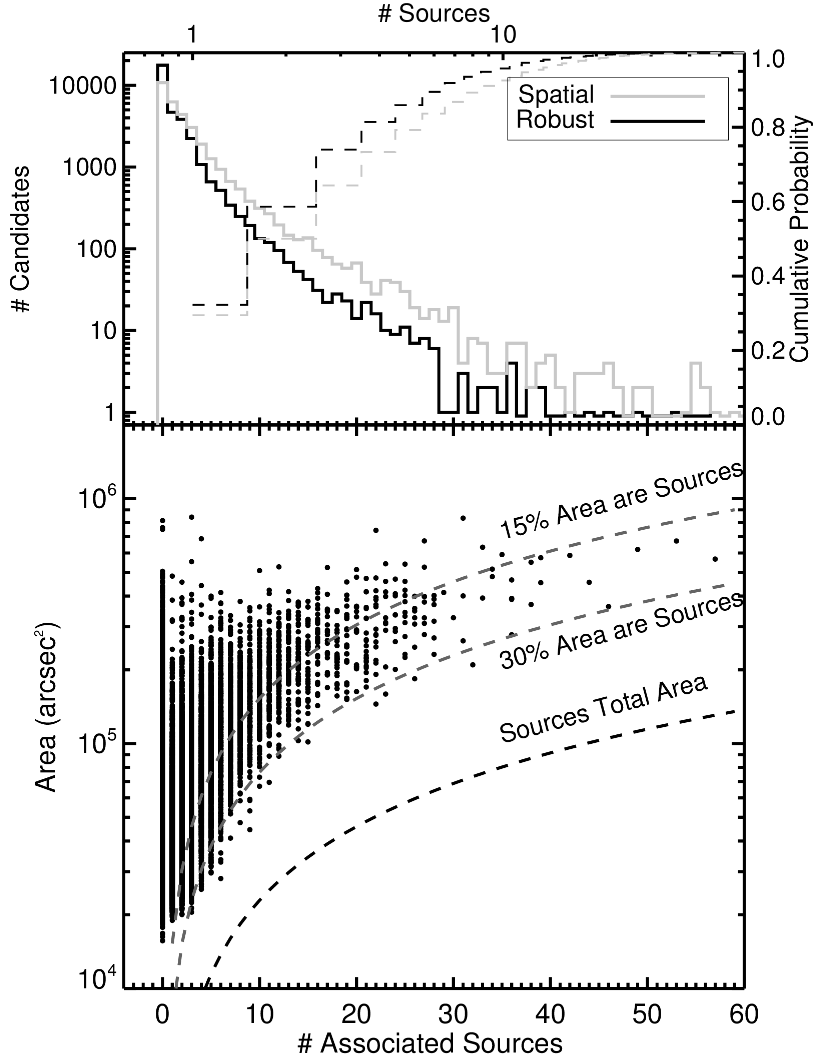}
\caption{  {\it Top panel}:  Candidate filament distribution in terms of the number of associated compact sources, assuming the simple spatial (light grey line) and robust (black line) associations. The cumulative distribution functions related to these histograms are drawn with dashed lines and refer to the axis scale at right and top (the logarithmic axis scale is assumed along the {\it x-axis} to highlight the behaviour where there is a low number of associated sources). The majority of filamentary candidates have a number of associated compact sources less than 5.  {\it Bottom panel }: Candidate filament area as a function of the number of objects present in the full Hi-GAL band-merged catalogue that fall within the candidate borders assuming the robust association criteria defined in the text. The dashed lines indicate an estimate of the total area covered by the sources (black dashed line) and the expected total area if sources cover only a fraction of the candidate area (the two cases of 15 per cent and 30 per cent are shown with a grey dashed line).  
}
\label{Fig:CumulativeAssociation}
\end{figure}

The features detected in the Hi-GAL column-density maps may be made up by groups of discrete sources aligned as chains along a main direction and mimicking the shape of an elongated filament.  To rule out this possibility, we estimated the area covered by the associated {\it Herschel} compact sources and compared with the area of our features.
The bottom panel of Fig.~\ref{Fig:CumulativeAssociation} shows the number of matched compact sources in relation of the area of the hosting filament. Structures that cover a larger area are associated with a larger number of sources. We computed the area covered by sources hosted in each filament, assuming that they are represented as non-overlapping discs with a diameter of $54 \arcsec$, derived from the modal value of the circularized sizes of the sources in the Hi-GAL catalogue at 500\,$\mu$m \citep{Molinari2016}. 
We found that candidate filaments in our catalogue always extend over a larger area than that covered only by the associated compact sources (black dashed line in the bottom panel of Fig.~\ref{Fig:CumulativeAssociation}). The filament areas are more extended than the total compact-source areas by a factor $\geq$3, as indicated by the grey dashed lines in Fig.~\ref{Fig:CumulativeAssociation} that represent the expected area that would have filaments if their associated compact sources cover a fraction of $15$ and $30$ per cent.
We conclude that most of the surface area ascribed to our candidates belongs to an underlying, more extended structure, i.e., the filament itself.
We do not find any filament consisting solely of strings of compact objects.

\subsection{Tenuous vs dense filaments }
\label{Sect:TenouosDenseFilaments}

\begin{figure}

\includegraphics[width=0.45\textwidth]{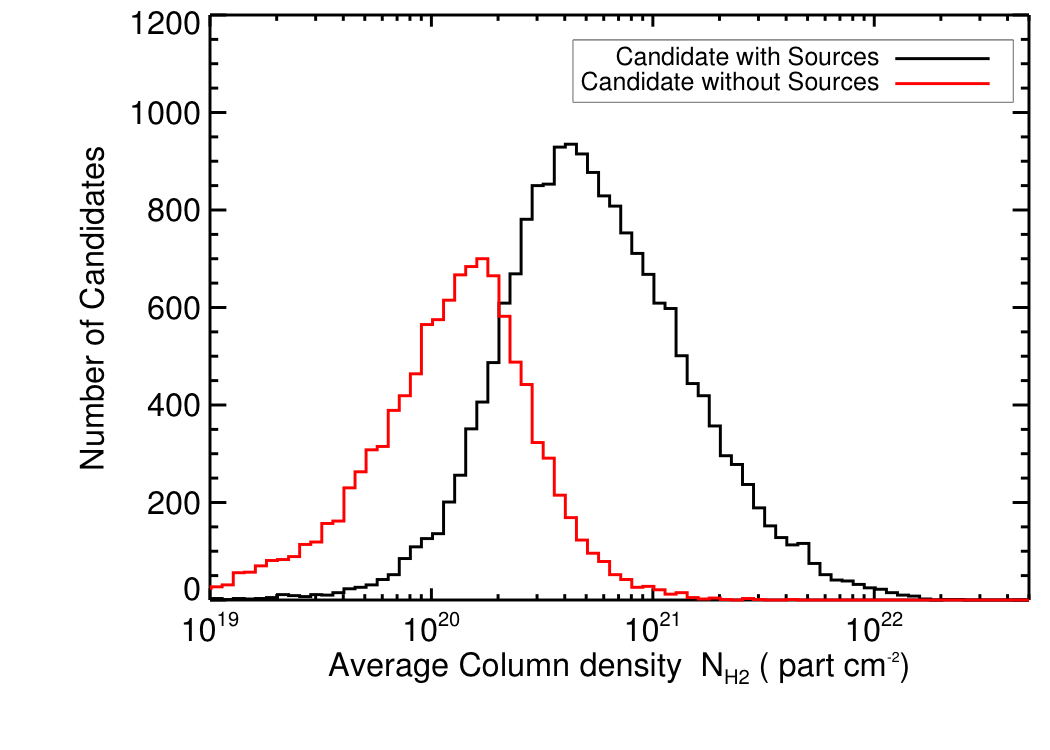}
\caption{Histograms of the average column density determined with the 2C2T model of the filamentary candidates with associated compact sources (black line) and without any association (red line).}
\label{Fig:HistogramColumnDensitiesSources}

\end{figure}

The identified filaments are split into two groups depending on whether there is an association with a compact source or not. In Figure\,\ref{Fig:HistogramColumnDensitiesSources} we show the distributions of the average column density estimated from the model 2C2T (see Sect.\,\ref{Sect:2C2Tmodel}), for those features associated with a clump and those that are not. There is a clear difference between the two samples: the filaments associated with compact sources are generally denser, with a typical average column density, $N_{\rm H_{2}}$, of $\sim8\times10^{20}$\,cm$^{-2}$, higher than $\sim0.5\times10^{20}$\,cm$^{-2}$, the mean of the sample without any association. 

This is indicative of the existence of two families of filamentary structures, one denser than the other. However, we notice that there is not a simple distinction between these two categories. 
In fact, filamentary structures with column densities in the  range $\sim4\times10^{19}$\,cm$^{-2}$\,$\leq\,N_{\rm H_{2}}\,\leq\,\sim2\times10^{21}$\,cm$^{-2}$ might belong to one or the other family. We note that by increasing the average column density $N_{\rm H_{2}}$ it is more probable to find a compact source associated with any filament. This result is partially biased by the fact that filaments hosting sources  should have larger average $N_{\rm H_{2}}$, caused by the presence of the sources within their boundaries. The associated sources are extracted from the Hi-GAL catalogue, so they are certainly detected at sub-mm wavelengths and are substantial overdensities with respect their surroundings \citep{Konyves2015}. 
On the other hand, we found above that they cover only a limited portion ($\leq$\,15 per cent) of the filament surface, so their impact on the average $N_{\rm H_{2}}$ should be minor.

Tenuous, low-density, non-self-gravitating filaments were already observed in translucent clouds \citep{Falgarone2001,HilyBlant2007,Andre2010}. These structures are also found in simulations, where they are preferentially aligned with the turbulent strain. This fact suggests that they are generated by the stretch induced by turbulence \citep{Hennebelle2013} or by the Galactic shear \citep{DuarteCabral2016}. In these works, star formations starts only when the filament density increases, possibly due a progressive stockpiling of material from the parent cloud, so gravity takes over. Anyway, we point out that our results indicate the presence of compact sources also in low-density structures. Even if it is still possible that our association includes mismatches (see discussion on the limits of our association in Sect.\,\ref{Sect:AssociationSources}), it is very unlikely that all the low-density features with sources derive  from projection effects along the line of sight. There are already several works reporting condensations detected on filaments that should not be dense enough to form cores and clumps \citep{Falgarone2001,Benedettini2015,HilyBlant2007,Hernandez2011}. These sources cannot be the result of filament fragmentation, therefore it is possible that the density, or the mass per unit length,  of the entire filamentary cloud might not be the only parameter governing the star formation. However, this result requires a more extensive analysis that should take into account the aforementioned uncertainty in the nature of the compact sources, some of which might reveal as spurious fragmentation of the filament emission. We leave this discussion to a future study, while here we focus on the on the ensemble properties of all the filaments in the Galaxy.

\subsection{Distances}
\label{Sect:Distances}

Distance estimates are fundamental to translate the measured geometric and photometric quantities into physical parameters like lengths and masses \citep{Heyer2015}.  A widespread method to estimate distances in the Milky Way relies on the gas kinematics. It adopts RV measurements and translate them into a heliocentric distance through a Galactic rotation model \citep{RomanDuval2009,Russeil2011, EllsworthBowers2013, Urquhart2014}. We used the RVs and the associated kinematic distance estimates available for the compact sources in the full HI-GAL catalogue (Mege et al., in preparation) to assign heliocentric distances, $d$, to the filaments in our catalogue. In total, we have these quantities available for $128,326$ compact sources spread almost uniformly over the entire GP, refining the results presented already in \cite{Elia2017} for $57,065$ clumps. 
This large dataset of RVs is measured from the data of all the major surveys of the GP available (Mege et al. in preparation). Most RVs are measured from $^{12}$CO and $^{13}$CO datacube from the Galactic Ring Survey \citep[GRS,][]{Jackson2006}, the Exeter-FCRAO Survey \citep{Brunt2003}, the MOPRA Galactic survey \citep{Burton2013}, ThrUMMS \citep{Barnes2015}, CHIMPS \citep{Rigby2016}, SEDIGISM \citep{Schuller2017}, NANTEN \citep{Onishi2005}, and the Forgotten Quadrant Survey (FQS, Benedettini et al., submitted). The results from CO were complemented with those from other molecular species, generally dense-gas tracers, from the surveys CHAMPS \citep{Barnes2011}, HOPS \citep{Walsh2011}, and MALT90 \citep{Jackson2013}, but the number cases where RV is confirmed by these dense-gas tracers is still limited.  Most of the distances associated to the compact sources are derived from the RV measurements by adopting the revised Galactic rotation curve presented by \citet{Russeil2017}, but in some cases they have been assigned through different criteria, like for example the spatial association with objects with an already known distance \citep{Russeil2011}.

Simulations have shown that rotation curve is very uncertain for objects inside the Galactic co-rotation radius, $R_{Gal}\,\lesssim\,4.5-5$\,kpc, where there is the strong influence of massive asymmetric structures presents in the central region of the Milky Way \citep{Chemin2015}. On the other hand, the adopted rotation curve of \citep{Russeil2017} is well constrained by data only for Galactocentric distances $R_{Gal}\,\leq\,22$\,kpc. Then we flagged any filament with $R_{Gal}\,\lesssim\,5$\,kpc and $R_{Gal}\,\gtrsim \,22$\,kpc, where the estimated kinematic distance might be affected by particularly large errors. 

We adopted the robust association to assign a distance estimate to  candidate filaments hosting compact sources (see Sect.~\ref{Sect:AssociationSources}). We assumed as filament distance the average of the associated source, paying attention to the cases affected by the KDA uncertainty as discussed in Sect.~\ref{Sect:AssociationSources}. We were able to assign distances to $18,389$ candidate filaments. We identified and flagged $1,528$ of these filaments matching with compact sources whose RVs exceed the expected tangent point velocity from the assumed Galactic rotation curve. We  assigned to these cases the distances derived from the tangent point velocities \citep{Russeil2011}, but we consider them highly uncertain. We further report that in $1,877$ cases the assigned distance is not derived from the Galactic curve rotation, but assigned from distance estimates of the sources obtained by other criteria, see Mege et al. in prep.

\subsection{Filaments and Galactic structure}
\label{Sect:GalacticStructureProbabilisticApproach}

\begin{figure*}

\includegraphics[width=0.90\textwidth]{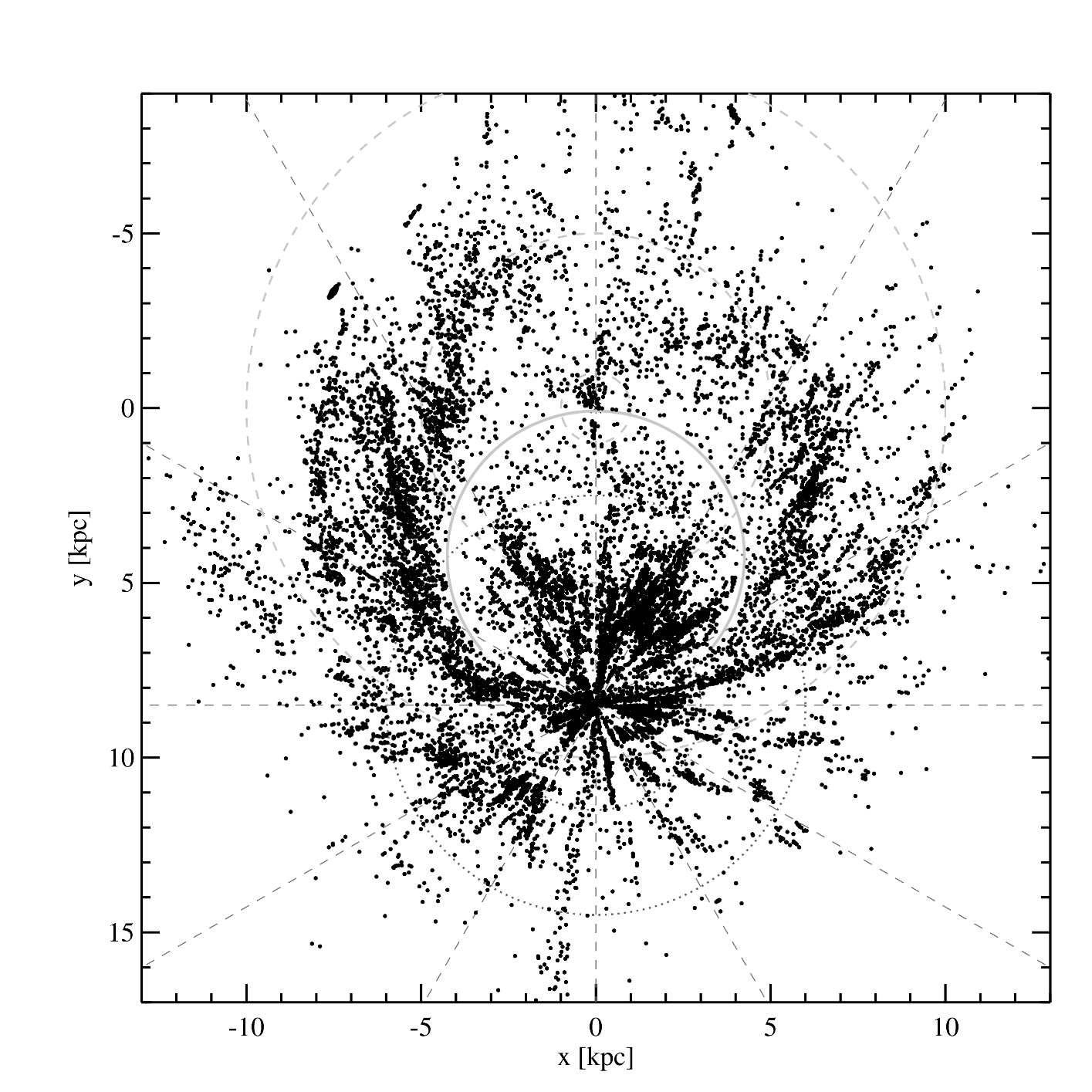}
\caption{Top-down view of Milky Way with black dots indicating  the location of $16,861$ candidate filaments with a distance estimate. The dashed lines indicate line of sights separated by 30 degrees in longitude. The concentric circles refer to heliocentric distances  equal to 1, 3, 6\,kpc and to galactocentric distances of 1, 5, 10\,kpc, with dotted and dashed lines respectively. The grey thick line identifies the position in the Galaxy where are located all the $1,528$ filaments with RVs reported to their tangent point velocity, see text.
}
\label{Fig:GalacticDistances}
\end{figure*}

The spatial distribution of filaments in the Galaxy is shown in Fig.~\ref{Fig:GalacticDistances}, where we plot the filaments with an assigned distance, including the objects with an uncertain distance located in the central region of the Galaxy at $R_{Gal}\,\leq\,4.5$\,kpc. The objects assigned to the tangent point distance are  not displayed in Fig.~\ref{Fig:GalacticDistances}, but are located along the grey arc. Fig.~\ref{Fig:GalacticDistances} shows that filaments are found to be spread all across the Galaxy. Despite in some regions there are a higher number of filaments, the filamnet distribution is rather contiguous across the Galaxy and agrees qualitatively with large-scale simulations \citep{Dobbs2006, Smith2014}. Therefore, we expect to find filamentary clouds lying close to or on a Galactic spiral arm, but also in a large number in the inter-arm space, as observed also in the simulations. The simulations predict that there is no noticeable differences between features located in arm and inter-arm environments \citep{DuarteCabral2016}. To test this prediction we associate our filament sample to the large-scale Galactic structure, issue that is severely limited by the uncertainties on the kinematic distances and/or by the spiral arm positions. Indeed, while it is feasible to infer to some extend the global Galactic structure from kinematic distances \citep{Gomez2006, Baba2009, Chemin2015}, simulations suggest that the derived location of spiral arms and of inter-arm regions can be distorted considerably with respect to their real position \citep{RamonFox2018}. Nevertheless, we attempted to define subsamples representative of arm and inter-arm regions, estimating an association probability to these Galactic regions for each object with RV measurements. To such aim, we determine for each filament a probability distribution for its location in the Galaxy that we compared with an assumed Galactic structure. 
 
\subsubsection{Uncertainties on kinematic distances}
\label{Sec:UncertainDistances}

The probability distribution of the filament positions depends on the uncertainties on the kinematic distances. These can be ascribed to three different sources of errors: uncertainties due to the association of the Hi-GAL clumps with the extended filamentary features, uncertainties on the estimate of the correct RV and, finally, uncertainties on the relation between RV and distance due from the rotation curve, mainly any departure from the assumed symmetrical shape and from the circular motions of molecular clouds. 

The robust association has been defined to limit the chance of mis-association between sources and filaments. In each filament the associated sources are spread in a narrow interval: the median of  standard deviation of the associated RVs is  $\sim0.4$\,km\,s$^{-1}$, that increases to  $1.7$\,km\,s$^{-1}$ when we restrict to filaments with more than 3 associated sources. 

Kinematic distances estimated with the tangent-point method are derived from the measurement of the velocity vector of a cloud in circular motion around the Galactic centre. The measured RV is assumed to be the projection of the circular velocity along the line of sight. The centroid velocity of the spectral line is generally determined with high accuracy, thanks to the spectral resolution of the recent surveys ($\sim0.1-0.2$\,km\,s$^{-1}$), but is limited by the gas velocity dispersion, depending on  adopted tracer.  Most of the available RVs are measured from CO and $^{13}$CO, where the typical cloud velocity dispersions are about $\sim3$\,km\,s$^{-1}$ \citep{RomanDuval2009}.
A further limitation, that arises by adopting low-density tracers in the GP, is the confusion produced by multiple emission components along the line of sight \citep{Russeil2011}. 
The solution adopted for Hi-GAL compact-sources was to selected the brightest component showing  similar morphology in the integrated CO  intensity map and in dust continuum Hi-GAL maps (Mege et al. in prep). This solution deal properly with the multiplicity of the emission components, but do not completely exclude cases where an uncorrected RV is assigned to a compact source and, from there, to a filament. RVs estimated from high-density tracers are generally more reliable since they have smaller velocity dispersions and lower probability of mis-association. In fact, NH$_{3}$, N$_{2}$H$^{+}$, and CS data have typical velocity dispersions of  $\sim1$\,km\,s$^{-1}$ \citep{Wienen2015}. The high density tracers allow average errors of the order of $\sim0.3$\,kpc \citep{Urquhart2018} when used with new Bayesian distance algorithms such as the one developed by \citet{Reid2016}. Nevertheless,  these data are available for a limited number of sources and the errors still depend on where the object is located.

The largest uncertainty on the kinematic distances are due to the departure from circular motion. Local streaming motions and velocity perturbations influence the measured RV with respect to the  velocity field expected from the Galactic rotation. For example, gravitational perturbations induced by massive star-forming complexes and spiral arms alter significantly the measured RV \citep{Baba2009}. Different prescriptions are trying to include these effects when estimating kinematic distances  \citep{Brand1993, Reid2009, Anderson2012, Wienen2015}, but the effective amplitudes of the streaming motions are still uncertain and possibly vary throughout the Galaxy.
These amplitudes have been estimated to be $\pm3-6$\,km\,s$^{-1}$ nearby the spiral arms \citep{Reid2012,Xu2013,Wu2014}. On the other hand, \citet{Anderson2012} and \citet{Wienen2015} quoted larger values of about $\pm7-8$\,km\,s$^{-1}$ to take into account the entire gravitational perturbations from arms and massive complexes. \citet{RomanDuval2009} assumed a maximum perturbation  of $\pm15$\,km\,s$^{-1}$ on the measured RVs towards the inner Galaxy.  Similar amplitudes have been determined also from observations by \citet{Reid2014} and from simulations by \citet{RamonFox2018}. These uncertainties on RVs introduce errors on the estimated distances that can be as large as $\sim1$\,kpc \citep{RomanDuval2009,Urquhart2018,RamonFox2018}.

\subsubsection{Association with Galactic structure}
\label{Sec:GalacticStructure}

We are not able to precisely locate filaments in the Galaxy due to the effects described in Sect.~\ref{Sec:UncertainDistances}. However, we can estimate for each object the distribution of its positions, $D^{fil}_{h}$, that are compatible with the observed RV. We computed the normalized $D^{fil}_{h}$ for each filament by generating a synthetic sample of $1,000$ RVs spread uniformly in the interval $\overline{v_{s}}\pm\sigma_{pec}$, with $\sigma_{pec}$  the amplitude of the peculiar motions present in the sample, and by determining the corresponding heliocentric distances from the Galactic rotation curve of \citet{Russeil2017}.
We assumed the same average velocity uncertainty  $\sigma_{pec}\,=\,\pm10$\,km\,s$^{-1}$ for all the filaments in our sample. We were able to proceed in this way for $11,643$ filaments that $a)$ have 4.5$\,\leq\,R_{Gal}\,\leq\,22\,$kpc, $b)$ have a $\overline{v_{s}}$ that does not exceed the tangent point terminal velocity, $c)$ do not exceeding the same terminal velocity when we sampled uniformly the interval $\overline{v_{s}}\pm\sigma_{pec}$.

\begin{figure*}

\includegraphics[width=0.90\textwidth]{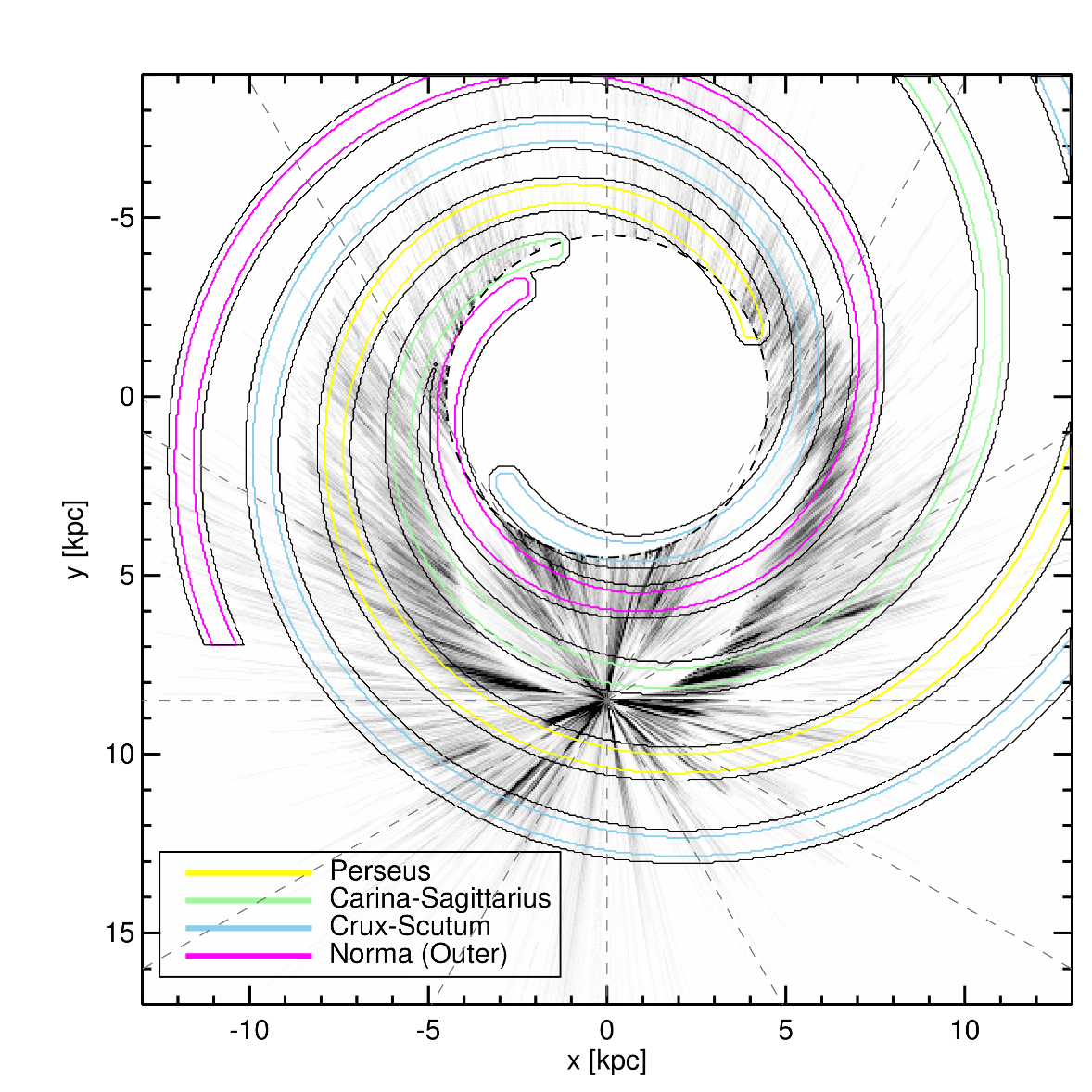}
\caption{ Top-down view of the overall density distribution of a subsample of $11,643$ filaments in the Galaxy for which it was possible to determine the probability distribution, $D^{fil}_{h}$, from their RV and the relative uncertainty. Each $D^{fil}_{h}$ is given from the heliocentric distances derived from $1,000$ synthetic RVs  obtained from an uniform sampling of the velocity interval $\overline{v_{s}}\pm10$\,km\,s$^{-1}$, with $\overline{v_{s}}$ equal to the RV assigned initially to the filament. The overall density distribution is compared with the main spiral arms in the Galaxy traced from the four-arm Milky Way prescription of \citet{Hou2009} shown with different colors. These arms are assumed to have two different sizes a full-width of $600$\,pc and $1$\,kpc, draw with coloured and black line respectively. Filaments located at $R_{Gal}\,\leq\,4.5$\,kpc are excluded from this plot due to the unreliability of the kinematic distances in central region of the Galaxy. 
} 

\label{Fig:DensityProbabilityDistributionFilaments}
\end{figure*}

The sum of the normalized distributions of the positions of all the $11,643$ filaments is shown in  Fig.~\ref{Fig:DensityProbabilityDistributionFilaments} with respect to the position of the main spiral arms derived from \cite{Hou2009}.  The probability of association of a filament to a spiral arm, $P_{sp}$, is given by the intersection between $D^{fil}_{h}$ and the region of the arms. The estimated probabilities, $P_{sp}$, depend on the adopted prescription for the Galactic structure, here  we adopted the spiral arms location  from \citet{Hou2009} and we assumed two different arm widths: a full-width of $600$\,pc, namely the upper limit  reported by \citet{Reid2014}, and 1\,kpc adopted in the study of \citet{Eden2013}. We exclude from the analysis the Local arm \citep{Xu2013}, a minor feature located between the Carina-Sagittarius and Perseus arms, since the uncertainties on the kinematic distances for nearby clouds are so large that it is not possible any clear association.

The overall probability distribution in Fig.~\ref{Fig:DensityProbabilityDistributionFilaments} suggests that the filaments in our catalogue fall preferentially in the inter-arm regions. These objects are indeed expected to be detected more easily than the ones located on the dense spiral arms due to their tenuous surroundings. We define as representative of the filaments lying on spiral arms and in the inter-arm regions the objects with  with $P_{sp}\,\geq\,0.8$ and $P_{sp}\,\leq\,0.2$ respectively. These subsets are composed by $1,178$ and $5,261$ objects, if we assume the width of the arms equal to $W\,=\,600$\,pc. Their relative ratio change significantly when we consider wider arms with $W\,=\,1$\,kpc, and we count $2,934$ and $3,168$ filaments in the two subsets respectively. We note that about $\sim5,000$ of filaments with associated distances have $0.2\,\leq\,P_{sp}\,\leq\,0.8$. We exclude these features from the two subsets adopted the comparison between arms and inter-arm features as we are not able to ascertain their association.

We found no substantial differences between the physical properties of filaments associated to arm or inter-arm regions. The distributions of average column densities and temperature  are similar in the two subsets, confirmed from the results of Kolmogorov-Smirnov (K-S) tests. We discuss the lengths and masses in the following, since they depend on the adopted distance. 

\subsection{Lengths of filaments }
\label{Sect:LengthFilaments}

\begin{figure}
\includegraphics[width=0.45\textwidth]{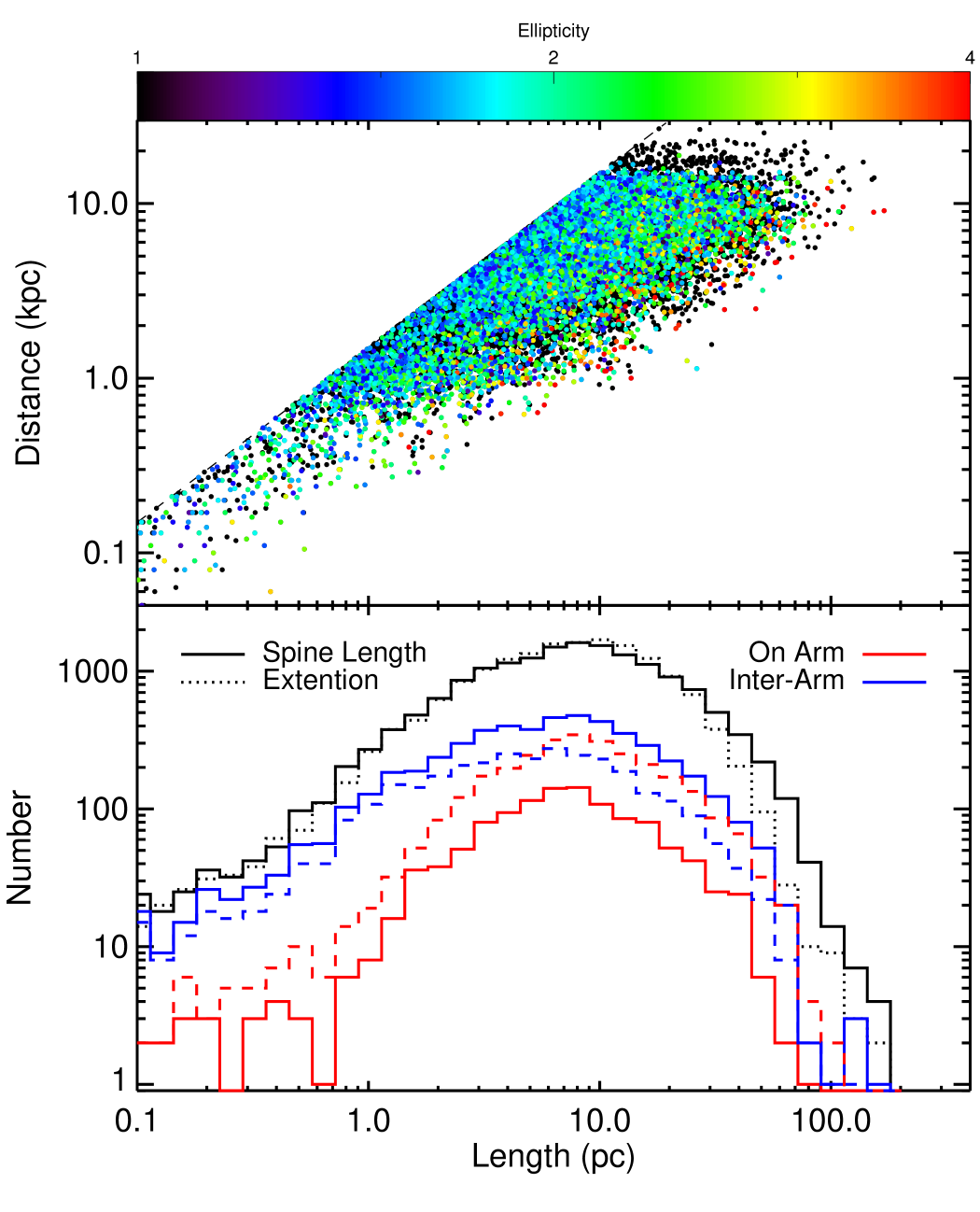}
\caption{ {\it Top panel}: Relation between the filament physical length and its assigned distance. The dashed line delimits the size corresponding to the cut-off criteria introduced in Sect.~\ref{Sect:SelCriteria}. Structures are shown with a colour code associated with their relative ellipticity $e$. 
{\it Lower panel}: Distributions of the physical sizes of the Hi-GAL filaments. The physical sizes are determined from two estimators (see Sect.~\ref{Sect:Length}): the length of the main axis, or angular length $l^{a}$, (continuous line) and the angular extension of the extended mask, $l^{e}$ (dotted line). The red and blue lines show the $l^{a}$ distributions for features associated to Galactic spiral arms and to  inter-arm regions, respectively, for arm sizes of $W\,=\,600$\,pc (continuous lines) and $W\,=\,1$\,kpc (dashed lines).}
\label{Fig:PhysLengthDistribution}
\end{figure}

We were able to determine lengths for the objects in our catalogue with a distance. The distributions of filament physical sizes, $L$, both from the angular length and from the extension (see Sect.~\ref{Sect:Length}) are shown in the bottom panel of Figure\,\ref{Fig:PhysLengthDistribution}. Our  filaments cover a wide range of sizes, ranging from a few times $\sim$0.1\,pc to over 100\,pc, with the majority of objects having $L \sim$\,0.5--70\,pc. The distributions of the subsets representative of the spiral arm and inter-arm features are  shown in Fig.\,\ref{Fig:PhysLengthDistribution} for the  two prescription of spiral arm widths. The two subset have similar sizes with median of $6.9$($7.9$)\,pc and $5.8$($4.7$)\,pc for arm and inter-arm features respectively assuming an arm width $W\,=\,600$\,pc\,pc ($1$\,kpc). There is a strong overlap between the two distributions as demonstrated by the interquartile ranges from $3.9$($4.5$)\,pc to $11.8$($14$)\,pc and filaments lying on spiral arms from $2.6$($2.0$)\,pc to $11.4$($10.$)\,pc for objects strongly associated to the inter-arm region. However, the KS statistical test could not confirm that these two distributions are different.

The top panel of Fig.\,\ref{Fig:PhysLengthDistribution} reports the filament sizes and distances, adopting a colour coding depending on their ellipticity and morphology (see Sect.~\ref{Sect:HiGALcatalog}). 
The identified objects span about an order of magnitude in physical sizes at fixed distances. The longer filaments are located farther from the Sun: objects in our catalogue with $L\,\geq\,10$ are typically located at $d\,\gtrsim\,1$\,kpc. 
The reported sample of filaments does not span uniformly the same interval of physical sizes at all the distances. On the one hand, this is caused to the cut-off criteria on $l^{a}$, introduced in Sect.~\ref{Sect:HiGALcatalog}. For example, there are no objects shorter than $\sim0.6$\,pc at distances $d\,\geq\,1$\,kpc. On the other hand, at fixed distance the number of detected  filaments quickly decreases for longer features.
One possible reason is due to the detection algorithm. In fact, we noticed that it splits a possible long structure into multiple distinct objects if the emission along its central region becomes weak, losing its global cylindrical-like appearance. This happens in particular for nearby objects, where the observations are able to resolve cloud substructures.   Nevertheless,  identifying the cases where features are located close to each other and are aligned as part of longer features is not doable using only the $Herschel$ data. The recognition of the underlying longer structure would require kinematic information from molecular line spectra as it was done for the case of the ``Nessie''  cloud \citep{Goodman2014}. Despite this limitation, our dataset allow to identify filaments with  $L\,\sim\,$10-20\,pc at distances of $1$-$2$\,kpc. 

We note a weak relation between object sizes and  ellipticities $e$: the longer features are typically more elongated as shown in Figure \,\ref{Fig:PhysLengthDistribution}. The elongated objects with $e\,\geq\,2$ include features with a linear and straight morphology, characteristic of simple filaments such as the Taurus B211/L1495 filament (physical size $\sim4$\,pc) \citep{Palmeirim2013}, the Orion Integral-Shaped (physical size $\sim$\,7\,pc) \citep{Bally1987, Johnstone1999}, or the already mentioned ``Nessie'' cloud (physical size $\sim$\,$80$--$150$\,pc) \citep{Jackson2010,Goodman2014}. 

Several recent studies aimed to search for long structures with linear morphology connected to the structure of the Milky Way: the Galactic ``bones'' \citep{Goodman2014, Wang2015, Zucker2015}, large and dense features representing the backbones of the spiral arms. We looked in our sample for possible ``bones'' candidate by selecting all the long features with $e\,\geq\,2$, for which we were able to estimate $P_{sp}$. We set our threshold for ``long filament'' to $20$\,pc to include the case of broken up structures and to avoid the selection of only objects at large distances, see top panel of Fig.\,\ref{Fig:PhysLengthDistribution}. There are $739$ objects in our sample with these properties, but they are not found to be associated preferentially to spiral arms. In fact, the number of long filament associated to spiral arm ranges from $66$ to $219$ for widths from $W$\,=\,$600$\,pc to $1$\,kpc, but for these arm-widths the inter-arm filaments drops from $124$ to $70$, showing that these features are not preferential associated to major structures in the Galaxy. Similar trends are found if we increase the  length threshold to define long structures. We conclude that ``long filaments'' cannot be only formed as Galactic ``bones'' by the gravitational potential well of spiral arms, but also by other phenomena in the inter-arm regions, like for example Galactic shear, that are able to stretch and reshape molecular clouds \citep{Koda2009,Ragan2014}. However, we point out that to draw more robust conclusions it would be necessary to determine a more complete sample of long filamentary clouds, solving the issue of the splitting of the cloud that could be present in our catalogue. This can be solved only with the additional informations granted by spectroscopic data.

\subsection{Physical properties: Masses and linear densities}
\label{Sect:MassMlin}

The distance association allows us to translate the measured total column density inside each extended region into a mass estimate for the candidate filament after the subtraction of the background contribution. The mass estimate is given by:

\begin{equation}
M^{\rm fil} =\, \mu_{\rm H_{2}} m_{\rm H} (\theta d)^{2} \sum^{mask}_{i,j} N^{\rm fil}_{\rm H_{2}}(i,j)
\end{equation}

\noindent
where $N^{\rm fil}_{\rm H_{2}}$(i,j) is the estimated column density associated with the filament in the pixel position $(i,j)$ and the sum is done over the entire filament extended mask, $\theta$ is the angular pixel size, $\mu_{\rm H_{2}}$ is the mean molecular weight of the interstellar medium with respect to the hydrogen molecules  which is assumed to be equal to 2.8; $m_{\rm H}$ is the mass of the hydrogen atom, and $d$ is the distance of the object. 

\begin{figure}
\includegraphics[width=0.45\textwidth]{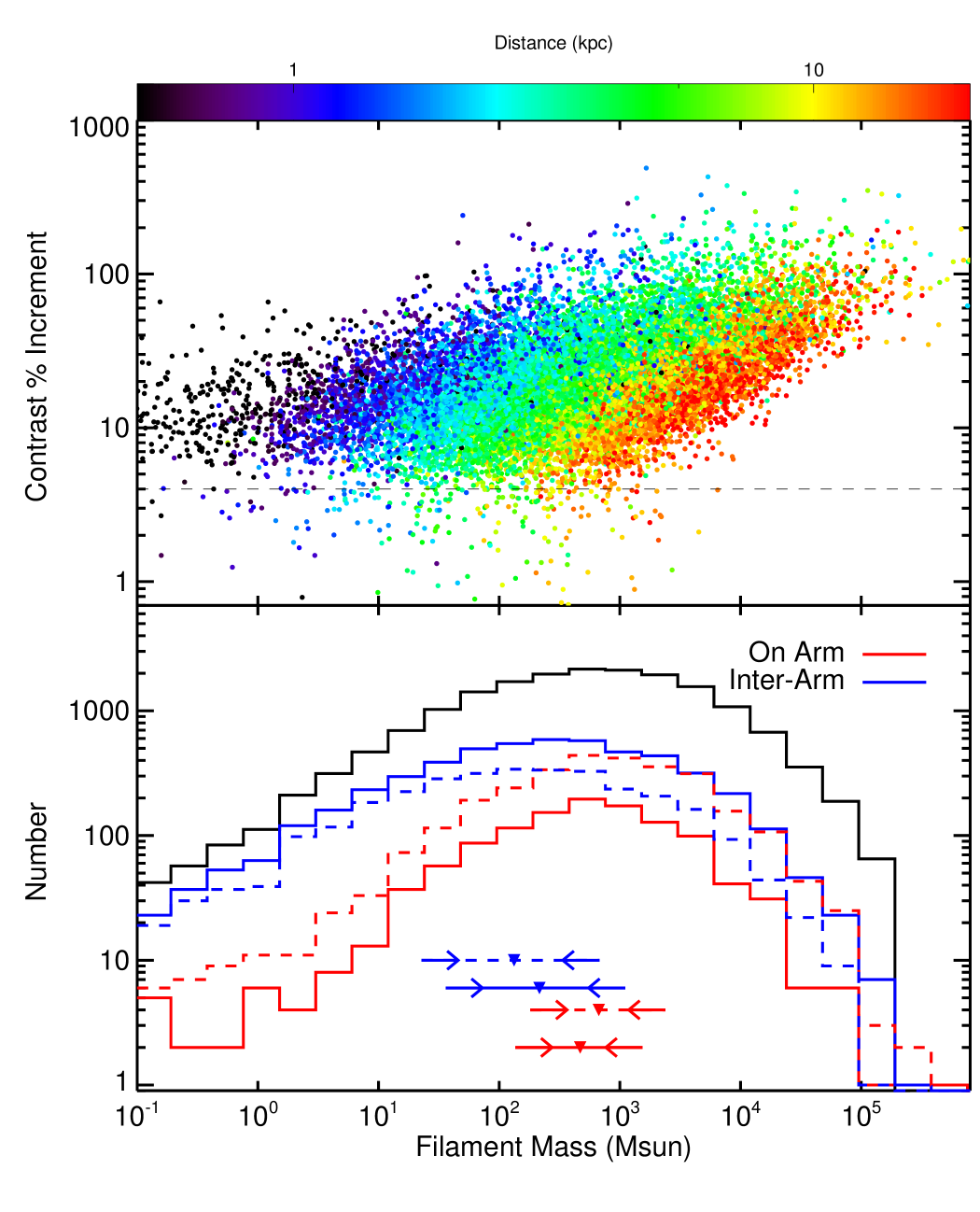}
\caption{{\it Top Panel}: Relation between the  contrast and the estimated mass of the features presented in this work. A colour scheme identifies features with similar assigned distance.
{\it Lower Panel}: Mass Distribution of  the candidate filaments in the Hi-GAL catalogue with an associated distance. The same distribution is shown for objects  associated to Galactic spiral arms (red line) or in the inter-arm regions (blue line), for two prescription of arm width,  $W\,=\,600$\,pc (continuous line) and  $W\,=\,1$\,kpc (dashed line). Triangles and arrows show the median and the interquartile ranges for these distributions. }
\label{Fig:MassDistribution}
\end{figure}

In the bottom panel of Figure~\ref{Fig:MassDistribution}, we show the distribution of the candidate filament masses for all the objects with an associated distance (see Sect.\,\ref{Sect:Distances}), using the estimated column density  $N^{\rm fil}_{\rm H_{2}}(i,j)$ as derived from the model 2C2T (see Sect.\,\ref{Sect:2C2Tmodel}).  
The estimated masses span an interval from few  M$_{\odot}$ to $\sim10^{5}$\,M$_{\odot}$ with typical values around $800$--$3,000$\,M$_{\odot}$. The top panel of Figure~\ref{Fig:MassDistribution} shows the contrast as function of the filament mass, draw in different colours depending on the associated distance. Features with different contrasts span different mass range. Objects with $M_{fil}\,\gtrsim\,1,000\,$M$_{\odot}$ are prominent features on Hi-GAL maps showing contrasts higher than $\sim10-15$\,per cent. Almost all the low mass candidates, $M_{fil}\,\leq\,10$\,M$_{\odot}$ are located within $\sim$1-2\,kpc, while more massive structures, $M_{fil}\,\geq\,10^{4}$\,M$_{\odot}$ have distances $d\,\gtrsim\,$5\,kpc.  We compared the mass distribution of the two subsets representative of filaments lying on arm and interarm regions, see Sect.~\ref{Sec:GalacticStructure}. 
Filaments associated to arms have a median mass of $470$($670$)\,M$_{\odot}$ with an interquartile range from $140$($180$) to $1,500$($2,400$)\,M$_{\odot}$ assuming the prescription of $W\,=\,600$\,pc ($1$\,kpc). On the contrary, the interarm feature have a median of $215$($130$)\,M$_{\odot}$ and interquartile range from $40$($30$)\,M$_{\odot}$ to $1,100$($680$)\,M$_{\odot}$, so they are typically less massive than the one associated to spiral arms. This difference in mass distribution is not excluded by the result of a K-S test that cannot rule out that these measurements belong to the the same distribution. A mass difference between arm and inter-arm filaments is not expected from simulations, but,  if confirmed, would indicate an influence of the environment on the structure mass and stability.

We compute the filament mass per unit length, $m_{\rm lin}$, as the ratio of the estimated masses and lengths. This quantity determines completely the stability of isothermal cylinders against their own gravity. Differently from the spherical case, the critical value over which the system becomes unstable, $m^{\rm crit}_{\rm lin}$, is independent of the structure central density, but depends only on the gas temperature and on the velocity dispersion in the central regions \citep{Inutsuka1992, Fiege2000}, $\sigma_{\rm c}$, for the case of pure thermal support and when turbulent motions are included.  Unstable filaments evolve extremely quickly, proceeding to collapse radially on free-fall timescales \citep{Pon2012} and may proceed to fragment into multiple cores.  

\begin{figure}
\includegraphics[width=0.45\textwidth]{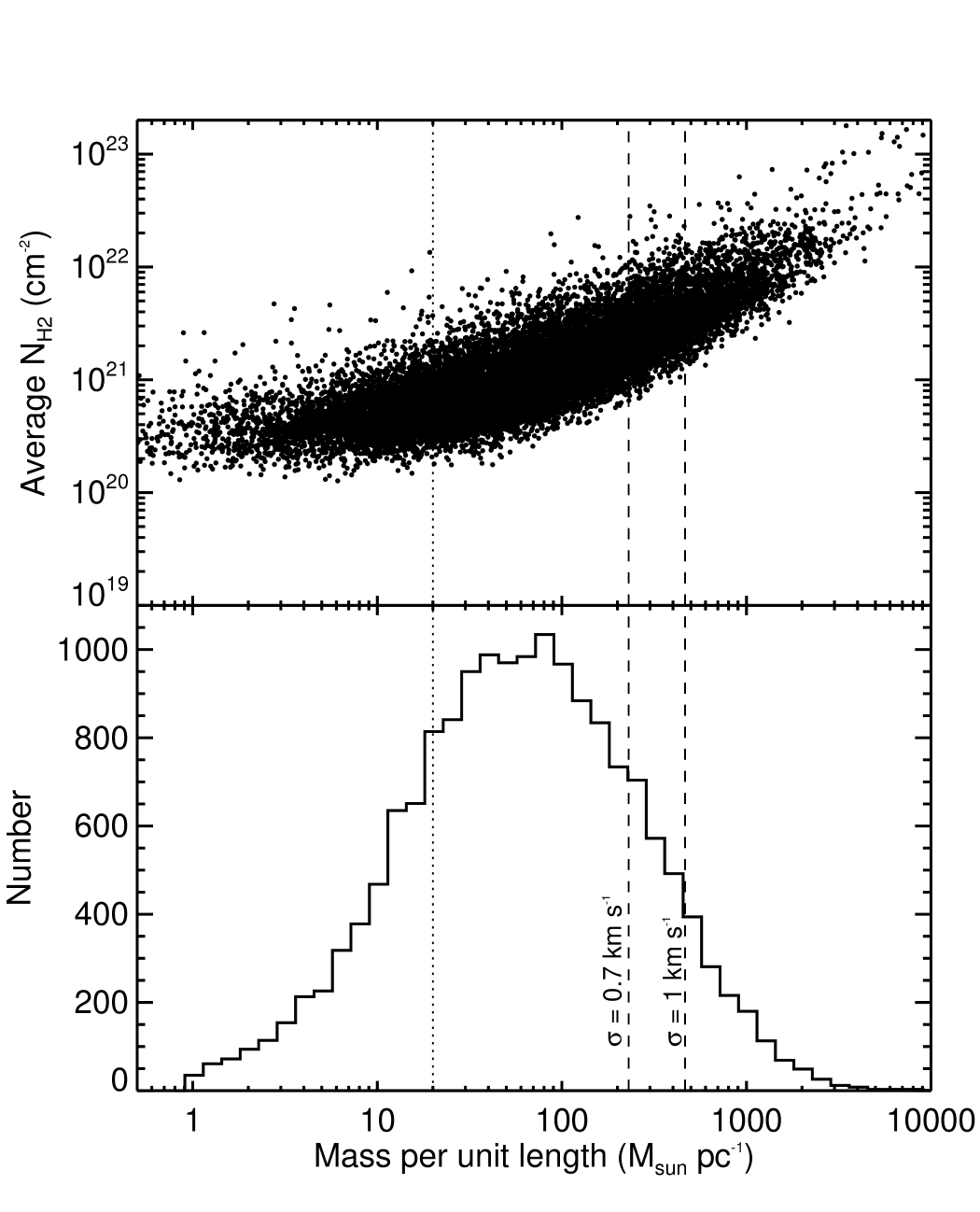}
\caption{
{\it Top Panel}: Relation between the measured mass per unit length and the average column density for all the features in the sample. {\it Bottom panel:}  Distribution of mass per unit length, $m_{\rm lin}$, of the  filaments in our sample with a distance assigned. The vertical lines trace the  critical values of $m_{\rm lin}$, over which an isothermal cylindrical structure is expected to be unstable against its self-gravity in the cases of support from only thermal pressure  (dotted line) and by turbulent motions of different strengths (dashed lines).  }     
\label{Fig:MLinDistribution}
\end{figure}

The distribution of $m_{\rm lin}$ in our sample is presented in the bottom panel of Fig.\,\ref{Fig:MLinDistribution}. We measured $m_{\rm lin}$ values ranging from $\sim$1 to $\sim$4000\,M$_{\odot}$\,pc$^{-1}$, with an average value of $\sim$250\,M$_{\odot}$\,pc$^{-1}$. The observed interval of $m_{\rm lin}$ indicates that our sample is composed by features in different dynamical states, including both subcritical,  $m_{\rm lin}\,\lesssim\,m^{\rm crit}_{\rm lin}$, and critical filaments, $m_{\rm lin}\,\gtrsim\,m^{\rm crit}_{\rm lin}$. In fact, we show typical values for $m^{\rm crit}_{\rm lin}$ in Fig.~\ref{Fig:MLinDistribution} reported as comparison. The smallest value is for the case of a pure thermal support, $m^{\rm crit}_{\rm lin}\,\approx\,16\,{\rm M_{\odot}\,pc^{-1}}\times(\frac{T}{10\,{\rm K}})$ \citep{Ostriker1964}, traced for the averages temperatures in our sample. Turbulent motions  grant a further support to the filamentary structure and their presence increase $m^{\rm crit}_{\rm lin}$ that scale as $m^{\rm crit}_{\rm lin}\,\approx\,470\,{\rm M_{\odot}\,pc^{-1} }\times (\frac{\sigma_{\rm c}}{1\,{\rm km s^{-1}}})^2$ \citep{Li2016},
where $\sigma_{\rm c}$ is the central line width measured from molecular line spectra. Studies of individual filaments report line widths with values $\sigma_{\rm c}\approx0.7-1.0$\,km\,s$^{-1}$, measured from different line tracers like NH$_{3}$ \citep{Sokolov2018} or C$^{18}$O \citep{Leurini2019},  corresponding to $m^{\rm crit}_{\rm lin}\approx230-470$\,M$_{\odot}$\,pc$^{-1}$ shown with dashed lines in Fig.\,\ref{Fig:MLinDistribution}. 

The critical filaments are from $\sim$10 per cent to $\leq$20 per cent of our sample and they are in average the features with higher densities, having $\overline{ N^{\rm fil}_{\rm H_{2}}}\geq\,1-2\times$10$^{21}$\,cm$^{-2}$. On the other hand,  thermally subcritical features ($m_{\rm lin}\,\lesssim\,24$\,M$_{\odot}$\,pc$^{-1}$) are typically tenuous filaments, with $\overline{ N^{\rm fil}_{\rm H_{2}}}\,\sim\,3-9\times\,10^{20}$\,cm$^{-2}$ and are $\sim$27 per cent of our sample.

\section{Comparison with other catalogues}\label{sect:comparison}

Other works exist in the literature dedicated to the identification of structures in the GP with a filamentary appearance \citep{Ragan2014,Wang2015,Li2016}. 
In this section, we compare our results with the catalogue of \citet{Li2016} that makes available the filamentary features from the 870-$\mu$m GP data of the ATLASGAL survey. We also compare to the IRDC catalogue of \citet{Peretto2009}, which, although it was not searching directly for filamentary morphologies, includes several well known filaments.

\subsection{ATLASGAL}
\label{Sect:ATLASGALassociated}

The ATLASGAL survey mapped the inner GP ($-60^{\circ} \leq l \leq +60^{\circ}$) with the APEX telescope at 870\,$\mu$m  at an angular resolution of 19\farcs2 \citep{Schuller2009}. From these data, \citet{Li2016} looked for dense filamentary features adopting the DisPerSE algorithm \citep{Sousbie2011} to trace their central regions. 
They identified a total of 517 filaments for which they report the derived properties (positions, sizes, position angles, total flux). In order to compare the two catalogues we require to match their contents. To perform this task, we first built a representative mask for each ATLASGAL filament to be compared to our extended masks. Objects are considered associated if their masks overlap.

Since the region contours adopted by \citet{Li2016} for their measurements are unavailable to us, we assumed, as a representative mask for each ATLASGAL source, an ellipse defined by the centre position, semi-axes and position angle reported in the \citet{Li2016} catalogue. We noticed that these ellipses are not the best representation of the ATLASGAL filaments, since they do not cover entirely the filament spines produced by \citet{Li2016}. This discrepancy is due to the criteria adopted by \citet{Li2016} to determine filament sizes as the eigenvalues of the second moments tensor of the pixel mask coordinates weighted by their intensity (see their Sect.\,3.3). On the other hand, we visually inspected all the ATLASGAL features (filament spines, ellipses and intensity maps) and recognized that doubling the reported semi-axis values allows us to encompass the spines in the large majority of cases. Therefore, we adopted these ellipses as a representation of the ATLASGAL filaments, even if they still do not reproduce the detailed shape of their contours. \\

We found that $491$ out of $517$ ATLASGAL filaments correspond to objects in our final catalogue. For the remaining $26$ objects, $10$ are actually identified from the $\lambda_{a}$ thresholding but are then filtered out by the adopted criteria concerning length ($4$ features have lengths between 70 and 110\,arcsec) or ellipticity ($6$ have $1.1\leq e \leq 1.3$), see Sect.\,\ref{Sect:HiGALcatalog}. Another $9$ features fall outside the area observed by {\it Herschel}, so only $7$ features are not confirmed from the analysis of the {\it Herschel} data, where they appear as multiple features, unconnected even at low contrast. In most cases, the match between the two catalogues is one-to-one. In these cases, the Hi-GAL feature generally extends over a larger area: in fact, in $77$ per cent of the cases ($379$ objects) the ATLASGAL representative mask is well within the borders of the Hi-GAL one. This implies that the filament sizes and lengths are typically larger in the Hi-GAL catalogue than in ATLASGAL. On other hand, there are cases where the feature association is one-to-many.  In these cases multiple ATLASGAL objects are just portions of a larger underlying structure recovered in $Herschel$ data. Indeed, in several cases the discrepancies between the two catalogues can be ascribed to the different appearance of the emission in the two datasets (Hi-GAL and ATLASGAL) and to the extraction methods adopted. In fact, the emission in Hi-GAL column-density maps appears to vary more smoothly than the ATLASGAL one. The latter filters out the diffuse emission on large spatial scales and thus shows abrupt variations. The {\it Herschel} data are more suitable to trace the filamentary structure emission even in its fainter portions, thanks to the lack of filtering. However, Hi-GAL pays the price of a harder definition for the edges of the structures with respect to the background, a difficulty that is not present in ATLASGAL data where the emission is truncated.
Moreover, it is worth noticing that the ATLASGAL filament catalogue was created by running the DisPerSE algorithm twice with different parameters \citep{Li2016}: first, a primary catalogue of reliable fragments of filaments was built, then the code was run with a lower threshold to connect these short filaments into larger coherent structures. All these processes were controlled through visual oversight of the final outputs. On the contrary, in our work, we run the code only once with an adaptive threshold. All the identified regions are left disconnected, even if there might be emission between them whose shape appears not to be filamentary. The merging of neighbour regions requires more information provided by additional data, such as molecular-line emission, and will be the subject of a future work.

\begin{figure} 
\includegraphics[width=0.45\textwidth]{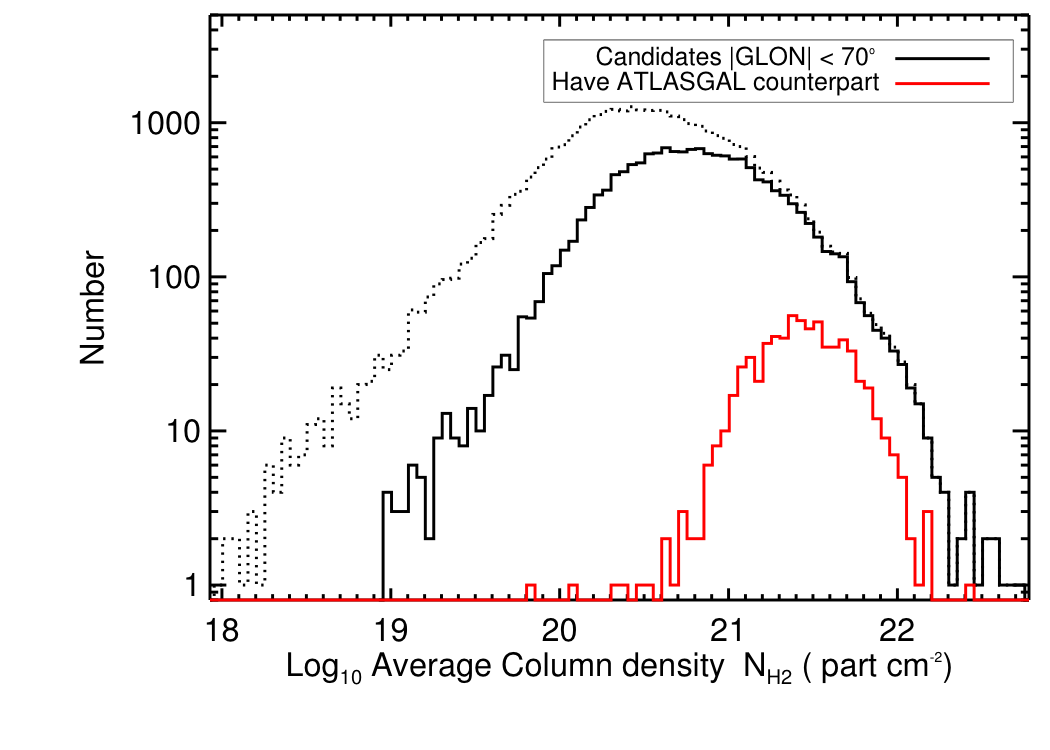}
\caption{Distribution of the average column densities of Hi-GAL candidate filaments for the entire sample (dashed line) and restricting to the inner Galaxy features (black). In red is the distribution of features associated with an ATLASGAL counterpart from the filament catalogue of \citet{Li2016}.}
\label{Fig:PropertiesAssociationWithATLASGAL}
\end{figure}

The ATLASGAL data are also affected by the spatial filtering usually present in the data acquired from ground-based telescopes, limiting the detectable features to the more compact and high-density ones. This is reflected in the  properties of the identified filaments, as  demonstrated in Figure\,\ref{Fig:PropertiesAssociationWithATLASGAL} where we show the distribution of the average column densities derived from Hi-GAL for the filaments reported by this work over the entire Galaxy (dashed line), the subset limited to the inner Galaxy with $|\,l\,| \leq 80^{\circ}$ (solid line) and the objects where we found an ATLASGAL counterpart (red line). Only denser objects are present in the ATLASGAL catalogue, with average column densities, $\overline{ N^{\rm fil}_{\rm H_{2}}}$, between 5\,$\times\,10^{20}$ and 2\,$\times\,10^{22}$\,cm$^{-2}$. 
Figure\,\ref{Fig:PropertiesAssociationWithATLASGAL} also shows that the catalogue of \citet{Li2016} is incomplete, missing several objects in the high-density regime. In fact, we identified 8,737 potential filaments in the inner Galaxy with $\overline{ N^{\rm fil}_{\rm H_{2}}} \geq 5 \times 10^{20}$\,cm$^{-2}$. Even limiting to those with $\overline{ N^{\rm fil}_{\rm H_{2}}} \geq 2.7 \times 10^{21}$\,cm$^{-2}$ (the mode of the distribution of features with an ATLASGAL counterpart) there are still $1,225$ structures. We inspected the intensity maps at $870$\,$\mu$m, and we found that, at the positions of the denser Hi-GAL filaments not present in the ATLASGAL catalogue, significant emission is indeed visible, suggesting that these features were detected, but then excluded from the catalogue due to the criteria adopted by \citet{Li2016}. 

\begin{figure} 
\includegraphics[width=0.5\textwidth]{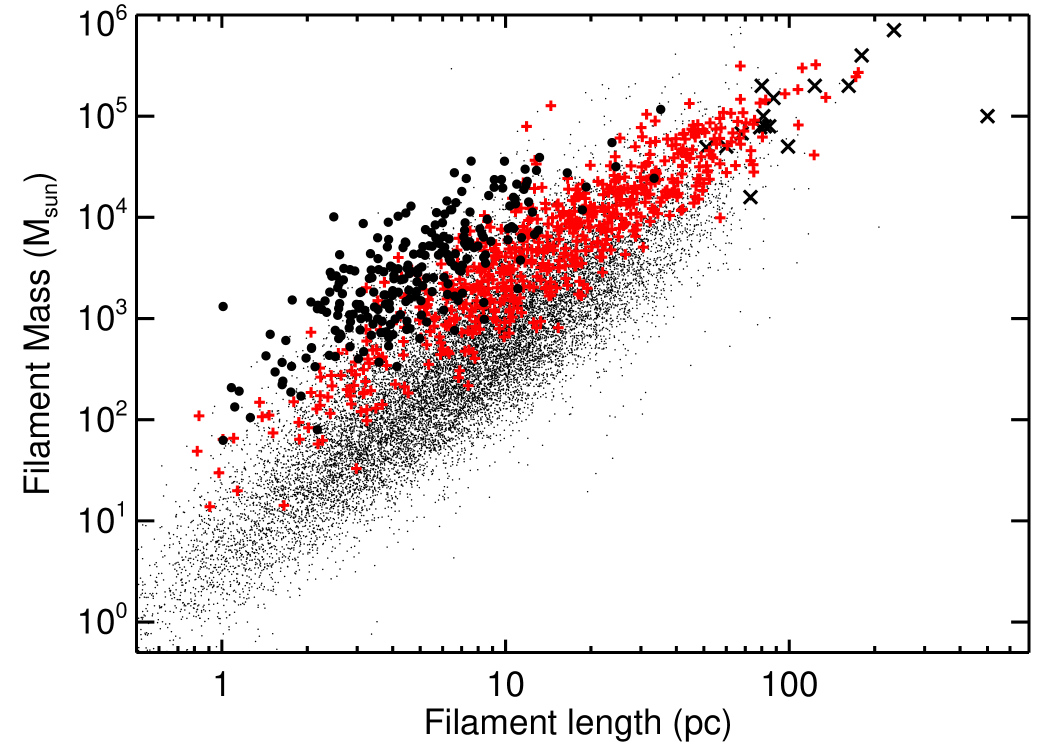}
\caption{Size-Mass diagram of the filamentary structures identified in ATLASGAL (large filled dots) and Hi-GAL (small dots and crosses). The red crosses mark the Hi-GAL objects that are matched to ATLASGAL sources. The black crosses are the location of already known giant molecular filaments identified in the literature and already collected by \citet{Li2016}.}
\label{Fig:SizeMassDiagrams}
\end{figure}

The ATLASGAL catalogue includes physical estimates for a limited sample of $241$ filaments for which they were able to estimate the distance and, consequently, lengths and masses. Most of them are matched with objects in our catalogue. In Figure\,\ref{Fig:SizeMassDiagrams}, we show the comparison of the properties as found in both catalogues: small dots show the properties of the objects in the whole Hi-GAL catalogue; black filled dots are the size and mass estimates as found by the ATLASGAL team; finally, red crosses give the values of the Hi-GAL measurements for the matched objects. Ideally, large black dots and red crosses should overlap, which is not the case.
The median measured sizes and masses are larger for Hi-GAL features by a factor $\sim4.2^{(7.5)}_{(2.3)}$ and $\sim4.6^{(14.0)}_{(1.5)}$, respectively (the upper and lower values are the first and third quartiles of the distribution). Since there are discrepancies in the distance estimates in the two catalogues, one may wonder to what level these differences are related to the distance mismatches: actually, if we restrict the comparison to objects whose distances agree within 20 per cent (75 in total), we find that the factors reduce to  $2.7^{(4.2)}_{(1.8)}$ and $1.6^{(2.9)}_{(0.8)}$, respectively, which are still not negligible differences. However, the different masses can be compatible each other after taking into account calibration uncertainties, the difficulties in estimating the backgrounds and the different dust opacity law assumed. The longer size of the Hi-GAL structures can be traced back, as said, to the spatial filtering of ground-based observations, whereas {\it Herschel} is able to recover the structures for their entire lengths. 

\subsection{IRDC}
\label{Sect:IRDCassociated}

InfraRed-Dark Clouds (IRDCs) are dark, high-extinction regions whose silhouette is identified against a sufficiently bright diffuse background emission \citep{Simon2005}. Searches for filaments throughout the Galaxy has been previously conducted by selecting IRDCs that appear filamentary at infrared wavelengths  \citep{Jackson2010,Wang2014, Ragan2014, Zucker2015}, with the consequence that the identified filamentary structures are mostly located nearby and towards the inner Galaxy. Only very dense filaments have been identified with this technique. The structures composed of condensations physically connected through lower-density regions are hardly detected in NIR/MIR maps, since they appear as a group of contiguous IRDC fragments, located where the density is high enough to produce extinction effects against the local background emission. The coherent and massive filament G24 \citep{Wang2015} is an example of such an apparent fragmentation: at near- to mid-infrared wavelengths it appears as $4$ slightly elongated IRDCs, the strongest peaks in column density being displaced along the structure. Instead, the whole connected structure of G24 is revealed by {\it Herschel} far-infrared/submm images and CO data \citep{Wang2015}.

In this section, we describe matching our catalogue with that of IRDCs by \citet{Peretto2009} and compare the observed properties of the objects that are found in common. Similarly to the methods used for the ATLASGAL filaments (see the previous section), we overlapped the extended masks of our sample with those associated with the IRDCs. In this case, we have the detailed contour defined by \citet{Peretto2009} available for each IRDC, and that corresponds to $\tau(8\,\mu m)$\,=\,0.35. Using the IRDC contour we have two advantages. First, we avoid any mis-association due to the elongated shape of large clouds that could extend beyond a circular association radius  (see \citealt{Peretto2016}). Second, we can better handle the large spatial resolution differencies between Spitzer 8-$\mu$m ($\sim2$\,arcsec) and {\it Herschel} column density ($\sim36$\,arcsec) datasets.

The \citet{Peretto2009} catalogue is limited to the longitude range $-70^{\rm o} \leq l \leq 70^{\rm o}$, where we identified 15,662 Hi-GAL candidate filaments, of which 3,785 ($\sim24$ per cent) include at least one IRDC. We found 1,496 features, $\sim40$ per cent of the matches, to be associated with an unique IRDC, while the remaining 60 per cent host multiple IRDC fragments, ($N_{\rm IRDC}$). On average, for each Hi-GAL filament there are $\sim$3 IRDCs, with $N_{\rm IRDC}$ typically ranging between 1 and 20 ($N_{\rm IRDC}\,\geq\,20$ only for 23 very extended filaments). A large majority of the cases ($\sim67$ per cent) shows the 8-$\mu$m contour completely inside the Hi-GAL one. For the other objects, we analysed  the fractions of the IRDC areas overlapping with the Hi-GAL contours, $f$, finding that they are uniformly distributed. Since the IRDCs with low $f$ have a high probability to be the result of a chance matching, we consider any feature not strictly included in the Hi-GAL filament contour as a possible mismatch.

Mismatches between the appearance in the IR and in column-density maps have been found also by \citet{Wilcock2012}, suggesting that a fraction of the IRDCs in \citet{Peretto2009} catalogue might not be due to extinction but dark features produced by lack of emission at 8\,$\mu$m.
This possibility has been also investigated by \citet{Peretto2016}, who verified that  the majority ($\sim76$ per cent) are real clouds. \citet{Peretto2016} found that most of the spurious features are extended clouds with size $R_\mathrm{eff} \geq 1$\,arcmin. Therefore, we revised the subsample of IRDCs left unmatched by our association (4,463, corresponding to $\sim28$ per cent, out of the 15,637 IRDCs). We noticed that about half of them have $R_\mathrm{eff} > 30$\,arcsec and that they are either not associated with any column density enhancement in our Hi-GAL dataset or they correspond to structures with a low elongation that we excluded from our catalogue. 
The remaining half ($\sim2,000$ features), are associated with features initially detected by the extraction algorithm, but then  removed from the Hi-GAL catalogue by the cut-off on their area (see Sect.\,\ref{Sect:HiGALcatalog}). These structures could be real density enhancements but the {\it Herschel} data cannot assess if they are filamentary, due to their spatial resolution.

\begin{figure} 
\includegraphics[width=0.45\textwidth]{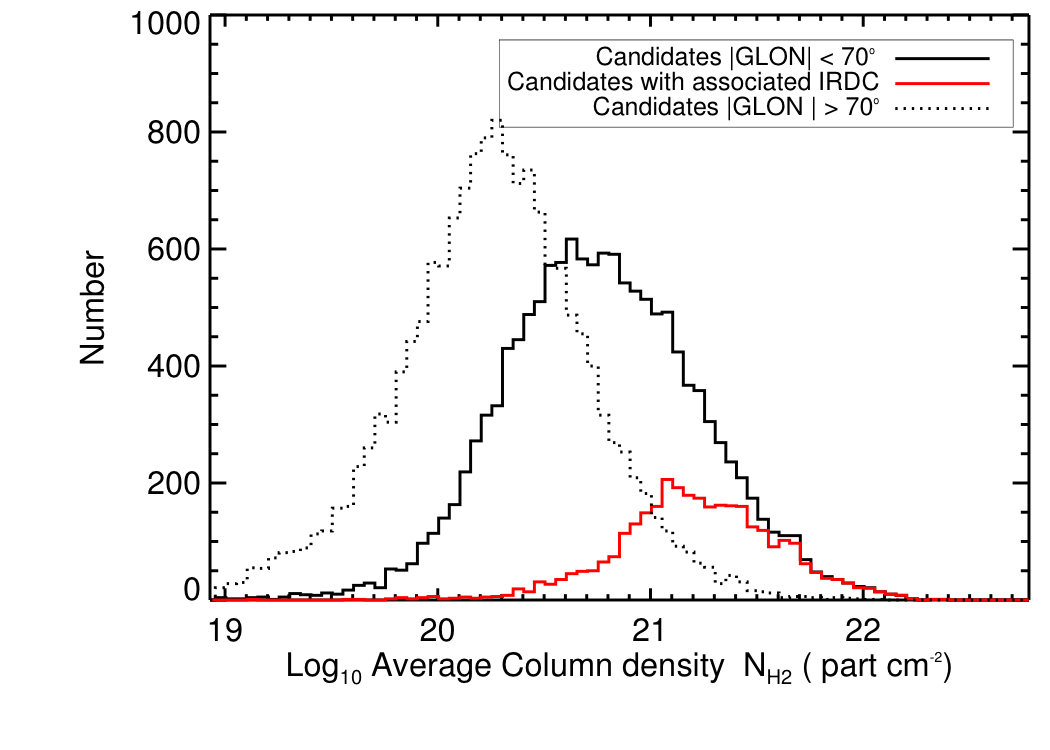}
\includegraphics[width=0.45\textwidth]{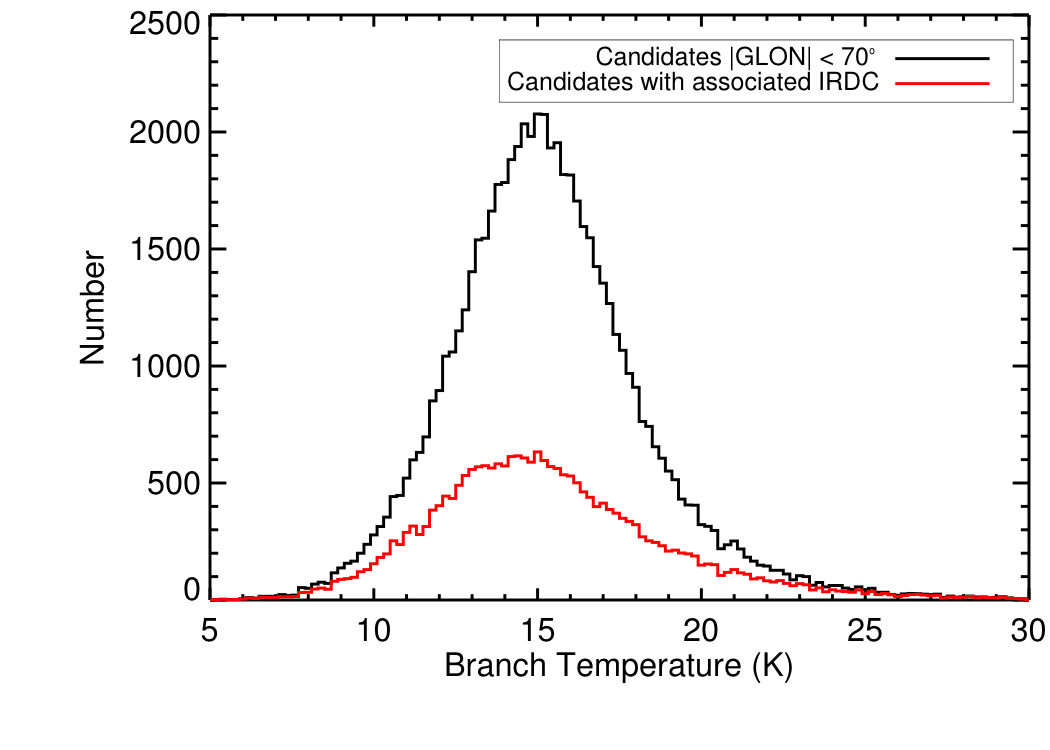}
\caption{{\it Top panel}: Average column-density distribution for candidate filaments separated in terms of their position on the Galactic Plane: the solid line shows the distribution for candidates in the inner Galaxy with Galactic longitudes $-70^{\circ}\,\leq\,l\,\leq\,70^{\circ}$, the dotted line is for candidates in the remainder of the plane. The distribution of inner-Galaxy sources that spatially overlap with IRDC contours from the \citet{Peretto2009} catalogue is drawn with a red solid line. 
{\it Bottom panel:}  Distribution of average temperature measured on the 1-D branches for candidates identified in the inner Galaxy with  Galactic longitudes $-70^{\circ}\,\leq\,l\,\leq\,70^{\circ}$. In red, the distribution limited to the subsample for which overlap is found with an IRDC.  }
\label{Fig:PropertiesAssociationWithIRDC}
\end{figure}

When comparing the properties of the filaments with the subsample associated with at least one IRDC, we found that they represent those with the highest average column density. This is shown in Figure\,\ref{Fig:PropertiesAssociationWithIRDC}, where we report both the distribution of average column density (top panel) and temperature over the branches (lower panel) for the sample of filaments,  separating features into those that fall in the same longitude range of the \citet{Peretto2009} catalogue and that do not. Any filament with an average column density $N_{\rm H_{2}} \geq3 \times 10^{21}$\,cm$^{-2}$ has at least one IRDC counterpart, while structures with $\sim3\times 10^{20} \leq N_{\rm H_{2}} \leq 3\times 10^{21}$\,cm$^{-2}$  can still be found associated with an IRDC with a probability decreasing with the average column density.  An interesting result of this comparison is that, while the Hi-GAL filaments in the inner Galaxy $|\,l\,| \leq 70^{\rm o}$ are typically denser than in the rest of the GP, we detect numerous structures in the outer Galaxy whose $N_{\rm H_{2}}$ would be compatible to the range typically measured for the IRDCs. The detection of IRDCs in $|\,l\,| \geq 70^{\rm o}$ is clearly limited, since the background behind which the cloud silhouette can appear is fainter than at other $|\,l\,|$ but still there are known IRDCs identified at these longitudes \citep{Frieswijk2008}. Therefore, the Hi-GAL candidate filamentary catalogue can also be used to search for IRDCs undetected so far. 

Although the IRDCs can be easily recognized, they still trace only the densest regions of the cloud, where opacity is so high as to extinguish the background emission. These regions are on average quite cold, T$\sim13-15$\,K \citep{Pillai2006}. On the other hand, as we discussed above, the {\it Herschel} data allows us to expand the detection contour to the entire, more extended cloud surrounding the IRDC. This includes portions of the cloud that are more tenuous, so they can be warmed up by the interstellar radiation field.  
This is reflected in the average temperature we measured along the branches of filaments with IRDC associations, where we found a wide range of values, from $7$ to $25$\,K. The colder branches are crossing over the associated IRDC, while the warmer ones generally extend beyond these areas. However, these features do not show any statistically significant difference with respect to the entire population as shown in the bottom panel of Figure\,\ref{Fig:PropertiesAssociationWithIRDC}. The clouds hosting IRDCs do not represent peculiar objects in the Galaxy, but they are just sites that are easily detectable by the previous observations due to their high column density and location in the Galaxy.
Therefore, given the good correlation between the IRDCs and the {\it Herschel} filamentary structures, the latter catalogue can be considered an extension of the previous studies, allowing us to take a census of all the crucial sites for star formation, including the massive ones traced by IRDCs.

\section{Summary and conclusions}\label{sect:conclusions}

In this paper, we present the first catalogue of filamentary structures identified from dust emission in the {\it Herschel} Hi-GAL survey over the entire Galactic Plane (GP).
A Hessian-based algorithm, designed to extract any elongated emission structure with sufficient contrast over its surroundings, was applied to column-density maps generated with a pixel-by-pixel grey body fitting to the {\it Herschel} Hi-GAL maps at $160$, $250$, $350$ and $500$\,$\mu$m. To avoid splitting any filamentary region over different Hi-GAL 2$^{\circ} \times 2^{\circ}$ tiles, we reprocessed the entire Hi-GAL dataset using the UNIMAP map maker in order to produce large mosaics. Each mosaic covers about $10$ degrees of Galactic longitude and overlaps with the adjacent ones approximately $2$ degrees. 

We introduced a definition of filament based on the intensity and geometrical shape in the map, aiming at being as general as possible. We extracted all the regions corresponding to such a definition and created a large catalogue of filamentary candidates. In order to identify a valid filament candidate, limiting the spurious detections and removing roundish clump-like structures, we introduced criteria on size ($5$ times the resolution of the column density maps, equivalent to 15 pixels), length (axis longer than $2$ arcmin), ellipticity ($e\,\geq\,1.3$ ) and filling factor ($f\,\leq\,0.85$). 

The final Hi-GAL Filament Catalog includes $32,059$ features spread over the entire GP. The sample shows a wide variety of morphologies, from rather isolated and straight features to complex networks composed by multiple nesting filaments. For each candidate filament we also trace and identify substructure/subfilaments branching off the main structure,  and included them in additional tables.
The entire catalogue is available at the the website \url{http://vialactea.iaps.inaf.it/} and it is part of the VIALACTEA Knowledge Base \citep{Molinaro2016}. The filaments and substructures of the catalogue can be visualized on the Hi-GAL data through the VIALACTEA Client that also allows queries of their physical properties.

The two catalogues, the candidate filaments and their relative branches, include general properties estimated from the {\it Herschel} data and two contrast-based quality parameters to allow the user to distinguish the more robust structures. We discuss different methods to measure the length of the structures, their average column density, $N_{\rm H_{2}}$, and temperature, $T$. The values of $N_{\rm H_{2}}$  and $T$ reported in the filaments catalogue are derived using a simple model assuming that the contribution in each pixel derives from two dust components, the filament itself and a background. For these components, we considered two cases $T^{\rm fil}\,=\,T^{\rm back}$ and  $T^{\rm fil}\,\neq\,T^{\rm back}$, discussed their differences and included the results from both cases in the catalogue. The $N^{\rm fil}_{\rm H_{2}}$ are systematically higher for the model with $T^{\rm fil}\,\neq\,T^{\rm back}$ and generally represent a more realistic estimate for filaments. 

We found that filamentary features in the Galaxy span a wide range of values in their physical properties. The catalogue includes filamentary regions whose length ranges from $2$ to $100$ arcmin, with a typical average length of $\sim 6$ arcmin. The filament-averaged $\overline{ N^{\rm fil}_{\rm H_{2}}}$ ranges from 10$^{20}$ to 10$^{23}$\,cm$^{-2}$, therefore including faint and tenuous ($A_{\rm V}\sim$ a few times $0.1$\,mag) features as well as several very dense $A_{\rm V}\geq 50$\,mag structures. We measured average temperatures along the main branches between $10$ and $25$\,K, with a few cases reaching T$\approx 30$-$35$\,K,  and found that $T$ may vary by several degrees, up to $10$--$15$\,K along a single structure, suggesting that isothermal models \citep{Inutsuka1992, Fiege2000} are not suited to describe the entire filamentary cloud structures.

We compared our catalogue with previous works available in literature: the ATLASGAL filamentary catalogue of \citet{Li2016} and the IRDC catalogue of \citet{Peretto2009} extracted with different methods and based on different data. Of the $517$ filaments  reported by \citet{Li2016}, only $26$ are found to be not associated with features in our catalogue, either because they fall outside the region observed by {\it Herschel}  or because they were filtered out by the criteria on length that we adopted in our work. The Hi-GAL filaments matching the ATLASGAL ones are on the high column density side of the Hi-GAL distribution, with $\overline{ N^{\rm fil}_{\rm H_{2}}}\geq\,$10$^{21}$\,cm$^{-2}$. However, ATLASGAL detects only some of the dense Hi-GAL features, which is not surprising, given the much higher {\it Herschel} sensitivity.  
 While the masses can be considered compatible for matching structures, once the diverse assumptions on the dust opacity and temperature of \citet{Li2016} are taken into account, the differences in lengths are a factor between $\sim 2$ and $\sim 4$. Filaments appear to be longer in {\it Herschel} maps, which again is not surprising since {\it Herschel} does not suffer from the atmospheric limitations in recovering large spatial scale emission. 

The comparison with the IRDC catalogue is less immediate because, while several IRDCs show a filamentary shape, a large number of them are quite roundish and not much extended and would have been therefore filtered out by our criteria. Moreover, IRDCs are associated only with the densest portion (A$_{v}\sim50-100$) of the molecular clouds and require a bright background emission to produce the observed extinction silhouette. In fact, we found that only 24 per cent of our filaments are associated with at least one IRDC, this leaves a large majority  of structures with multiple IRDC associations. From the IRDC viewpoint, $\sim 72$ per cent of them fall well within Hi-GAL filament edges and, even if they are round shaped, they are still associated with the densest portions of the Hi-GAL filaments. Indeed, all the Hi-GAL filaments with $\overline{ N^{\rm fil}_{\rm H_{2}}}\geq\,3\times\,10^{21}$\,cm$^{-2}$ include an IRDC. On other hand, IRDCs are still found to be associated with filaments also with  $3\times\,10^{20}\,\leq\,\overline{ N^{\rm fil}_{\rm H_{2}}}\leq\,3\times\,10^{21}$\,cm$^{-2}$, even if with a decreasing occurrence rate for decreasing $\overline{ N^{\rm fil}_{\rm H_{2}}}$. 

We crossmatched the candidate filaments with the {\it Herschel} compact-source lists \citep{Molinari2016,Elia2017} to determine the relevance of the filamentary structure for the star-forming clumps. We discuss two different criteria to match the two catalogues: simple angular association and a more robust criterion that takes into account also the radial velocity (only for a subsample of objects for which this information is available). This association also allows us to assign radial velocities and distances to a subsample of $18,389$ filaments, for which we determined the physical sizes, masses and mass per unit lengths. For this subsample with determined distances, we find filament lengths between $1$ and $70$\,pc and masses between $\sim 1$ and $10^{5}$\,M$_{\odot}$, corresponding to mass per unit length from $\sim$1 to $2,500$\,M$_{\odot}$\,pc$^{-1}$.
We find that a significant fraction ($44-66$ per cent, depending on the clump-filament association criterion) of the candidate filaments hosts at least one compact clump, with a median value of $\sim 3$ clumps per filament. We excluded the possibility that the filaments we detect in 36-arcsec resolution maps are simple chains of compact condensations. Indeed, the area ascribed to the filamentary structures is always larger than the typical area covered by sources. 

The distributions of average column density for filaments hosting and not hosting clumps is bimodal, with clump-hosting structures being denser. However, there is a large range of overlap and significant fractions of dense filaments are found without clumps, as well as clumps associated with relatively low-density filaments. This confirms our previous finding in \citet{Schisano2014} that there is no evidence of a column-density threshold for the formation of dense clumps in filaments, and also suggests that filament density (or mass per unit length) is not the only parameter governing filament fragmentation.

Finally, the census of filaments in the entire Galaxy allows us to study their Galactic distribution and to perform a comparison with the large-scale Galactic structure.
We quantified for the first time the idea widely recognized in the literature \citep{Molinari2010, Contreras2013}   that filaments are ubiquitously found  in the Galaxy. Their number density varies from $\sim 60$ to $\sim 25$ objects per square degree, going from the inner to the outer Galaxy, with an asymmetrical distribution with respect the Galactic Centre. This number-density drop is smooth and regular for  $l \geq 0$ while, for $l \leq 0$, it shows a rapid jump caused by the presence of a  interarm region covering several degrees in Galactic longitude. 

We identified for the catalogue two subsets representative of the features lying in different Galactic environments: filaments that can be associated with high probability to the spiral arms and filaments that are sufficiently distant from these that can be considered inter-arm region. We compared the global properties of these filaments finding similar distributions for average column densities, temperature and lengths. We measure a weak difference on the mass distribution depending on the surrounding environment, with filaments associated to spiral arms being more massive than the ones in the inter-arm region. However, the K-S test cannot exclude that these measurements belong to the same underlying distribution. These results are in agreement with the predictions from simulations, where the global properties of the clouds were found to be independent from the environment \citep{DuarteCabral2016}.

The Hi-GAL candidate filaments represent the widest catalogue of structures with a filamentary shape rigorously defined in terms of their observed morphology. It covers the entire Galactic plane and all the possible environments observed in our Galaxy. It extends previous works refining the estimates for the filament physical properties in view of the results from {\it Herschel} observations. The catalogue not only includes previously unexplored Galactic longitudes, but also relies on much higher sensitivities, allowing the inclusion of low-density structures. The catalogue is a precious tool to connect the processes of star formation to the large Galactic structures. 

\section*{Acknowledgements}

The authors thank the anonymous referee for her/his careful reading of the manuscript and all the constructive comments that improved the original manuscript.
This work is part of the VIALACTEA Project, a Collaborative Project under Framework Programme 7 of the European Union, funded under Contract \#607380 that is hereby acknowledged. E.S. acknowledge financial support from the aforementioned VIALACTEA Project. The data processing  {\it Herschel} Hi-GAL is the part of a multi-year effort funded thanks to Contracts I/038/080/0 and I/029/12/0 from ASI, Agenzia Spaziale Italiana. MM acknowledges support from CONICYT Programa de Astronom{\'i}a Fondo ALMA-CONICYT, project 3119AS0001.
{\it Herschel} is an ESA space observatory with science instruments provided by European-led Principal Investigator consortia and with important participation from NASA. PACS has been developed by a consortium of institutes led by MPE (Germany) and including UVIE (Austria); KUL, CSL, IMEC (Belgium); CEA, OAMP (France); MPIA (Germany); IAPS, OAP/OAT, OAA/CAISMI, LENS, SISSA (Italy); and IAC (Spain). This development has been supported by the funding agencies BMVIT
(Austria), ESA-PRODEX (Belgium), CEA/CNES (France), DLR (Germany), ASI (Italy) and CICYT/MCYT (Spain). SPIRE has been developed by a consortium of institutes led by Cardiff University (UK) and including the University of Lethbridge (Canada); NAOC (China); CEA, LAM (France); IAPS, University of Padua (Italy); IAC (Spain); Stockholm Observatory (Sweden); Imperial College London, RAL, UCL-MSSL, UKATC, University of Sussex (UK); and Caltech, JPL, NHSC, University of Colorado (USA). This development has also been supported by national funding agencies: CSA (Canada); NAOC (China); CEA, CNES, CNRS (France); ASI (Italy); MCINN (Spain); Stockholm Observatory (Sweden); STFC (UK); and NASA (USA).

\bibliographystyle{mnras}
\bibliography{BiblioBiBTex} 

\begin{thebibliography}{}
\makeatletter
\relax
\def\mn@urlcharsother{\let\do\@makeother \do\$\do\&\do\#\do\^\do\_\do\%\do\~}
\def\mn@doi{\begingroup\mn@urlcharsother \@ifnextchar [ {\mn@doi@}
  {\mn@doi@[]}}
\def\mn@doi@[#1]#2{\def\@tempa{#1}\ifx\@tempa\@empty \href
  {http://dx.doi.org/#2} {doi:#2}\else \href {http://dx.doi.org/#2} {#1}\fi
  \endgroup}
\def\mn@eprint#1#2{\mn@eprint@#1:#2::\@nil}
\def\mn@eprint@arXiv#1{\href {http://arxiv.org/abs/#1} {{\tt arXiv:#1}}}
\def\mn@eprint@dblp#1{\href {http://dblp.uni-trier.de/rec/bibtex/#1.xml}
  {dblp:#1}}
\def\mn@eprint@#1:#2:#3:#4\@nil{\def\@tempa {#1}\def\@tempb {#2}\def\@tempc
  {#3}\ifx \@tempc \@empty \let \@tempc \@tempb \let \@tempb \@tempa \fi \ifx
  \@tempb \@empty \def\@tempb {arXiv}\fi \@ifundefined
  {mn@eprint@\@tempb}{\@tempb:\@tempc}{\expandafter \expandafter \csname
  mn@eprint@\@tempb\endcsname \expandafter{\@tempc}}}

\bibitem[\protect\citeauthoryear{{Anderson}, {Bania}, {Balser}  \&
  {Rood}}{{Anderson} et~al.}{2012}]{Anderson2012}
{Anderson} L.~D.,  {Bania} T.~M.,  {Balser} D.~S.,   {Rood} R.~T.,  2012,
  \mn@doi [\apj] {10.1088/0004-637X/754/1/62}, \href
  {https://ui.adsabs.harvard.edu/abs/2012ApJ...754...62A} {754, 62}

\bibitem[\protect\citeauthoryear{{Andr{\'e}} et~al.,}{{Andr{\'e}}
  et~al.}{2010}]{Andre2010}
{Andr{\'e}} P.,  et~al., 2010, \mn@doi [\aap] {10.1051/0004-6361/201014666},
  \href {http://adsabs.harvard.edu/abs/2010A%26A...518L.102A} {518, L102}

\bibitem[\protect\citeauthoryear{{Andr{\'e}}, {Di Francesco}, {Ward-Thompson},
  {Inutsuka}, {Pudritz}  \& {Pineda}}{{Andr{\'e}} et~al.}{2014}]{Andre2014}
{Andr{\'e}} P.,  {Di Francesco} J.,  {Ward-Thompson} D.,  {Inutsuka} S.-I.,
  {Pudritz} R.~E.,   {Pineda} J.~E.,  2014, Protostars and Planets VI, \href
  {http://adsabs.harvard.edu/abs/2014prpl.conf...27A} {pp 27--51}

\bibitem[\protect\citeauthoryear{{Arzoumanian} et~al.,}{{Arzoumanian}
  et~al.}{2011}]{Arzoumanian2011}
{Arzoumanian} D.,  et~al., 2011, \mn@doi [\aap] {10.1051/0004-6361/201116596},
  \href {http://adsabs.harvard.edu/abs/2011A%26A...529L...6A} {529, L6}

\bibitem[\protect\citeauthoryear{{Arzoumanian} et~al.,}{{Arzoumanian}
  et~al.}{2019}]{Arzoumanian2019}
{Arzoumanian} D.,  et~al., 2019, \mn@doi [\aap] {10.1051/0004-6361/201832725},
  \href {https://ui.adsabs.harvard.edu/abs/2019A&A...621A..42A} {621, A42}

\bibitem[\protect\citeauthoryear{{Baba}, {Asaki}, {Makino}, {Miyoshi}, {Saitoh}
   \& {Wada}}{{Baba} et~al.}{2009}]{Baba2009}
{Baba} J.,  {Asaki} Y.,  {Makino} J.,  {Miyoshi} M.,  {Saitoh} T.~R.,   {Wada}
  K.,  2009, \mn@doi [\apj] {10.1088/0004-637X/706/1/471}, \href
  {https://ui.adsabs.harvard.edu/abs/2009ApJ...706..471B} {706, 471}

\bibitem[\protect\citeauthoryear{{Bally}, {Langer}, {Stark}  \&
  {Wilson}}{{Bally} et~al.}{1987}]{Bally1987}
{Bally} J.,  {Langer} W.~D.,  {Stark} A.~A.,   {Wilson} R.~W.,  1987, \mn@doi
  [\apjl] {10.1086/184817}, \href
  {http://adsabs.harvard.edu/abs/1987ApJ...312L..45B} {312, L45}

\bibitem[\protect\citeauthoryear{{Barnes} et~al.,}{{Barnes}
  et~al.}{2011}]{Barnes2011}
{Barnes} P.~J.,  et~al., 2011, \mn@doi [\apjs] {10.1088/0067-0049/196/1/12},
  \href {http://adsabs.harvard.edu/abs/2011ApJS..196...12B} {196, 12}

\bibitem[\protect\citeauthoryear{{Barnes}, {Muller}, {Indermuehle},
  {O'Dougherty}, {Lowe}, {Cunningham}, {Hernandez}  \& {Fuller}}{{Barnes}
  et~al.}{2015}]{Barnes2015}
{Barnes} P.~J.,  {Muller} E.,  {Indermuehle} B.,  {O'Dougherty} S.~N.,  {Lowe}
  V.,  {Cunningham} M.,  {Hernandez} A.~K.,   {Fuller} G.~A.,  2015, \mn@doi
  [\apj] {10.1088/0004-637X/812/1/6}, \href
  {http://adsabs.harvard.edu/abs/2015ApJ...812....6B} {812, 6}

\bibitem[\protect\citeauthoryear{{Barrow}, {Bhavsar}  \& {Sonoda}}{{Barrow}
  et~al.}{1985}]{Barrow1985}
{Barrow} J.~D.,  {Bhavsar} S.~P.,   {Sonoda} D.~H.,  1985, \mn@doi [\mnras]
  {10.1093/mnras/216.1.17}, \href
  {http://adsabs.harvard.edu/abs/1985MNRAS.216...17B} {216, 17}

\bibitem[\protect\citeauthoryear{{Battersby}, {Bally}, {Dunham}, {Ginsburg},
  {Longmore}  \& {Darling}}{{Battersby} et~al.}{2014}]{Battersby2014}
{Battersby} C.,  {Bally} J.,  {Dunham} M.,  {Ginsburg} A.,  {Longmore} S.,
  {Darling} J.,  2014, \mn@doi [\apj] {10.1088/0004-637X/786/2/116}, \href
  {http://adsabs.harvard.edu/abs/2014ApJ...786..116B} {786, 116}

\bibitem[\protect\citeauthoryear{{Benedettini} et~al.,}{{Benedettini}
  et~al.}{2015}]{Benedettini2015}
{Benedettini} M.,  et~al., 2015, \mn@doi [\mnras] {10.1093/mnras/stv1750},
  \href {http://adsabs.harvard.edu/abs/2015MNRAS.453.2036B} {453, 2036}

\bibitem[\protect\citeauthoryear{{Bernard} et~al.,}{{Bernard}
  et~al.}{2010}]{Bernard2010}
{Bernard} J.-P.,  et~al., 2010, \mn@doi [\aap] {10.1051/0004-6361/201014540},
  \href {http://adsabs.harvard.edu/abs/2010A%26A...518L..88B} {518, L88}

\bibitem[\protect\citeauthoryear{{Boulanger}, {Abergel}, {Bernard}, {Burton},
  {Desert}, {Hartmann}, {Lagache}  \& {Puget}}{{Boulanger}
  et~al.}{1996}]{Boulanger1996}
{Boulanger} F.,  {Abergel} A.,  {Bernard} J.-P.,  {Burton} W.~B.,  {Desert}
  F.-X.,  {Hartmann} D.,  {Lagache} G.,   {Puget} J.-L.,  1996, \aap, \href
  {http://adsabs.harvard.edu/abs/1996A%26A...312..256B} {312, 256}

\bibitem[\protect\citeauthoryear{{Brand} \& {Blitz}}{{Brand} \&
  {Blitz}}{1993}]{Brand1993}
{Brand} J.,  {Blitz} L.,  1993, \aap, \href
  {http://adsabs.harvard.edu/abs/1993A%26A...275...67B} {275, 67}

\bibitem[\protect\citeauthoryear{{Brunt}, {Kerton}  \& {Pomerleau}}{{Brunt}
  et~al.}{2003}]{Brunt2003}
{Brunt} C.~M.,  {Kerton} C.~R.,   {Pomerleau} C.,  2003, \mn@doi [\apjs]
  {10.1086/344245}, \href {http://adsabs.harvard.edu/abs/2003ApJS..144...47B}
  {144, 47}

\bibitem[\protect\citeauthoryear{{Burton} et~al.,}{{Burton}
  et~al.}{2013}]{Burton2013}
{Burton} M.~G.,  et~al., 2013, \mn@doi [\pasa] {10.1017/pasa.2013.22}, \href
  {http://adsabs.harvard.edu/abs/2013PASA...30...44B} {30, e044}

\bibitem[\protect\citeauthoryear{{Cambr{\'e}sy}, {Boulanger}, {Lagache}  \&
  {Stepnik}}{{Cambr{\'e}sy} et~al.}{2001}]{Cambresy2001}
{Cambr{\'e}sy} L.,  {Boulanger} F.,  {Lagache} G.,   {Stepnik} B.,  2001,
  \mn@doi [\aap] {10.1051/0004-6361:20010930}, \href
  {http://adsabs.harvard.edu/abs/2001A%26A...375..999C} {375, 999}

\bibitem[\protect\citeauthoryear{{Chemin}, {Renaud}  \& {Soubiran}}{{Chemin}
  et~al.}{2015}]{Chemin2015}
{Chemin} L.,  {Renaud} F.,   {Soubiran} C.,  2015, \mn@doi [\aap]
  {10.1051/0004-6361/201526040}, \href
  {https://ui.adsabs.harvard.edu/abs/2015A&A...578A..14C} {578, A14}

\bibitem[\protect\citeauthoryear{{Contreras} et~al.,}{{Contreras}
  et~al.}{2013}]{Contreras2013}
{Contreras} Y.,  et~al., 2013, \mn@doi [\aap] {10.1051/0004-6361/201220155},
  \href {http://adsabs.harvard.edu/abs/2013A%26A...549A..45C} {549, A45}

\bibitem[\protect\citeauthoryear{{Dave}, {Hellinger}, {Primack}, {Nolthenius}
  \& {Klypin}}{{Dave} et~al.}{1997}]{Dave1997}
{Dave} R.,  {Hellinger} D.,  {Primack} J.,  {Nolthenius} R.,   {Klypin} A.,
  1997, \mn@doi [\mnras] {10.1093/mnras/284.3.607}, \href
  {http://adsabs.harvard.edu/abs/1997MNRAS.284..607D} {284, 607}

\bibitem[\protect\citeauthoryear{{Dobbs} \& {Bonnell}}{{Dobbs} \&
  {Bonnell}}{2006}]{Dobbs2006}
{Dobbs} C.~L.,  {Bonnell} I.~A.,  2006, \mn@doi [\mnras]
  {10.1111/j.1365-2966.2006.10146.x}, \href
  {http://adsabs.harvard.edu/abs/2006MNRAS.367..873D} {367, 873}

\bibitem[\protect\citeauthoryear{{Dobbs} \& {Pringle}}{{Dobbs} \&
  {Pringle}}{2013}]{Dobbs2013}
{Dobbs} C.~L.,  {Pringle} J.~E.,  2013, \mn@doi [\mnras]
  {10.1093/mnras/stt508}, \href
  {http://adsabs.harvard.edu/abs/2013MNRAS.432..653D} {432, 653}

\bibitem[\protect\citeauthoryear{{Duarte-Cabral} \& {Dobbs}}{{Duarte-Cabral} \&
  {Dobbs}}{2016}]{DuarteCabral2016}
{Duarte-Cabral} A.,  {Dobbs} C.~L.,  2016, \mn@doi [\mnras]
  {10.1093/mnras/stw469}, \href
  {http://adsabs.harvard.edu/abs/2016MNRAS.458.3667D} {458, 3667}

\bibitem[\protect\citeauthoryear{{Eden}, {Moore}, {Morgan}, {Thompson}  \&
  {Urquhart}}{{Eden} et~al.}{2013}]{Eden2013}
{Eden} D.~J.,  {Moore} T.~J.~T.,  {Morgan} L.~K.,  {Thompson} M.~A.,
  {Urquhart} J.~S.,  2013, \mn@doi [\mnras] {10.1093/mnras/stt279}, \href
  {http://adsabs.harvard.edu/abs/2013MNRAS.431.1587E} {431, 1587}

\bibitem[\protect\citeauthoryear{{Elia} et~al.,}{{Elia}
  et~al.}{2013}]{Elia2013}
{Elia} D.,  et~al., 2013, \mn@doi [\apj] {10.1088/0004-637X/772/1/45}, \href
  {http://adsabs.harvard.edu/abs/2013ApJ...772...45E} {772, 45}

\bibitem[\protect\citeauthoryear{{Elia} et~al.,}{{Elia}
  et~al.}{2017}]{Elia2017}
{Elia} D.,  et~al., 2017, \mn@doi [\mnras] {10.1093/mnras/stx1357}, \href
  {http://adsabs.harvard.edu/abs/2017MNRAS.471..100E} {471, 100}

\bibitem[\protect\citeauthoryear{{Ellsworth-Bowers} et~al.,}{{Ellsworth-Bowers}
  et~al.}{2013}]{EllsworthBowers2013}
{Ellsworth-Bowers} T.~P.,  et~al., 2013, \mn@doi [\apj]
  {10.1088/0004-637X/770/1/39}, \href
  {http://adsabs.harvard.edu/abs/2013ApJ...770...39E} {770, 39}

\bibitem[\protect\citeauthoryear{{Falgarone}, {Pety}  \&
  {Phillips}}{{Falgarone} et~al.}{2001}]{Falgarone2001}
{Falgarone} E.,  {Pety} J.,   {Phillips} T.~G.,  2001, \mn@doi [\apj]
  {10.1086/321483}, \href {http://adsabs.harvard.edu/abs/2001ApJ...555..178F}
  {555, 178}

\bibitem[\protect\citeauthoryear{{Federrath} \& {Klessen}}{{Federrath} \&
  {Klessen}}{2013}]{Federrath2013}
{Federrath} C.,  {Klessen} R.~S.,  2013, \mn@doi [\apj]
  {10.1088/0004-637X/763/1/51}, \href
  {http://adsabs.harvard.edu/abs/2013ApJ...763...51F} {763, 51}

\bibitem[\protect\citeauthoryear{{Fiege} \& {Pudritz}}{{Fiege} \&
  {Pudritz}}{2000}]{Fiege2000}
{Fiege} J.~D.,  {Pudritz} R.~E.,  2000, \mn@doi [\mnras]
  {10.1046/j.1365-8711.2000.03066.x}, \href
  {http://adsabs.harvard.edu/abs/2000MNRAS.311...85F} {311, 85}

\bibitem[\protect\citeauthoryear{{Finkbeiner}, {Davis}  \&
  {Schlegel}}{{Finkbeiner} et~al.}{1999}]{Finkbeiner1999}
{Finkbeiner} D.~P.,  {Davis} M.,   {Schlegel} D.~J.,  1999, \mn@doi [\apj]
  {10.1086/307852}, \href {http://adsabs.harvard.edu/abs/1999ApJ...524..867F}
  {524, 867}

\bibitem[\protect\citeauthoryear{{Flagey} et~al.,}{{Flagey}
  et~al.}{2009}]{Flagey2009}
{Flagey} N.,  et~al., 2009, \mn@doi [\apj] {10.1088/0004-637X/701/2/1450},
  \href {http://adsabs.harvard.edu/abs/2009ApJ...701.1450F} {701, 1450}

\bibitem[\protect\citeauthoryear{{Frieswijk}, {Spaans}, {Shipman}, {Teyssier},
  {Carey}  \& {Tielens}}{{Frieswijk} et~al.}{2008}]{Frieswijk2008}
{Frieswijk} W.~F.,  {Spaans} M.,  {Shipman} R.~F.,  {Teyssier} D.,  {Carey}
  S.~J.,   {Tielens} A.~G.~G.~M.,  2008, \mn@doi [\apjl] {10.1086/592382},
  \href {http://adsabs.harvard.edu/abs/2008ApJ...685L..51F} {685, L51}

\bibitem[\protect\citeauthoryear{{Goldsmith}, {Heyer}, {Narayanan}, {Snell},
  {Li}  \& {Brunt}}{{Goldsmith} et~al.}{2008}]{Goldsmith2008}
{Goldsmith} P.~F.,  {Heyer} M.,  {Narayanan} G.,  {Snell} R.,  {Li} D.,
  {Brunt} C.,  2008, \mn@doi [\apj] {10.1086/587166}, \href
  {http://adsabs.harvard.edu/abs/2008ApJ...680..428G} {680, 428}

\bibitem[\protect\citeauthoryear{{G{\'o}mez}}{{G{\'o}mez}}{2006}]{Gomez2006}
{G{\'o}mez} G.~C.,  2006, \mn@doi [\aj] {10.1086/508412}, \href
  {https://ui.adsabs.harvard.edu/abs/2006AJ....132.2376G} {132, 2376}

\bibitem[\protect\citeauthoryear{{G{\'o}mez} \&
  {V{\'a}zquez-Semadeni}}{{G{\'o}mez} \&
  {V{\'a}zquez-Semadeni}}{2014}]{Gomez2014}
{G{\'o}mez} G.~C.,  {V{\'a}zquez-Semadeni} E.,  2014, \mn@doi [\apj]
  {10.1088/0004-637X/791/2/124}, \href
  {http://adsabs.harvard.edu/abs/2014ApJ...791..124G} {791, 124}

\bibitem[\protect\citeauthoryear{Gonzalez \& Woods}{Gonzalez \&
  Woods}{2006}]{Gonzalez2006}
Gonzalez R.~C.,  Woods R.~E.,  2006, Digital Image Processing (3rd Edition).
Prentice-Hall, Inc., Upper Saddle River, NJ, USA

\bibitem[\protect\citeauthoryear{{Goodman} et~al.,}{{Goodman}
  et~al.}{2014}]{Goodman2014}
{Goodman} A.~A.,  et~al., 2014, \mn@doi [\apj] {10.1088/0004-637X/797/1/53},
  \href {http://adsabs.harvard.edu/abs/2014ApJ...797...53G} {797, 53}

\bibitem[\protect\citeauthoryear{{Griffin} et~al.,}{{Griffin}
  et~al.}{2010}]{Griffin2010}
{Griffin} M.~J.,  et~al., 2010, \mn@doi [\aap] {10.1051/0004-6361/201014519},
  \href {http://adsabs.harvard.edu/abs/2010A%26A...518L...3G} {518, L3}

\bibitem[\protect\citeauthoryear{{Gutermuth} \& {Heyer}}{{Gutermuth} \&
  {Heyer}}{2015}]{Gutermuth2015}
{Gutermuth} R.~A.,  {Heyer} M.,  2015, \mn@doi [\aj]
  {10.1088/0004-6256/149/2/64}, \href
  {http://adsabs.harvard.edu/abs/2015AJ....149...64G} {149, 64}

\bibitem[\protect\citeauthoryear{{Hacar} \& {Tafalla}}{{Hacar} \&
  {Tafalla}}{2011}]{Hacar2011}
{Hacar} A.,  {Tafalla} M.,  2011, \mn@doi [\aap] {10.1051/0004-6361/201117039},
  \href {http://adsabs.harvard.edu/abs/2011A%26A...533A..34H} {533, A34}

\bibitem[\protect\citeauthoryear{{Hartmann} \& {Burkert}}{{Hartmann} \&
  {Burkert}}{2007}]{Hartmann2007}
{Hartmann} L.,  {Burkert} A.,  2007, \mn@doi [\apj] {10.1086/509321}, \href
  {http://adsabs.harvard.edu/abs/2007ApJ...654..988H} {654, 988}

\bibitem[\protect\citeauthoryear{{Heitsch}, {Hartmann}, {Slyz}, {Devriendt}  \&
  {Burkert}}{{Heitsch} et~al.}{2008}]{Heitsch2008}
{Heitsch} F.,  {Hartmann} L.~W.,  {Slyz} A.~D.,  {Devriendt} J.~E.~G.,
  {Burkert} A.,  2008, \mn@doi [\apj] {10.1086/523697}, \href
  {http://adsabs.harvard.edu/abs/2008ApJ...674..316H} {674, 316}

\bibitem[\protect\citeauthoryear{{Hennebelle}}{{Hennebelle}}{2013}]{Hennebelle2013}
{Hennebelle} P.,  2013, \mn@doi [\aap] {10.1051/0004-6361/201321292}, \href
  {http://adsabs.harvard.edu/abs/2013A%26A...556A.153H} {556, A153}

\bibitem[\protect\citeauthoryear{{Hennebelle}, {Banerjee},
  {V{\'a}zquez-Semadeni}, {Klessen}  \& {Audit}}{{Hennebelle}
  et~al.}{2008}]{Hennebelle2008}
{Hennebelle} P.,  {Banerjee} R.,  {V{\'a}zquez-Semadeni} E.,  {Klessen} R.~S.,
   {Audit} E.,  2008, \mn@doi [\aap] {10.1051/0004-6361:200810165}, \href
  {http://adsabs.harvard.edu/abs/2008A%26A...486L..43H} {486, L43}

\bibitem[\protect\citeauthoryear{{Hennemann} et~al.,}{{Hennemann}
  et~al.}{2012}]{Hennemann2012}
{Hennemann} M.,  et~al., 2012, \mn@doi [\aap] {10.1051/0004-6361/201219429},
  \href {http://adsabs.harvard.edu/abs/2012A%26A...543L...3H} {543, L3}

\bibitem[\protect\citeauthoryear{{Hernandez}, {Tan}, {Caselli}, {Butler},
  {Jim{\'e}nez-Serra}, {Fontani}  \& {Barnes}}{{Hernandez}
  et~al.}{2011}]{Hernandez2011}
{Hernandez} A.~K.,  {Tan} J.~C.,  {Caselli} P.,  {Butler} M.~J.,
  {Jim{\'e}nez-Serra} I.,  {Fontani} F.,   {Barnes} P.,  2011, \mn@doi [\apj]
  {10.1088/0004-637X/738/1/11}, \href
  {http://adsabs.harvard.edu/abs/2011ApJ...738...11H} {738, 11}

\bibitem[\protect\citeauthoryear{{Heyer} \& {Dame}}{{Heyer} \&
  {Dame}}{2015}]{Heyer2015}
{Heyer} M.,  {Dame} T.~M.,  2015, \mn@doi [\araa]
  {10.1146/annurev-astro-082214-122324}, \href
  {https://ui.adsabs.harvard.edu/abs/2015ARA&A..53..583H} {53, 583}

\bibitem[\protect\citeauthoryear{{Hildebrand}}{{Hildebrand}}{1983}]{Hildebrand1983}
{Hildebrand} R.~H.,  1983, \qjras, \href
  {http://adsabs.harvard.edu/abs/1983QJRAS..24..267H} {24, 267}

\bibitem[\protect\citeauthoryear{{Hill} et~al.,}{{Hill}
  et~al.}{2011}]{Hill2011}
{Hill} T.,  et~al., 2011, \mn@doi [\aap] {10.1051/0004-6361/201117315}, \href
  {http://adsabs.harvard.edu/abs/2011A%26A...533A..94H} {533, A94}

\bibitem[\protect\citeauthoryear{{Hily-Blant} \& {Falgarone}}{{Hily-Blant} \&
  {Falgarone}}{2007}]{HilyBlant2007}
{Hily-Blant} P.,  {Falgarone} E.,  2007, \mn@doi [\aap]
  {10.1051/0004-6361:20054565}, \href
  {http://adsabs.harvard.edu/abs/2007A%26A...469..173H} {469, 173}

\bibitem[\protect\citeauthoryear{{Hou}, {Han}  \& {Shi}}{{Hou}
  et~al.}{2009}]{Hou2009}
{Hou} L.~G.,  {Han} J.~L.,   {Shi} W.~B.,  2009, \mn@doi [\aap]
  {10.1051/0004-6361/200809692}, \href
  {http://adsabs.harvard.edu/abs/2009A%26A...499..473H} {499, 473}

\bibitem[\protect\citeauthoryear{{Inutsuka} \& {Miyama}}{{Inutsuka} \&
  {Miyama}}{1992}]{Inutsuka1992}
{Inutsuka} S.-I.,  {Miyama} S.~M.,  1992, \mn@doi [\apj] {10.1086/171162},
  \href {http://adsabs.harvard.edu/abs/1992ApJ...388..392I} {388, 392}

\bibitem[\protect\citeauthoryear{{Jackson} et~al.,}{{Jackson}
  et~al.}{2006}]{Jackson2006}
{Jackson} J.~M.,  et~al., 2006, \mn@doi [\apjs] {10.1086/500091}, \href
  {http://adsabs.harvard.edu/abs/2006ApJS..163..145J} {163, 145}

\bibitem[\protect\citeauthoryear{{Jackson}, {Finn}, {Chambers}, {Rathborne}  \&
  {Simon}}{{Jackson} et~al.}{2010}]{Jackson2010}
{Jackson} J.~M.,  {Finn} S.~C.,  {Chambers} E.~T.,  {Rathborne} J.~M.,
  {Simon} R.,  2010, \mn@doi [\apjl] {10.1088/2041-8205/719/2/L185}, \href
  {http://adsabs.harvard.edu/abs/2010ApJ...719L.185J} {719, L185}

\bibitem[\protect\citeauthoryear{{Jackson} et~al.,}{{Jackson}
  et~al.}{2013}]{Jackson2013}
{Jackson} J.~M.,  et~al., 2013, \mn@doi [\pasa] {10.1017/pasa.2013.37}, \href
  {http://adsabs.harvard.edu/abs/2013PASA...30...57J} {30, e057}

\bibitem[\protect\citeauthoryear{{Johnstone} \& {Bally}}{{Johnstone} \&
  {Bally}}{1999}]{Johnstone1999}
{Johnstone} D.,  {Bally} J.,  1999, \mn@doi [\apjl] {10.1086/311792}, \href
  {http://adsabs.harvard.edu/abs/1999ApJ...510L..49J} {510, L49}

\bibitem[\protect\citeauthoryear{{Kirk}, {Myers}, {Bourke}, {Gutermuth},
  {Hedden}  \& {Wilson}}{{Kirk} et~al.}{2013}]{Kirk2013}
{Kirk} H.,  {Myers} P.~C.,  {Bourke} T.~L.,  {Gutermuth} R.~A.,  {Hedden} A.,
  {Wilson} G.~W.,  2013, \mn@doi [\apj] {10.1088/0004-637X/766/2/115}, \href
  {http://adsabs.harvard.edu/abs/2013ApJ...766..115K} {766, 115}

\bibitem[\protect\citeauthoryear{{Koch} \& {Rosolowsky}}{{Koch} \&
  {Rosolowsky}}{2015}]{Koch2015}
{Koch} E.~W.,  {Rosolowsky} E.~W.,  2015, \mn@doi [\mnras]
  {10.1093/mnras/stv1521}, \href
  {http://adsabs.harvard.edu/abs/2015MNRAS.452.3435K} {452, 3435}

\bibitem[\protect\citeauthoryear{{Koda} et~al.,}{{Koda}
  et~al.}{2009}]{Koda2009}
{Koda} J.,  et~al., 2009, \mn@doi [\apjl] {10.1088/0004-637X/700/2/L132}, \href
  {http://adsabs.harvard.edu/abs/2009ApJ...700L.132K} {700, L132}

\bibitem[\protect\citeauthoryear{{K{\"o}nyves} et~al.,}{{K{\"o}nyves}
  et~al.}{2015}]{Konyves2015}
{K{\"o}nyves} V.,  et~al., 2015, \mn@doi [\aap] {10.1051/0004-6361/201525861},
  \href {http://adsabs.harvard.edu/abs/2015A%26A...584A..91K} {584, A91}

\bibitem[\protect\citeauthoryear{{Koyama} \& {Inutsuka}}{{Koyama} \&
  {Inutsuka}}{2000}]{Koyama2000}
{Koyama} H.,  {Inutsuka} S.-I.,  2000, \mn@doi [\apj] {10.1086/308594}, \href
  {http://adsabs.harvard.edu/abs/2000ApJ...532..980K} {532, 980}

\bibitem[\protect\citeauthoryear{{Lada}, {Alves}  \& {Lombardi}}{{Lada}
  et~al.}{2007}]{Lada2007}
{Lada} C.~J.,  {Alves} J.~F.,   {Lombardi} M.,  2007, Protostars and Planets V,
  \href {http://adsabs.harvard.edu/abs/2007prpl.conf....3L} {pp 3--15}

\bibitem[\protect\citeauthoryear{{Larson}}{{Larson}}{2005}]{Larson2005}
{Larson} R.~B.,  2005, \mn@doi [\mnras] {10.1111/j.1365-2966.2005.08881.x},
  \href {http://adsabs.harvard.edu/abs/2005MNRAS.359..211L} {359, 211}

\bibitem[\protect\citeauthoryear{{Leurini} et~al.,}{{Leurini}
  et~al.}{2019}]{Leurini2019}
{Leurini} S.,  et~al., 2019, \mn@doi [\aap] {10.1051/0004-6361/201833612},
  \href {https://ui.adsabs.harvard.edu/abs/2019A&A...621A.130L} {621, A130}

\bibitem[\protect\citeauthoryear{{Li}, {Wyrowski}, {Menten}  \&
  {Belloche}}{{Li} et~al.}{2013}]{Li2013}
{Li} G.-X.,  {Wyrowski} F.,  {Menten} K.,   {Belloche} A.,  2013, \mn@doi
  [\aap] {10.1051/0004-6361/201322411}, \href
  {http://adsabs.harvard.edu/abs/2013A%26A...559A..34L} {559, A34}

\bibitem[\protect\citeauthoryear{{Li}, {Urquhart}, {Leurini}, {Csengeri},
  {Wyrowski}, {Menten}  \& {Schuller}}{{Li} et~al.}{2016}]{Li2016}
{Li} G.-X.,  {Urquhart} J.~S.,  {Leurini} S.,  {Csengeri} T.,  {Wyrowski} F.,
  {Menten} K.~M.,   {Schuller} F.,  2016, \mn@doi [\aap]
  {10.1051/0004-6361/201527468}, \href
  {http://adsabs.harvard.edu/abs/2016A%26A...591A...5L} {591, A5}

\bibitem[\protect\citeauthoryear{{Low} et~al.,}{{Low} et~al.}{1984}]{Low1984}
{Low} F.~J.,  et~al., 1984, \mn@doi [\apjl] {10.1086/184213}, \href
  {http://adsabs.harvard.edu/abs/1984ApJ...278L..19L} {278, L19}

\bibitem[\protect\citeauthoryear{{Lumsden}, {Hoare}, {Urquhart}, {Oudmaijer},
  {Davies}, {Mottram}, {Cooper}  \& {Moore}}{{Lumsden}
  et~al.}{2013}]{Lumsden2013}
{Lumsden} S.~L.,  {Hoare} M.~G.,  {Urquhart} J.~S.,  {Oudmaijer} R.~D.,
  {Davies} B.,  {Mottram} J.~C.,  {Cooper} H.~D.~B.,   {Moore} T.~J.~T.,  2013,
  \mn@doi [\apjs] {10.1088/0067-0049/208/1/11}, \href
  {http://adsabs.harvard.edu/abs/2013ApJS..208...11L} {208, 11}

\bibitem[\protect\citeauthoryear{{May}, {Murphy}  \& {Thaddeus}}{{May}
  et~al.}{1988}]{May1988}
{May} J.,  {Murphy} D.~C.,   {Thaddeus} P.,  1988, \aaps, \href
  {http://adsabs.harvard.edu/abs/1988A%26AS...73...51M} {73, 51}

\bibitem[\protect\citeauthoryear{{McClure-Griffiths}, {Dickey}, {Gaensler},
  {Green}  \& {Haverkorn}}{{McClure-Griffiths}
  et~al.}{2006}]{McClureGriffiths2006}
{McClure-Griffiths} N.~M.,  {Dickey} J.~M.,  {Gaensler} B.~M.,  {Green} A.~J.,
   {Haverkorn} M.,  2006, \mn@doi [\apj] {10.1086/508706}, \href
  {http://adsabs.harvard.edu/abs/2006ApJ...652.1339M} {652, 1339}

\bibitem[\protect\citeauthoryear{{Men'shchikov}}{{Men'shchikov}}{2013}]{Menshchikov2013}
{Men'shchikov} A.,  2013, \mn@doi [\aap] {10.1051/0004-6361/201321885}, \href
  {http://adsabs.harvard.edu/abs/2013A%26A...560A..63M} {560, A63}

\bibitem[\protect\citeauthoryear{{Miville-Desch{\^e}nes}
  et~al.,}{{Miville-Desch{\^e}nes} et~al.}{2010}]{Miville-Deschenes2010}
{Miville-Desch{\^e}nes} M.-A.,  et~al., 2010, \mn@doi [\aap]
  {10.1051/0004-6361/201014678}, \href
  {http://adsabs.harvard.edu/abs/2010A%26A...518L.104M} {518, L104}

\bibitem[\protect\citeauthoryear{{Molinari} et~al.,}{{Molinari}
  et~al.}{2010}]{Molinari2010}
{Molinari} S.,  et~al., 2010, \mn@doi [\pasp] {10.1086/651314}, \href
  {http://adsabs.harvard.edu/abs/2010PASP..122..314M} {122, 314}

\bibitem[\protect\citeauthoryear{{Molinari}, {Schisano}, {Faustini},
  {Pestalozzi}, {di Giorgio}  \& {Liu}}{{Molinari} et~al.}{2011}]{Molinari2011}
{Molinari} S.,  {Schisano} E.,  {Faustini} F.,  {Pestalozzi} M.,  {di Giorgio}
  A.~M.,   {Liu} S.,  2011, \mn@doi [\aap] {10.1051/0004-6361/201014752}, \href
  {http://adsabs.harvard.edu/abs/2011A%26A...530A.133M} {530, A133}

\bibitem[\protect\citeauthoryear{{Molinari} et~al.,}{{Molinari}
  et~al.}{2014}]{Molinari2014}
{Molinari} S.,  et~al., 2014, \mn@doi [Protostars and Planets VI]
  {10.2458/azu_uapress_9780816531240-ch006}, \href
  {http://adsabs.harvard.edu/abs/2014prpl.conf..125M} {pp 125--148}

\bibitem[\protect\citeauthoryear{{Molinari} et~al.,}{{Molinari}
  et~al.}{2016}]{Molinari2016}
{Molinari} S.,  et~al., 2016, \mn@doi [\aap] {10.1051/0004-6361/201526380},
  \href {http://adsabs.harvard.edu/abs/2016A%26A...591A.149M} {591, A149}

\bibitem[\protect\citeauthoryear{{Molinaro} et~al.,}{{Molinaro}
  et~al.}{2016}]{Molinaro2016}
{Molinaro} M.,  et~al., 2016, in Software and Cyberinfrastructure for Astronomy
  IV. p. 99130H (\mn@eprint {arXiv} {1608.04526}), \mn@doi{10.1117/12.2231674}

\bibitem[\protect\citeauthoryear{{Motte}, {Andre}  \& {Neri}}{{Motte}
  et~al.}{1998}]{Motte1998}
{Motte} F.,  {Andre} P.,   {Neri} R.,  1998, \aap, \href
  {http://adsabs.harvard.edu/abs/1998A%26A...336..150M} {336, 150}

\bibitem[\protect\citeauthoryear{{Motte} et~al.,}{{Motte}
  et~al.}{2010}]{Motte2010}
{Motte} F.,  et~al., 2010, \mn@doi [\aap] {10.1051/0004-6361/201014690}, \href
  {http://adsabs.harvard.edu/abs/2010A%26A...518L..77M} {518, L77}

\bibitem[\protect\citeauthoryear{{Myers}}{{Myers}}{2009}]{Myers2009}
{Myers} P.~C.,  2009, \mn@doi [\apj] {10.1088/0004-637X/700/2/1609}, \href
  {http://adsabs.harvard.edu/abs/2009ApJ...700.1609M} {700, 1609}

\bibitem[\protect\citeauthoryear{{Nagai}, {Inutsuka}  \& {Miyama}}{{Nagai}
  et~al.}{1998}]{Nagai1998}
{Nagai} T.,  {Inutsuka} S.-i.,   {Miyama} S.~M.,  1998, \mn@doi [\apj]
  {10.1086/306249}, \href {http://adsabs.harvard.edu/abs/1998ApJ...506..306N}
  {506, 306}

\bibitem[\protect\citeauthoryear{{Onishi}, {Mizuno}, {Mizuno}, {Fukui}  \&
  {Nanten Team}}{{Onishi} et~al.}{2005}]{Onishi2005}
{Onishi} T.,  {Mizuno} N.,  {Mizuno} A.,  {Fukui} Y.,   {Nanten Team} 2005, in
  Protostars and Planets V Posters. p.~8301

\bibitem[\protect\citeauthoryear{{Ostriker}}{{Ostriker}}{1964}]{Ostriker1964}
{Ostriker} J.,  1964, \mn@doi [\apj] {10.1086/148057}, \href
  {http://adsabs.harvard.edu/abs/1964ApJ...140.1529O} {140, 1529}

\bibitem[\protect\citeauthoryear{{Padoan}, {Nordlund}, {R{\"o}gnvaldsson}  \&
  {Goodman}}{{Padoan} et~al.}{2001}]{Padoan2001}
{Padoan} P.,  {Nordlund} {\AA}.,  {R{\"o}gnvaldsson} {\"O}.~E.,   {Goodman} A.,
   2001, in {Montmerle} T.,  {Andr{\'e}} P.,  eds,  Astronomical Society of the
  Pacific Conference Series Vol. 243, From Darkness to Light: Origin and
  Evolution of Young Stellar Clusters. p.~279

\bibitem[\protect\citeauthoryear{{Padoan}, {Nordlund}, {Kritsuk}, {Norman}  \&
  {Li}}{{Padoan} et~al.}{2007}]{Padoan2007}
{Padoan} P.,  {Nordlund} {\AA}.,  {Kritsuk} A.~G.,  {Norman} M.~L.,   {Li}
  P.~S.,  2007, \mn@doi [\apj] {10.1086/516623}, \href
  {http://adsabs.harvard.edu/abs/2007ApJ...661..972P} {661, 972}

\bibitem[\protect\citeauthoryear{{Palmeirim} et~al.,}{{Palmeirim}
  et~al.}{2013}]{Palmeirim2013}
{Palmeirim} P.,  et~al., 2013, \mn@doi [\aap] {10.1051/0004-6361/201220500},
  \href {http://adsabs.harvard.edu/abs/2013A%26A...550A..38P} {550, A38}

\bibitem[\protect\citeauthoryear{{Paradis}, {Bernard}, {M{\'e}ny}  \&
  {Gromov}}{{Paradis} et~al.}{2011}]{Paradis2011}
{Paradis} D.,  {Bernard} J.-P.,  {M{\'e}ny} C.,   {Gromov} V.,  2011, \mn@doi
  [\aap] {10.1051/0004-6361/201116862}, \href
  {http://adsabs.harvard.edu/abs/2011A%26A...534A.118P} {534, A118}

\bibitem[\protect\citeauthoryear{{Peretto} \& {Fuller}}{{Peretto} \&
  {Fuller}}{2009}]{Peretto2009}
{Peretto} N.,  {Fuller} G.~A.,  2009, \mn@doi [\aap]
  {10.1051/0004-6361/200912127}, \href
  {http://adsabs.harvard.edu/abs/2009A%26A...505..405P} {505, 405}

\bibitem[\protect\citeauthoryear{{Peretto} et~al.,}{{Peretto}
  et~al.}{2010}]{Peretto2010}
{Peretto} N.,  et~al., 2010, \mn@doi [\aap] {10.1051/0004-6361/201014652},
  \href {http://adsabs.harvard.edu/abs/2010A%26A...518L..98P} {518, L98}

\bibitem[\protect\citeauthoryear{{Peretto} et~al.,}{{Peretto}
  et~al.}{2012}]{Peretto2012}
{Peretto} N.,  et~al., 2012, \mn@doi [\aap] {10.1051/0004-6361/201118663},
  \href {http://adsabs.harvard.edu/abs/2012A%26A...541A..63P} {541, A63}

\bibitem[\protect\citeauthoryear{{Peretto}, {Lenfestey}, {Fuller},
  {Traficante}, {Molinari}, {Thompson}  \& {Ward-Thompson}}{{Peretto}
  et~al.}{2016}]{Peretto2016}
{Peretto} N.,  {Lenfestey} C.,  {Fuller} G.~A.,  {Traficante} A.,  {Molinari}
  S.,  {Thompson} M.~A.,   {Ward-Thompson} D.,  2016, \mn@doi [\aap]
  {10.1051/0004-6361/201527064}, \href
  {http://adsabs.harvard.edu/abs/2016A%26A...590A..72P} {590, A72}

\bibitem[\protect\citeauthoryear{{Piazzo}, {Calzoletti}, {Faustini},
  {Pestalozzi}, {Pezzuto}, {Elia}, {di Giorgio}  \& {Molinari}}{{Piazzo}
  et~al.}{2015}]{Piazzo2015}
{Piazzo} L.,  {Calzoletti} L.,  {Faustini} F.,  {Pestalozzi} M.,  {Pezzuto} S.,
   {Elia} D.,  {di Giorgio} A.,   {Molinari} S.,  2015, \mn@doi [\mnras]
  {10.1093/mnras/stu2453}, \href
  {http://adsabs.harvard.edu/abs/2015MNRAS.447.1471P} {447, 1471}

\bibitem[\protect\citeauthoryear{{Pilbratt} et~al.,}{{Pilbratt}
  et~al.}{2010}]{Pilbratt2010}
{Pilbratt} G.~L.,  et~al., 2010, \mn@doi [\aap] {10.1051/0004-6361/201014759},
  \href {http://adsabs.harvard.edu/abs/2010A%26A...518L...1P} {518, L1}

\bibitem[\protect\citeauthoryear{{Pillai}, {Wyrowski}, {Carey}  \&
  {Menten}}{{Pillai} et~al.}{2006}]{Pillai2006}
{Pillai} T.,  {Wyrowski} F.,  {Carey} S.~J.,   {Menten} K.~M.,  2006, \mn@doi
  [\aap] {10.1051/0004-6361:20054128}, \href
  {http://adsabs.harvard.edu/abs/2006A%26A...450..569P} {450, 569}

\bibitem[\protect\citeauthoryear{{Planck Collaboration} et~al.,}{{Planck
  Collaboration} et~al.}{2011}]{Planck2011XXV}
{Planck Collaboration} et~al., 2011, \mn@doi [\aap]
  {10.1051/0004-6361/201116483}, \href
  {http://adsabs.harvard.edu/abs/2011A%26A...536A..25P} {536, A25}

\bibitem[\protect\citeauthoryear{{Planck Collaboration} et~al.,}{{Planck
  Collaboration} et~al.}{2014}]{Planck2014XIV}
{Planck Collaboration} et~al., 2014, \mn@doi [\aap]
  {10.1051/0004-6361/201322367}, \href
  {http://adsabs.harvard.edu/abs/2014A%26A...564A..45P} {564, A45}

\bibitem[\protect\citeauthoryear{{Planck Collaboration} et~al.,}{{Planck
  Collaboration} et~al.}{2016}]{Planck2016XXXII}
{Planck Collaboration} et~al., 2016, \mn@doi [\aap]
  {10.1051/0004-6361/201425044}, \href
  {http://adsabs.harvard.edu/abs/2016A%26A...586A.135P} {586, A135}

\bibitem[\protect\citeauthoryear{{Poglitsch} et~al.,}{{Poglitsch}
  et~al.}{2010}]{Poglitsch2010}
{Poglitsch} A.,  et~al., 2010, \mn@doi [\aap] {10.1051/0004-6361/201014535},
  \href {http://adsabs.harvard.edu/abs/2010A%26A...518L...2P} {518, L2}

\bibitem[\protect\citeauthoryear{{Pon}, {Toal{\'a}}, {Johnstone},
  {V{\'a}zquez-Semadeni}, {Heitsch}  \& {G{\'o}mez}}{{Pon}
  et~al.}{2012}]{Pon2012}
{Pon} A.,  {Toal{\'a}} J.~A.,  {Johnstone} D.,  {V{\'a}zquez-Semadeni} E.,
  {Heitsch} F.,   {G{\'o}mez} G.~C.,  2012, \mn@doi [\apj]
  {10.1088/0004-637X/756/2/145}, \href
  {http://adsabs.harvard.edu/abs/2012ApJ...756..145P} {756, 145}

\bibitem[\protect\citeauthoryear{{Ragan}, {Henning}, {Tackenberg}, {Beuther},
  {Johnston}, {Kainulainen}  \& {Linz}}{{Ragan} et~al.}{2014}]{Ragan2014}
{Ragan} S.~E.,  {Henning} T.,  {Tackenberg} J.,  {Beuther} H.,  {Johnston}
  K.~G.,  {Kainulainen} J.,   {Linz} H.,  2014, \mn@doi [\aap]
  {10.1051/0004-6361/201423401}, \href
  {http://adsabs.harvard.edu/abs/2014A%26A...568A..73R} {568, A73}

\bibitem[\protect\citeauthoryear{{Ram{\'o}n-Fox} \& {Bonnell}}{{Ram{\'o}n-Fox}
  \& {Bonnell}}{2018}]{RamonFox2018}
{Ram{\'o}n-Fox} F.~G.,  {Bonnell} I.~A.,  2018, \mn@doi [\mnras]
  {10.1093/mnras/stx2866}, \href
  {https://ui.adsabs.harvard.edu/abs/2018MNRAS.474.2028R} {474, 2028}

\bibitem[\protect\citeauthoryear{{Reid}}{{Reid}}{2013}]{Reid2012}
{Reid} M.~J.,  2013, in {de Grijs} R.,  ed.,  IAU Symposium Vol. 289, Advancing
  the Physics of Cosmic Distances. pp 188--193,
  \mn@doi{10.1017/S1743921312021369}

\bibitem[\protect\citeauthoryear{{Reid} \& {Dame}}{{Reid} \&
  {Dame}}{2016}]{Reid2016}
{Reid} M.~J.,  {Dame} T.~M.,  2016, \mn@doi [\apj]
  {10.3847/0004-637X/832/2/159}, \href
  {https://ui.adsabs.harvard.edu/abs/2016ApJ...832..159R} {832, 159}

\bibitem[\protect\citeauthoryear{{Reid} et~al.,}{{Reid}
  et~al.}{2009}]{Reid2009}
{Reid} M.~J.,  et~al., 2009, \mn@doi [\apj] {10.1088/0004-637X/700/1/137},
  \href {https://ui.adsabs.harvard.edu/abs/2009ApJ...700..137R} {700, 137}

\bibitem[\protect\citeauthoryear{{Reid} et~al.,}{{Reid}
  et~al.}{2014}]{Reid2014}
{Reid} M.~J.,  et~al., 2014, \mn@doi [\apj] {10.1088/0004-637X/783/2/130},
  \href {http://adsabs.harvard.edu/abs/2014ApJ...783..130R} {783, 130}

\bibitem[\protect\citeauthoryear{{Rigby} et~al.,}{{Rigby}
  et~al.}{2016}]{Rigby2016}
{Rigby} A.~J.,  et~al., 2016, \mn@doi [\mnras] {10.1093/mnras/stv2808}, \href
  {http://adsabs.harvard.edu/abs/2016MNRAS.456.2885R} {456, 2885}

\bibitem[\protect\citeauthoryear{{Roman-Duval}, {Jackson}, {Heyer}, {Johnson},
  {Rathborne}, {Shah}  \& {Simon}}{{Roman-Duval} et~al.}{2009}]{RomanDuval2009}
{Roman-Duval} J.,  {Jackson} J.~M.,  {Heyer} M.,  {Johnson} A.,  {Rathborne}
  J.,  {Shah} R.,   {Simon} R.,  2009, \mn@doi [\apj]
  {10.1088/0004-637X/699/2/1153}, \href
  {http://adsabs.harvard.edu/abs/2009ApJ...699.1153R} {699, 1153}

\bibitem[\protect\citeauthoryear{{Russeil} et~al.,}{{Russeil}
  et~al.}{2011}]{Russeil2011}
{Russeil} D.,  et~al., 2011, \mn@doi [\aap] {10.1051/0004-6361/201015852},
  \href {http://adsabs.harvard.edu/abs/2011A%26A...526A.151R} {526, A151}

\bibitem[\protect\citeauthoryear{{Russeil}, {Zavagno}, {M{\`e}ge}, {Poulin},
  {Molinari}  \& {Cambresy}}{{Russeil} et~al.}{2017}]{Russeil2017}
{Russeil} D.,  {Zavagno} A.,  {M{\`e}ge} P.,  {Poulin} Y.,  {Molinari} S.,
  {Cambresy} L.,  2017, \mn@doi [\aap] {10.1051/0004-6361/201730540}, \href
  {https://ui.adsabs.harvard.edu/abs/2017A&A...601L...5R} {601, L5}

\bibitem[\protect\citeauthoryear{{Salji} et~al.,}{{Salji}
  et~al.}{2015}]{Salji2015}
{Salji} C.~J.,  et~al., 2015, \mn@doi [\mnras] {10.1093/mnras/stv369}, \href
  {http://adsabs.harvard.edu/abs/2015MNRAS.449.1782S} {449, 1782}

\bibitem[\protect\citeauthoryear{{Schisano} et~al.,}{{Schisano}
  et~al.}{2014}]{Schisano2014}
{Schisano} E.,  et~al., 2014, \mn@doi [\apj] {10.1088/0004-637X/791/1/27},
  \href {http://adsabs.harvard.edu/abs/2014ApJ...791...27S} {791, 27}

\bibitem[\protect\citeauthoryear{{Schlegel}, {Finkbeiner}  \&
  {Davis}}{{Schlegel} et~al.}{1998}]{Schlegel1998}
{Schlegel} D.~J.,  {Finkbeiner} D.~P.,   {Davis} M.,  1998, \mn@doi [\apj]
  {10.1086/305772}, \href {http://adsabs.harvard.edu/abs/1998ApJ...500..525S}
  {500, 525}

\bibitem[\protect\citeauthoryear{{Schneider} \& {Elmegreen}}{{Schneider} \&
  {Elmegreen}}{1979}]{Schneider1979}
{Schneider} S.,  {Elmegreen} B.~G.,  1979, \mn@doi [\apjs] {10.1086/190609},
  \href {http://adsabs.harvard.edu/abs/1979ApJS...41...87S} {41, 87}

\bibitem[\protect\citeauthoryear{{Schneider} et~al.,}{{Schneider}
  et~al.}{2012}]{Schneider2012}
{Schneider} N.,  et~al., 2012, \mn@doi [\aap] {10.1051/0004-6361/201118566},
  \href {http://adsabs.harvard.edu/abs/2012A%26A...540L..11S} {540, L11}

\bibitem[\protect\citeauthoryear{{Schneider} et~al.,}{{Schneider}
  et~al.}{2013}]{Schneider2013}
{Schneider} N.,  et~al., 2013, \mn@doi [\apjl] {10.1088/2041-8205/766/2/L17},
  \href {http://adsabs.harvard.edu/abs/2013ApJ...766L..17S} {766, L17}

\bibitem[\protect\citeauthoryear{{Schuller} et~al.,}{{Schuller}
  et~al.}{2009}]{Schuller2009}
{Schuller} F.,  et~al., 2009, \mn@doi [\aap] {10.1051/0004-6361/200811568},
  \href {http://adsabs.harvard.edu/abs/2009A%26A...504..415S} {504, 415}

\bibitem[\protect\citeauthoryear{{Schuller} et~al.,}{{Schuller}
  et~al.}{2017}]{Schuller2017}
{Schuller} F.,  et~al., 2017, \mn@doi [\aap] {10.1051/0004-6361/201628933},
  \href {http://adsabs.harvard.edu/abs/2017A%26A...601A.124S} {601, A124}

\bibitem[\protect\citeauthoryear{{Shandarin} \& {Yess}}{{Shandarin} \&
  {Yess}}{1998}]{Shandarin1998}
{Shandarin} S.~F.,  {Yess} C.,  1998, \mn@doi [\apj] {10.1086/306135}, \href
  {http://adsabs.harvard.edu/abs/1998ApJ...505...12S} {505, 12}

\bibitem[\protect\citeauthoryear{{Simon}, {Jackson}, {Bania}, {Clemens}  \&
  {Heyer}}{{Simon} et~al.}{2005}]{Simon2005}
{Simon} R.,  {Jackson} J.~M.,  {Bania} T.~M.,  {Clemens} D.~P.,   {Heyer}
  M.~H.,  2005, Astronomische Nachrichten, \href
  {http://adsabs.harvard.edu/abs/2005AN....326..668S} {326, 668}

\bibitem[\protect\citeauthoryear{{Smith}, {Glover}, {Clark}, {Klessen}  \&
  {Springel}}{{Smith} et~al.}{2014}]{Smith2014}
{Smith} R.~J.,  {Glover} S.~C.~O.,  {Clark} P.~C.,  {Klessen} R.~S.,
  {Springel} V.,  2014, \mn@doi [\mnras] {10.1093/mnras/stu616}, \href
  {http://adsabs.harvard.edu/abs/2014MNRAS.441.1628S} {441, 1628}

\bibitem[\protect\citeauthoryear{{Sokolov} et~al.,}{{Sokolov}
  et~al.}{2018}]{Sokolov2018}
{Sokolov} V.,  et~al., 2018, \mn@doi [\aap] {10.1051/0004-6361/201832746},
  \href {http://adsabs.harvard.edu/abs/2018A%26A...611L...3S} {611, L3}

\bibitem[\protect\citeauthoryear{{Sousbie}}{{Sousbie}}{2011}]{Sousbie2011}
{Sousbie} T.,  2011, \mn@doi [\mnras] {10.1111/j.1365-2966.2011.18394.x}, \href
  {http://adsabs.harvard.edu/abs/2011MNRAS.414..350S} {414, 350}

\bibitem[\protect\citeauthoryear{{Steinacker}, {Bacmann}, {Henning}  \&
  {Heigl}}{{Steinacker} et~al.}{2016}]{Steinacker2016}
{Steinacker} J.,  {Bacmann} A.,  {Henning} T.,   {Heigl} S.,  2016, \mn@doi
  [\aap] {10.1051/0004-6361/201628815}, \href
  {http://adsabs.harvard.edu/abs/2016A%26A...593A...6S} {593, A6}

\bibitem[\protect\citeauthoryear{{Stepnik} et~al.,}{{Stepnik}
  et~al.}{2003}]{Stepnik2003}
{Stepnik} B.,  et~al., 2003, \mn@doi [\aap] {10.1051/0004-6361:20021309}, \href
  {http://adsabs.harvard.edu/abs/2003A%26A...398..551S} {398, 551}

\bibitem[\protect\citeauthoryear{{Traficante} et~al.,}{{Traficante}
  et~al.}{2011}]{Traficante2011}
{Traficante} A.,  et~al., 2011, \mn@doi [\mnras]
  {10.1111/j.1365-2966.2011.19244.x}, \href
  {http://adsabs.harvard.edu/abs/2011MNRAS.416.2932T} {416, 2932}

\bibitem[\protect\citeauthoryear{{Traficante}, {Fuller}, {Peretto}, {Pineda}
  \& {Molinari}}{{Traficante} et~al.}{2015}]{Traficante2015}
{Traficante} A.,  {Fuller} G.~A.,  {Peretto} N.,  {Pineda} J.~E.,   {Molinari}
  S.,  2015, \mn@doi [\mnras] {10.1093/mnras/stv1158}, \href
  {http://adsabs.harvard.edu/abs/2015MNRAS.451.3089T} {451, 3089}

\bibitem[\protect\citeauthoryear{{Ungerechts} \& {Thaddeus}}{{Ungerechts} \&
  {Thaddeus}}{1987}]{Ungerechts1987}
{Ungerechts} H.,  {Thaddeus} P.,  1987, \mn@doi [\apjs] {10.1086/191176}, \href
  {http://adsabs.harvard.edu/abs/1987ApJS...63..645U} {63, 645}

\bibitem[\protect\citeauthoryear{{Urquhart}, {Figura}, {Moore}, {Hoare},
  {Lumsden}, {Mottram}, {Thompson}  \& {Oudmaijer}}{{Urquhart}
  et~al.}{2014}]{Urquhart2014}
{Urquhart} J.~S.,  {Figura} C.~C.,  {Moore} T.~J.~T.,  {Hoare} M.~G.,
  {Lumsden} S.~L.,  {Mottram} J.~C.,  {Thompson} M.~A.,   {Oudmaijer} R.~D.,
  2014, \mn@doi [\mnras] {10.1093/mnras/stt2006}, \href
  {http://adsabs.harvard.edu/abs/2014MNRAS.437.1791U} {437, 1791}

\bibitem[\protect\citeauthoryear{{Urquhart} et~al.,}{{Urquhart}
  et~al.}{2018}]{Urquhart2018}
{Urquhart} J.~S.,  et~al., 2018, \mn@doi [\mnras] {10.1093/mnras/stx2258},
  \href {https://ui.adsabs.harvard.edu/abs/2018MNRAS.473.1059U} {473, 1059}

\bibitem[\protect\citeauthoryear{{V{\'a}zquez-Semadeni}, {G{\'o}mez},
  {Jappsen}, {Ballesteros-Paredes}, {Gonz{\'a}lez}  \&
  {Klessen}}{{V{\'a}zquez-Semadeni} et~al.}{2007}]{VazquezSemaden2007}
{V{\'a}zquez-Semadeni} E.,  {G{\'o}mez} G.~C.,  {Jappsen} A.~K.,
  {Ballesteros-Paredes} J.,  {Gonz{\'a}lez} R.~F.,   {Klessen} R.~S.,  2007,
  \mn@doi [\apj] {10.1086/510771}, \href
  {http://adsabs.harvard.edu/abs/2007ApJ...657..870V} {657, 870}

\bibitem[\protect\citeauthoryear{{V{\'a}zquez-Semadeni}, {Banerjee},
  {G{\'o}mez}, {Hennebelle}, {Duffin}  \& {Klessen}}{{V{\'a}zquez-Semadeni}
  et~al.}{2011}]{VazquezSemadeni2011}
{V{\'a}zquez-Semadeni} E.,  {Banerjee} R.,  {G{\'o}mez} G.~C.,  {Hennebelle}
  P.,  {Duffin} D.,   {Klessen} R.~S.,  2011, \mn@doi [\mnras]
  {10.1111/j.1365-2966.2011.18569.x}, \href
  {http://adsabs.harvard.edu/abs/2011MNRAS.414.2511V} {414, 2511}

\bibitem[\protect\citeauthoryear{{V{\'a}zquez}, {May}, {Carraro}, {Bronfman},
  {Moitinho}  \& {Baume}}{{V{\'a}zquez} et~al.}{2008}]{Vazquez2008}
{V{\'a}zquez} R.~A.,  {May} J.,  {Carraro} G.,  {Bronfman} L.,  {Moitinho} A.,
   {Baume} G.,  2008, \mn@doi [\apj] {10.1086/524003}, \href
  {http://adsabs.harvard.edu/abs/2008ApJ...672..930V} {672, 930}

\bibitem[\protect\citeauthoryear{{Walsh} et~al.,}{{Walsh}
  et~al.}{2011}]{Walsh2011}
{Walsh} A.~J.,  et~al., 2011, \mn@doi [\mnras]
  {10.1111/j.1365-2966.2011.19115.x}, \href
  {http://adsabs.harvard.edu/abs/2011MNRAS.416.1764W} {416, 1764}

\bibitem[\protect\citeauthoryear{{Wang} et~al.,}{{Wang}
  et~al.}{2014}]{Wang2014}
{Wang} K.,  et~al., 2014, \mn@doi [\mnras] {10.1093/mnras/stu127}, \href
  {http://adsabs.harvard.edu/abs/2014MNRAS.439.3275W} {439, 3275}

\bibitem[\protect\citeauthoryear{{Wang}, {Testi}, {Ginsburg}, {Walmsley},
  {Molinari}  \& {Schisano}}{{Wang} et~al.}{2015}]{Wang2015}
{Wang} K.,  {Testi} L.,  {Ginsburg} A.,  {Walmsley} C.~M.,  {Molinari} S.,
  {Schisano} E.,  2015, \mn@doi [\mnras] {10.1093/mnras/stv735}, \href
  {http://adsabs.harvard.edu/abs/2015MNRAS.450.4043W} {450, 4043}

\bibitem[\protect\citeauthoryear{{Wang}, {Testi}, {Burkert}, {Walmsley},
  {Beuther}  \& {Henning}}{{Wang} et~al.}{2016}]{Wang2016}
{Wang} K.,  {Testi} L.,  {Burkert} A.,  {Walmsley} C.~M.,  {Beuther} H.,
  {Henning} T.,  2016, \mn@doi [\apjs] {10.3847/0067-0049/226/1/9}, \href
  {http://adsabs.harvard.edu/abs/2016ApJS..226....9W} {226, 9}

\bibitem[\protect\citeauthoryear{{Wienen} et~al.,}{{Wienen}
  et~al.}{2015}]{Wienen2015}
{Wienen} M.,  et~al., 2015, \mn@doi [\aap] {10.1051/0004-6361/201424802}, \href
  {https://ui.adsabs.harvard.edu/abs/2015A&A...579A..91W} {579, A91}

\bibitem[\protect\citeauthoryear{{Wilcock} et~al.,}{{Wilcock}
  et~al.}{2012}]{Wilcock2012}
{Wilcock} L.~A.,  et~al., 2012, \mn@doi [\mnras]
  {10.1111/j.1365-2966.2012.20680.x}, \href
  {http://adsabs.harvard.edu/abs/2012MNRAS.422.1071W} {422, 1071}

\bibitem[\protect\citeauthoryear{{Wu} et~al.,}{{Wu} et~al.}{2014}]{Wu2014}
{Wu} Y.~W.,  et~al., 2014, \mn@doi [\aap] {10.1051/0004-6361/201322765}, \href
  {https://ui.adsabs.harvard.edu/abs/2014A&A...566A..17W} {566, A17}

\bibitem[\protect\citeauthoryear{{Xu} et~al.,}{{Xu} et~al.}{2013}]{Xu2013}
{Xu} Y.,  et~al., 2013, \mn@doi [\apj] {10.1088/0004-637X/769/1/15}, \href
  {http://adsabs.harvard.edu/abs/2013ApJ...769...15X} {769, 15}

\bibitem[\protect\citeauthoryear{{Zucker}, {Battersby}  \& {Goodman}}{{Zucker}
  et~al.}{2015}]{Zucker2015}
{Zucker} C.,  {Battersby} C.,   {Goodman} A.,  2015, \mn@doi [\apj]
  {10.1088/0004-637X/815/1/23}, \href
  {http://adsabs.harvard.edu/abs/2015ApJ...815...23Z} {815, 23}

\makeatother
\end{thebibliography}

\appendix

\section{The Hi-GAL mosaics}\label{App:mosaics}

Here we describe the strategy adopted to create the Hi-GAL mosaics, and to compute the column density and temperature maps used for this work. 
The Hi-GAL survey was designed to map the entire Galactic plane in single blocks, named tiles, each one covering a region of $2\fdg 2 \times 2\fdg 2$ and observed separately by scanning in two orthogonal directions \citep{Molinari2010}. The tiles were selected to overcome the offset between the fields of view of the two photometric instruments, PACS and SPIRE, due to their different position on the focal plane that reduces the effective area mapped by all five bands in a single tile.

We then divided the entire Hi-GAL survey into large blocks, the mosaic footprints, each one spanning about 10 degrees of Galactic longitude, and we used such footprints for the runs with UNIMAP.  The mosaic borders are set in order to include an integer number of Hi-GAL tiles in order to process data from an observing run altogether. 
We selected the footprints to duplicate any tile lying at the mosaic borders and also in the neighbour mosaics to minimize the possibility of missing extend structures. With such a choice, the Hi-GAL mosaics built for this work include a number of single tiles that range between $5$ and $6$. Table~\ref{Tab:Mosaics} shows a summary of the mosaics computed, with their extension in Galactic longitude and the single-tile dataset used to create each of them. 

While, in theory, it is possible to use UNIMAP to process together any group of Hi-GAL tiles, there are limits imposed by computation resources and time. Both of them vastly grow when processing the shorter-wavelength data, i.e., the PACS bands. A further problem is produced by the peculiarity of the Hi-GAL dataset that shows a huge dynamic intensity range all along the Galactic plane. 
In fact, all along the plane there are several bright regions of very
high emission embedded within more tenuous ones. This indeed is the case for the outer Galaxy where the background itself is faint. Such a large discrepancy in emission level between different portions of the map raises difficulties for any map-making algorithm, particularly when computing the final synthetic map from which the correlated noise has to be removed. Such is the case of UNIMAP map maker, which is based on an iterative solver that is used to reconstruct a synthetic ``de-noised'' image, starting from an initial guess \citep{Piazzo2015}. UNIMAP has different prescriptions for the initial guess map and the number of iterations needed to reach the convergence of the GLS iterative solver, depending on the overall emission level of the map. Such a problem is obviously amplified when several data covering extended portions of the plane have to be treated simultaneously, as is the case in the large mosaics that have been computed. In our first attempt to produce mosaics processing the Hi-GAL tiles altogether in a single UNIMAP run, we often found distortions. These appeared as large-scale intensity gradients spanning whole mosaics, and prior to the recalibration step using the IRAS/{\it Planck} data \citep{Bernard2010}. We determined that the distortions are produced by a poor convergence of the UNIMAP iterative solver and they can be generally overcome with an increase in the number of iterations performed by the GLS solver or by a more suitable choice of the initial guess. Nevertheless, these choices, when practicable, vastly increase the already long computation time even on a cluster machine. 

For such reasons,  we adopted a different strategy in computing our Hi-GAL large mosaics. We first determine the final mosaic footprint as the region spanned by the overall group of Hi-GAL tiles to be combined together. 
Then, we perform separate UNIMAP runs, each one combining only the data of two adjacent Hi-GAL tiles and projecting them on a common footprint. We call these maps produced from a single UNIMAP run {\it texels}. If the mosaic spans over {\it N} tiles, we compute a total of {\it N-1} texels with the data of each single Hi-GAL tile processed at most in two independent runs. For example, to compute the mosaic {\it ``m321330''} we processed together, respectively, the HIGAL tiles l321 with l323, l323 with l325, l325 with l327 and l327 with l330. We found that such a solution represents the best compromise between computational time and high quality of the final product and,
in fact, UNIMAP reaches a convergence for almost all the texels with 300 -- 400 iterations of the GLS solver. Then we absolutely calibrated each {\it texel} by applying a linear transformation with gains
and offsets values determined by comparing {\em Herschel} with IRAS and {\it Planck} data, following the prescription of \citet{Bernard2010}. Finally, we merged together all {\it texels} in the output mosaic, computing a weighted mean in their overlapping regions. 

\begin{table*}
\caption{List of the Hi-GAL mosaics produced with the UNIMAP mapmaker in this work. We list the raw Hi-GAL data used and the final coverage of the footprint  for each mosaic.}
\label{Tab:Mosaics}
\begin{center}
\space
\begin{tabular}{c l c c c c }
\hline
Mosaic Name  & Single Hi-GAL tiles  &  Number of Hi-GAL tiles & Galactic longitude & $l_{min}$ & $l_{max}$ \\ 
  & & & extension & & \\
  & &  & (degrees) & (degrees) & (degrees) \\
\hline
m349358   &      l349 l352 l354 l356 l358 &  5 &      & & \\   
m341352   &      l341 l343 l345 l347 l349 l352 & 6  &  13.8  & 339.6  & 353.4   \\ 
m330341   &      l330 l332 l334 l336 l338 l341 &  6  &  13.8  & 328.6  & 342.4   \\ 
m321330   &      l321 l323 l325 l327 l330 & 5  &   11.4  & 319.9  & 331.4   \\ 
m310321   &     l310 l312 l314 l316 l319 l321 & 6 &  13.5  & 309.0  & 322.5   \\ 
m301310   &      l301 l303 l305 l308 l310 & 5 & 11.2  &  300.2  & 311.4   \\ 
m290301   &      l290 l292 l294 l297 l299 l301 & 6 & 13.4  &  289.2  & 302.6   \\ 
m281290   &     l281 l283 l286 l288 l290 &  5 &  11.3  & 280.3  & 291.6   \\ 
m270281   &      l270 l272 l275 l277 l279 l281 & 6 &  13.5  & 269.3  & 282.8   \\ 
m261270   &      l261 l264 l266 l268 l270 & 5 & 11.3  &  260.6  & 271.9   \\ 
m250261   &      l250 l253 l255 l257 l259 l261 & 6 &  13.4  & 249.6  & 263.0   \\ 
m239250   &      l239 l242 l244 l246 l248 l250 & 6 & 13.4  &  238.6  & 252.0   \\ 
m231242   &      l231 l233 l235 l237 l239 l242 &  6 &  13.4  & 229.8  & 243.2   \\ 
m220231   &      l220 l222 l224 l226 l228 l231 & 6 &  13.3  &  218.8  & 232.2   \\ 
m211220   &      l211 l213 l215 l217 l220 & 5  & 11.2  & 210.0  & 221.2   \\ 
m200211   &      l200 l202 l204 l206 l209 l211 & 6 &  13.6  & 198.8  & 212.4   \\ 
m191200   &      l191 l193 l195 l198 l200 & 5  & 11.6  & 190.0  & 201.6   \\ 
m180191   &      l180 l182 l184 l187 l189 l191 & 6 & 13.8  & 179.0  & 192.8   \\ 
m171180   &      l171 l173 l176 l178 l180 & 5  & 11.6  &170.2  & 181.8   \\ 
m160171   &     l160 l162 l165 l167 l169 l171 & 6 & 13.7  & 159.3  & 173.0   \\ 
m151160   &       l151 l154 l156 l158 l160 & 5 &  11.3  & 150.6  & 161.9   \\
m140151   &      l140 l143 l145 l147 l149 l151 & 6 & 13.5  & 139.6  & 153.0   \\ 
m129140   &      l129 l132 l134 l136 l138 l140 & 6  & 13.4  & 128.6  & 142.0   \\ 
m121132   &      l121 l123 l125 l127 l129 l132 & 6 &  13.4  & 119.8  & 133.2   \\ 
m110121   &      l110 l112 l114 l116 l118 l121 & 6 & 13.5  &   108.7  & 122.2   \\ 
m101110   &      l101 l103 l105 l107 l110 & 5 & 11.3  &  99.9  & 111.2   \\ 
m090101  &       l090 l092 l094 l096 l099 l101 & 6 &13.6  &  88.9  & 102.5   \\
m081090  &       l081 l083 l085 l088 l090 & 5 &  11.5  & 80.0  & 91.5   \\ 
m070081  &       l070 l072 l074 l077 l079 l081 & 6 &  13.8  & 69.0  & 82.8   \\ 
m060070  &       l060 l061 l063 l066 l068 l070 & 6 &  12.0  & 59.8  & 71.8   \\ 
m050060  &       l050 l052 l055 l057 l059 l060 & 6 & 11.3  &  49.4  & 60.7   \\ 
m041050  &       l041 l044 l046 l048 l050 & 5  & 11.2  &  40.5  & 51.8   \\ 
m030041  &       l030 l031 l033 l035 l037 l039 l041 & 7  & 14.4  &   28.7  & 43.1   \\ 
m019030  &       l019 l022 l024 l026 l028 l030 & 6 &  12.9  &  18.4  & 31.3   \\ 
m011022  &       l011 l013 l015 l017 l019 l022 & 6 & 13.8  &  9.6  & 23.4   \\
m000011  &       l000 l002 l004 l006 l008 l011 & 6 &      & & \\
\hline 
\end{tabular}
\end{center}
\footnotesize{}
\end{table*}

\newpage

\section{Parameters for filament extraction}\label{App:Parameters}
\label{Sect:FeatureExtraction}

We discuss here our choice of the parameters adopted to run the filament extraction algorithm and to identify the regions from which the candidate filaments are finally selected. 
The two parameters required by the algorithm are the threshold level, $T$, to be applied to the eigenvalue $\lambda_{a}$ map, and the dilation parameter, $D$, to extend the initial mask to ensure that the entire filament area is included its borders. 

\subsection{The choice of threshold level}

The adopted algorithm extracts the candidate regions by thresholding the map of  $\lambda_{a}$ (in absolute value) above a certain level $T$. The cut-off defines the total number of candidate regions: the lower is its value $T$, the larger is the final number of candidate regions. Moreover, $T$ can influence the total area (and the shape) of each initial mask. The choice of $T$ is a first critical step to define a sample of candidate regions that is the most complete and, at same time, has the least number of false detection as possible. Obviously, the threshold cannot be lowered indefinitely: under a certain limit the neighbouring regions start merging together until they cover the entire map. 

We point out that it is not possible to directly connect the thresholding of the $\lambda_{a}(x,y)$ map to the same operation on the intensity map $I(x,y)$, since the first is obtained through a non-linear transformation (the diagonalization) applied to the latter. In other words, changes in $T$ do not correspond directly to cuts in $I(x,y)$ in similar proportion, as happens in the more familiar case of thresholding of intensity maps. The reason behind this is that $\lambda_{a}(x,y)$ is a measurement (through the second derivative) of the variations of $I(x,y)$ with respect to the immediate neighbourhood. This implies that a lower threshold detects features where the intensity varies more smoothly (with smaller variation) with respect to their surroundings instead of fainter regions. The detections depend only partially on the region absolute intensity. The brightness of the region can still be a factor in whether the threshold is passed or not, but the fact that a lower $T$ translates to the identification of fainter objects is strictly true only for features showing the same {\it relative} variations.
There is a subtle link between the variation of $I(x,y)$ and the contrast that we introduce in Sect.\,\ref{Sect:Contrast}, that we attempt to exploit to characterize the output of the detection. We expect that, in general, small values of $T$ are able to detect any slight variation in the map, including cosmetic artefacts or random fluctuations of $I(x,y)$. However, these  should be composed of a small number of pixels and theoretically can be excluded.

We aim to determine a consistent threshold $T$ able to: 1) reduce the impact of the random fluctuations; 2) adapt to the variable properties of the Hi-GAL maps that show  changes in average intensity and noise depending on their Galactic longitude.  Therefore we require to connect $T$ to the local fluctuations of the map and to estimate the probability that a pixel is above $T$ due to random effects. This step would be trivial if the random noise present in $\lambda_{a}$ map follows a known probability distribution.  However, we discussed above that the transformation $I(x,y)\rightarrow\lambda_{a}$ is not linear, hence we expect that the noise probability distribution is not preserved.

Our first step is to characterize how the diagonalization of $H(x,y)$ impacts the noise distribution $N(x,y)$ present in the data. With this aim, we run our algorithm on several simulated maps composed of pure Gaussian noise $N^{G}(x,y)$, assuming different amplitudes and standard deviations $\sigma_{\rm noise}$. We analyse the distributions $D(\lambda^{N}_{a})$ of $\lambda^{N}_{a}$ for different noise parameters, finding that in all cases they are well approximated by Gaussian functions, despite the aforementioned non-linearity. The major important effect of the transformation is to broadene the distribution  of $D(\lambda^{N}_{a})$.  We also noticed marginal differences for low negative values of $\lambda^{N}_{a}$, where the $D(\lambda^{N}_{a})$ exceeds the Gaussian distribution by $\sim$5 per cent. This analysis indicates that the algorithm changes the Gaussian noise present in an image by increasing its dispersion, $\sigma_{noise}^{\lambda_{a}}$. We measured that the dispersion on $\lambda_{a}$ map is related to the initial noise by  $\sigma_{noise}^{\lambda_{a}}\approx\,2.49\,\times\,\sigma_{\rm noise}$.

This analysis induced us to assume that, to a good approximation, $\lambda^{N}_{a}$ is normally distributed, therefore we could adopt $T'\,=\,3\times\,\sigma_{\lambda_{a}}$ as a sufficiently robust cut-off, avoiding an excessive number of false detections due to random fluctuations. We verify that the probability $p'$ for a single pixel to be above the threshold $T'$ due to random fluctuations is small, despite the presence of minor differences between the $D(\lambda^{N}_{a})$ and a Gaussian function. Assuming that $D(\lambda^{N}_{a})$ is the probability distribution function for $\lambda_{a}$, $p'$ is found to be $\leq 0.19$ per cent, a value that is only marginally higher than the case of a pure Gaussian distribution ($p^{G}\approx 0.135$ per cent). 
This measured probability means that we should expect about $\sim$6000 pixels above $T'$ for each Hi-GAL mosaic, composed of about $\sim\,3\,\times\,10^{6}$ pixels. However, we point out that these false detections should be randomly distributed over the entire mosaic and we expect that they would be mostly isolated due to their limited number.

Strictly speaking,  $D(\lambda_{a})$ should not be considered as a normal distribution and neither should it be adopted as a probability distribution function. In fact, the  $\lambda_{a}$ in any position $(x',y')$ of map is not statistically independent from other values, since it is derived by combining the $I(x,y)$ of the closest neighbouring pixels $[x'\pm\Delta, y'\pm\Delta]$. This fact  implies that a random fluctuation of $I$ in an individual pixel $(x',y')$ will modify the probability of $\lambda_{a}$ in all the adjacent pixels, generally increasing it. In other words, a single random event at a single position, if strong enough, could produce candidate regions wider than a pixel. We recovered this in our noise simulations where we extracted regions with a typical size between 3 and 5 pixels by assuming a threshold of $T'\,=\,3\times\,\sigma_{\lambda_{a}}$. The extracted regions are rarely large; we estimate that regions larger than 15 pixels are found in less than $\leq$\,2 per cent of all the cases that overcome the threshold $T'$. If we apply this probability to the case of the Hi-GAL mosaic discussed above, the $\sim$6000 pixels above $T'$
potentially aggregate into $\sim$400 regions, thus we expect $\leq$8 candidate regions that are artefacts caused by random fluctuation.

We conclude from this discussion that we can adopt for our purposes a threshold $T\,=\,3\times\,\sigma_{\lambda_{a}}$, with $\sigma_{\lambda_{a}}$ a proper estimate of the fluctuations of $\lambda_{a}$, as long we filter out regions with a small area from the resulting extraction. 

Furthermore, we empirically tested different threshold levels on mosaics in the crowded and bright inner Galaxy and in the low-brightness regions of the GP by visually inspecting the results. Despite this approach not being rigorous and quite subjective, we determined that  thresholds between 2.8 and 3.2 times the $\sigma_{\lambda_{a}}$ were able to identify the majority of the structures. The threshold should not be selected beyond these values in our opinion.  For lower values we noticed a rapid increase of dubious cases, while for higher values we start missing 
features that look like true filaments by eye. These tests confirm the indications obtained from the statistical analysis described above, so it strengthens our decision to process the dataset with $T\,=\,3\times\,\sigma_{\lambda^{l}_{a}}$, removing any candidate with an area smaller than 15 pixels.

\subsection{Estimation of the fluctuation of $\lambda_{a}$}

\begin{figure} 
\includegraphics[width=0.5\textwidth]{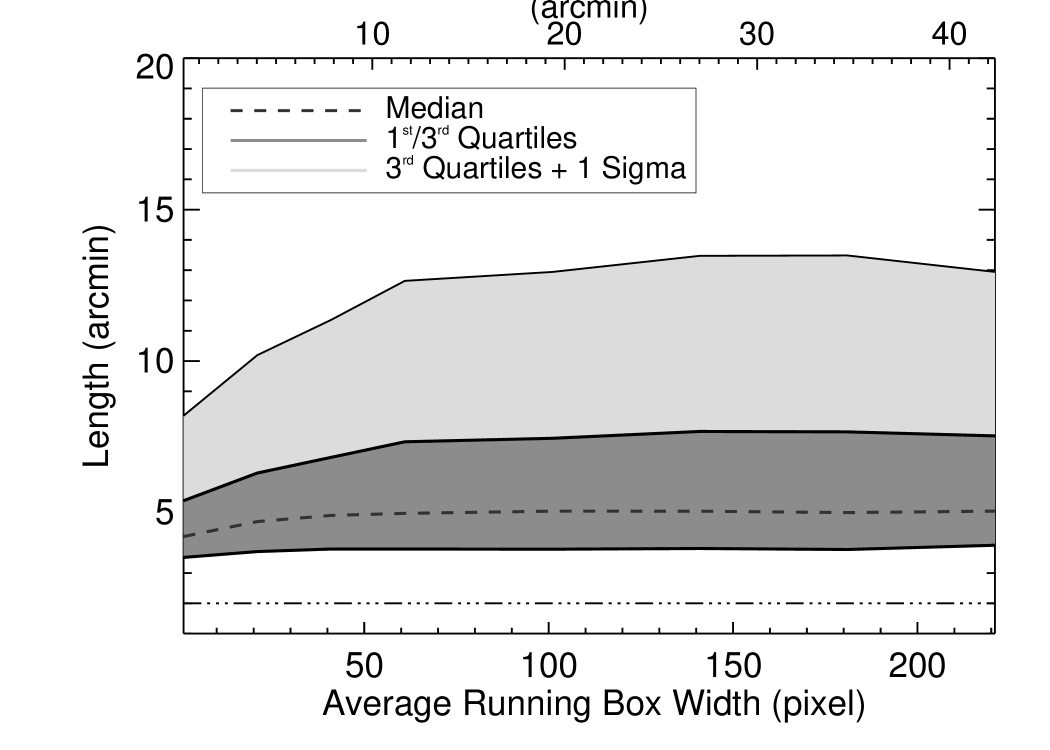}
\caption{Distribution of the angular size of the candidates extracted with the same threshold ($T\,=\,3\times\,\sigma_{\lambda^{l}_{a}}$) in the region of the GP between $50^{\circ}\,\leq\,l\,\leq\,60^{\circ}$ as a function of the width of the running box where $\sigma_{\lambda^{l}_{a}(x,y)}$ is estimated. The dark grey area shows the amplitude, defined as the distance between the $1^{st}$ and $3^{rd}$ quartiles,  of the distribution changes with the different $W$. The light grey area shows how the tail of the distribution extends towards larger objects. The median of the angular sizes and the adopted cut-off on region sizes (see Sect.~\ref{Sect:SelCriteria}) are shown by dashed and dot-dashed line, respectively.}
\label{Fig:LengthCutsWidthBox}
\end{figure}

The statistical analysis presented above proves that the threshold $T$ should be determined from an estimate of the fluctuation of  $\sigma_{\lambda_{a}}(x,y)$. However, the emission and its fluctuations vary a lot along the entire GP, even when one restricts to a single mosaic with extent $\sim10$ degrees of Galactic longitude.  The large dynamic range and variability of the emission has severe consequences for the observed distribution of $\lambda_{a}$. In particular, we notice that  it is not appropriate to adopt a single  $\sigma_{\lambda_{a}}$ for an entire map. Instead, we decided to adopt a local value for $\sigma_{\lambda^{l}_{a}}(x,y)$ estimated over running "boxes" within the map. With this approach, the threshold level $T(x,y)$ adapts to the surrounding emission and its local variations. The running box should be large enough to keep the statistical significance of $\sigma_{\lambda_{a}}(x,y)$ and to cover an area larger than filaments.  The size of the running box, $W$, can potentially influence the outputs of the extraction. In particular, it is possible that a feature is split into multiple objects if the threshold changes abruptly along its length. We tested this potential bias by comparing the results of the extractions while changing the width $W$ from $20$ to $240$ pixels, equal to $\sim4$-$25$\,arcmin. The results are shown in Fig.\,\ref{Fig:LengthCutsWidthBox}, where we present how the distribution of angular size varies as function of $W$ for the objects extracted in one field. As expected, this distribution is influenced when $\sigma_{\lambda_{a}}(x,y)$ is estimated in small boxes: the median and third quartile of the distribution increase for $W < 60$ pixels,  i.e., $\sim$ 0.2\,degree. However, this trend flattens for $W > 60$, where the statistical properties of the distribution are not affected anymore. Therefore we decided to adopt $W\,=\,61$ pixels to generate our catalogue. We stress that the width of $W$ does not introduce an exact cut-off on the sizes of the detected  feature. Indeed, we recover about $\sim$2,300 objects extending more than the adopted $W$, i.e., 0.2\,degree, as shown in  Fig.\,\ref{Fig:AngularLength}.

\subsection{Extension of the inner mask: the dilation parameter }

\begin{figure} 
\includegraphics[width=0.4\textwidth]{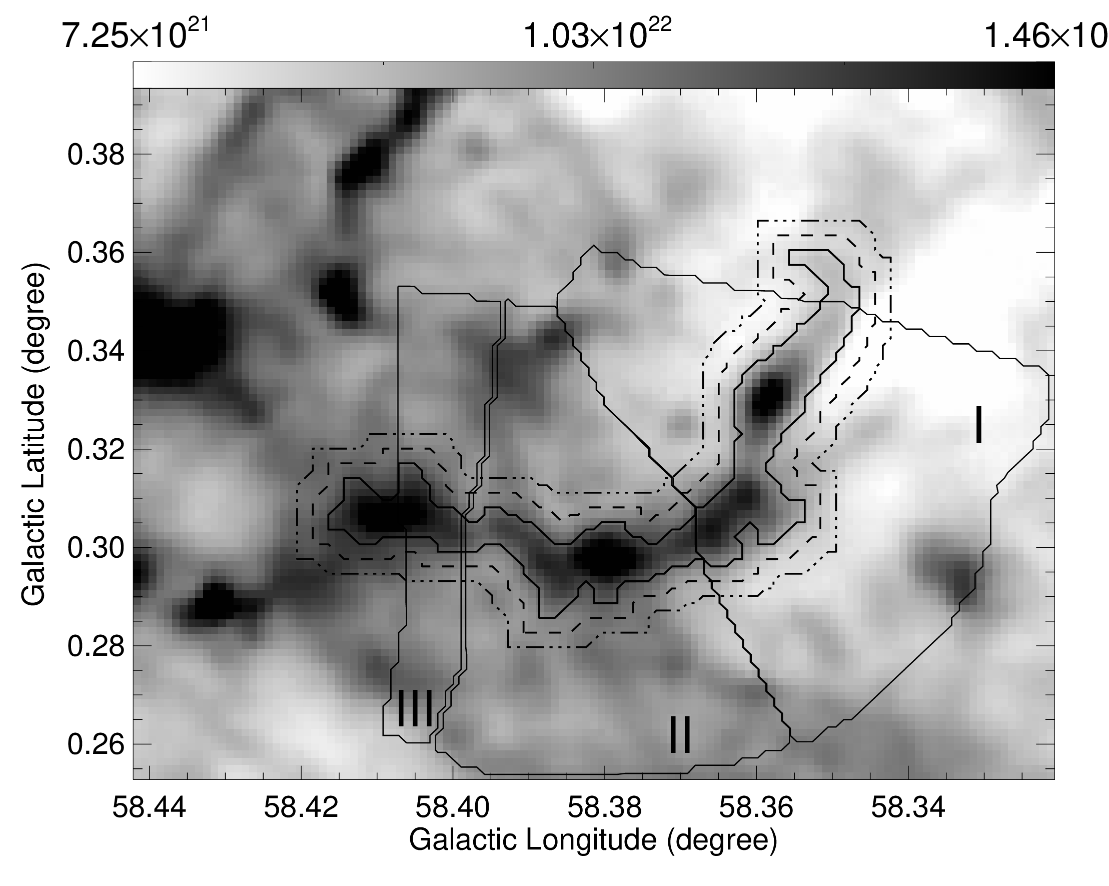}
\caption{Example of one of the candidate filaments used to determine the best choice for the dilation parameter $D$. The border of the extended mask relative to $D\,=1,3,5$ pixel are draw with black full, dashed and dot-dashed line, respectively. The entire region is split into three sections used to compute the the radial profiles shown in Fig.\,\ref{Fig:ProfilesDfact}. }
\label{Fig:TestFilDfact}
\end{figure}

\begin{figure*} 
\includegraphics[width=0.98\textwidth]{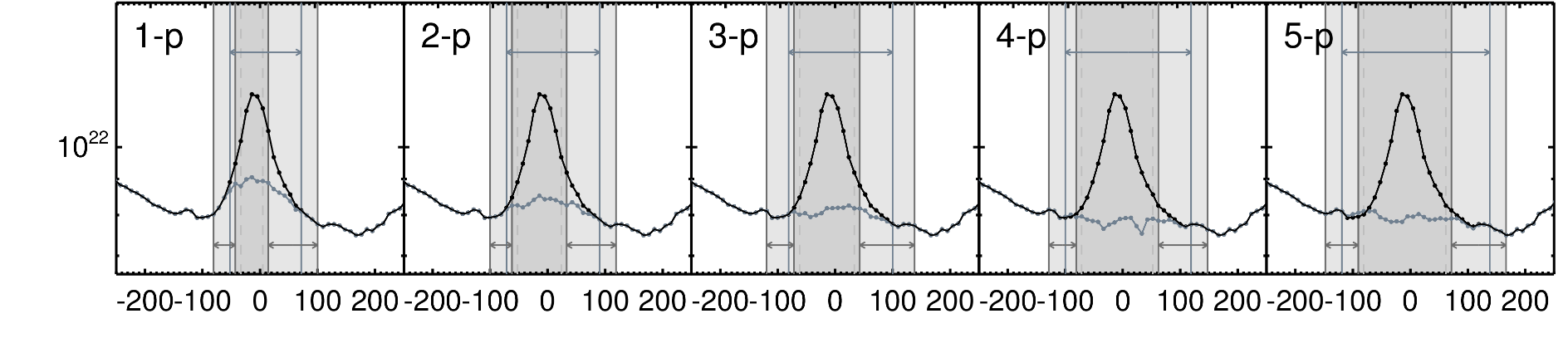}
\includegraphics[width=0.98\textwidth]{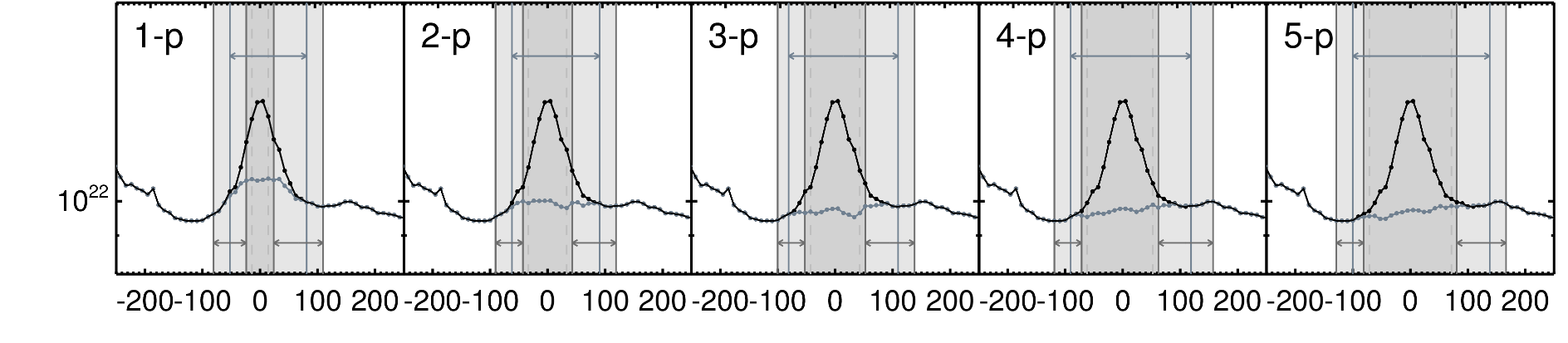}
\includegraphics[width=0.98\textwidth]{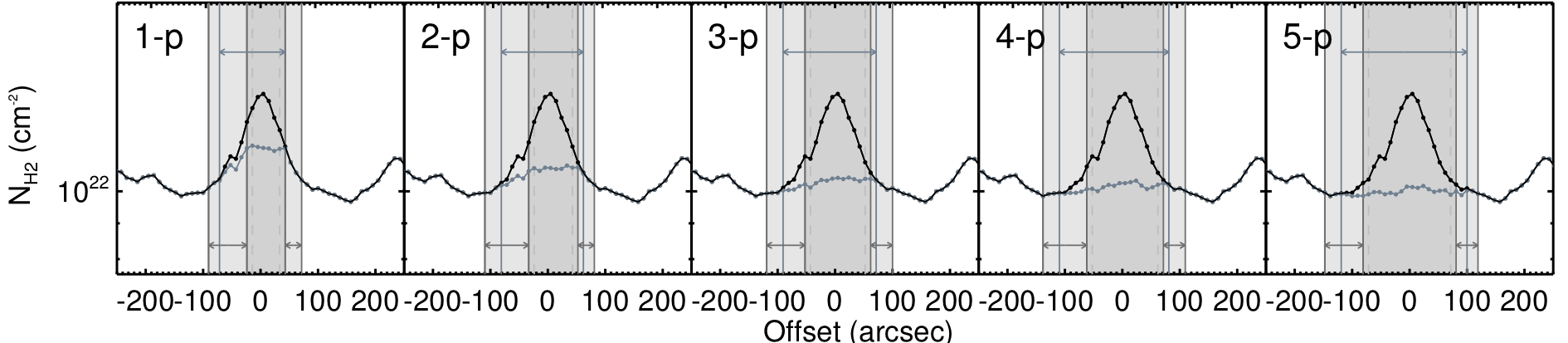}
\caption{Average radial profile of the sections I ({\it top panel}), II  ({\it middle panel}) and III ({\it lower panel}) of the object shown in Fig.\,\ref{Fig:TestFilDfact}. The filament emission profile and the estimated background are shown with black solid and grey line, respectively. The panels from left to right refer to the results with a dilation parameter $D$ ranging from $1$ to $5$. The black line is the measured emission profile. The shaded rectangles trace the region ascribed to the filament ({\it dark grey}) and the positions along the profile adopted for the estimate of the background ({\it light grey}). The radial extension of the rectangles is also drawn by segments delimited by the arrows both for the filament area ({\it top)} and for the background measurement ({\it bottom}).  The irregular shape of the mask produces the overlap between the two regions;  the mask is crossed at these radial distances for while running along the feature.}
\label{Fig:ProfilesDfact}
\end{figure*}

The initial mask obtained from the thresholding of $\lambda_{a}$ ($\lambda_{a}\,\leq\,T\,<,0$) does not cover the entire extension of the filament, but traces only its central portion where the intensity profile retains a downward concavity. We widen the initial mask to encompass the entire area of the feature, identifying an extended mask adopted to measure the filament column density and to determine the underlying background emission. To this end, we follow the prescription suggested by \citet{Schisano2014}, expanding the initial mask by a preset number of pixels, hereafter dilation size $D$, in all directions. The same method is also used by \citet{Li2016} on ATLASGAL data  starting from the one pixel wide segment  identified by the DisPerSE algorithm \citep{Sousbie2011}.  The idea behind this approach is to add further pixels until the mask includes the position where the filament merges into the surrounding background. After that, any additional pixel introduces a small contribution to the total emission if the background is properly estimated. We assume that the background emission varies less than the filament one, so we can estimate it from the pixels surrounding the extended mask and remove it from the total. 
Therefore, a different value for $D$ changes the extended mask and shifts the positions where used to estimate the background. Large $D$ values would be preferable to ensure that the entire emission from the filament feature is included, but they are unfeasible for two reasons. Firstly, the background emission still varies on the Hi-GAL maps, even if with a smaller amplitude and over larger spatial scales than the filaments. Secondly, the high density of features identified in the GP implies that they can start overlapping with large values of $D$.

We selected the dilation size $D$ that is more suitable for our sample by testing its effect on one mosaic of the GP. We ran our extraction with different $D$ ranging from $1$ to $5$ pixels, equivalent to angular expansions from $11.5$ to $57.5$\,arcsec. We evaluated the filament radial profiles and the relative backgrounds for a subsample of features to verify that the filament emission is correctly estimated.  Fig.\,\ref{Fig:ProfilesDfact} shows the profiles corresponding to the different filament sections drawn in Fig.\,\ref{Fig:TestFilDfact} for different $D$ values from left to right. The area assigned to the filament (dark-grey shaded area) increases with $D$ and shifts the regions from which the  background is estimated to larger radial distances (light-grey shaded area).

Fig.\,\ref{Fig:ProfilesDfact} shows that extending the initial mask by $1$ or $2$ pixels is not sufficient to include the emission from the profile wings. The filament emission not included in the mask introduce an overestimation of the background. The results largely improve for $D\,>3$ where the filament seems to be completely included in the mask and the background converges to similar values. However, a dilation  $D\,=\,5$ pixels ($\sim60$\,arcsec) would overlap neighbour features in crowded region. Moreover, large $D$ would extend excessively the filament mask by including several pixels with only  background emission. These larger area ascribed to the filament systematically reduce the measured average column density. To avoid these effects we adopted $D=3$ pixels and assumed that the expanded contour effectively traces the area of the filament.

\begin{figure} 
\includegraphics[width=0.5\textwidth]{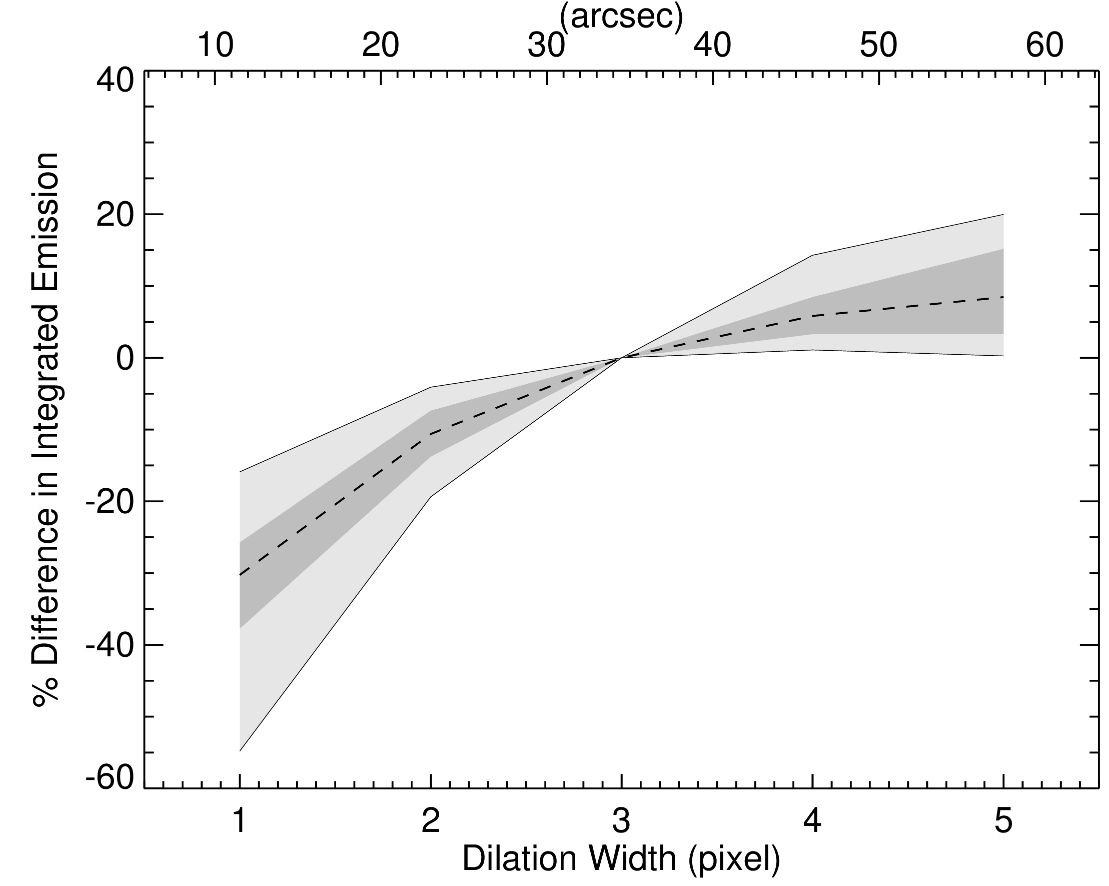}
\caption{Percentage change of the total integrated emission as a function of the dilation width $D$. The shaded area delimit the area where fall all the curves relative to the filaments in the test field with $C\,>1.1$, The integrated emission is normalize to the values measured for $D\,=\,3$. The median of the distribution is shown as a black dashed line, while the dark-grey area identify the interquartile ($3^{rd}$ to $1^{st}$) range.}
\label{Fig:DilationTotal}
\end{figure}

The radial profiles in Fig.\,\ref{Fig:ProfilesDfact} suggest that the filament integrated emission measured with  $D\,=\,3$ can be slightly underestimated in some portion of the filament (see lower panel). We quantify this possible systematic by selecting all the high-contrast ($C>1.1$) objects in the test field and measuring the integrated emission as function of $D$. These are a set of curves, one for each filament, that typically increase with $D$, but flatten for $D\,\geq\,3$. We show the distribution of these curves in Fig.\,\ref{Fig:DilationTotal} normalized to the  integrated fluxes measured for $D=3$.  

Fig.\,\ref{Fig:DilationTotal} confirms the results obtained above: dilation sizes with $D\,\leq\,2$ typically underestimate the filament emission. A dilation equal to $D\,=\,3$ leaving only a residual filament emission outside the mask. The integrated fluxes increase only marginally for larger $D$. Measurements with $D\,=\,3$ underestimate on average the ones with $D\,=\,5$ only by $\sim10$ per cent, a fraction that is comparable to the uncertainties introduced by calibration. The discrepancy on the integrated intensities measured in the two cases $D\,=\,3$ and $D\,=\,5$ is found to be always smaller than $20$ per cent.

\newpage

\section{Description of the Hi-GAL catalog of candidate filaments }\label{App:catalog}

The  Hi-GAL filamentary feature catalogue is hosted in the VIALACTEA knowledge base \citep{Molinaro2016}\footnote{The VIALACTEA Knowledge Base (VLKB) is a database accessible through the VIALACTEA application downloadable at  \url{http://vialactea.iaps.inaf.it/}} and is composed of three tables, one for each of the features defined in the article: entire candidate regions, branches and singular points.

Each table has a different column structure that is described here starting for the candidate regions: 
\begin{itemize}
\item Column [1], \texttt{NAMEID}: Unique Designation  for the filament. The designation is built by naming the candidate as \emph{'HiGALFilNNN.NNNNsM.MMMM'} where \emph{NNN.NNNN} and \emph{sM.MMMM} are respectively the Galactic longitude and the Galactic latitude of the centroid of the extended mask with $4$ decimal digits with the character {\it s} that is equal to \emph{'+'} or \emph{'-'} depending if the latitude is positive or negative. 
\item Column [2], \texttt{IDMOS}: Long integer that identifies univocally the candidate. 
\item Column [3], \texttt{NAMEMOS}: String defining the mosaic from which the feature has been extracted.
\item Column [4], \texttt{GLON} and  Column [5], \texttt{GLAT}: Galactic longitude and latitude, respectively, assigned to the feature defined as the centroid position from the pixels in the extended mask.
\item Columns from [6] to [9], \texttt{MINGLON}, \texttt{MAXGLON}, \texttt{MINGLAT} and \texttt{MAXGLAT}: Maximum and minimum Galactic longitude and latitude, respectively, of the filament contour. They define a rectangular shape including the entire extended mask of the candidate.
\item Columns [10] and [11], \texttt{DELTAGLON} and \texttt{DELTAGLAT}: Angular extension in Galactic latitude and longitude, respectively, of the candidate filament associated mask.
\item Column [12], \texttt{LENGTH}: Filament angular length in arcseconds measured along the spine.
\item Column [13], \texttt{AREA}: Total area of the candidate in arcmin$^{2}$ computed from the sum of the pixels covering the extended mask covering of the extracted structure. 
\item Column [14], \texttt{ELLIPTICITY}: Ellipticity defined by the ratio of the  major and minor axis ellipse best fitting the initial mask.
\item Column [15], \texttt{FILLINGFACT}: Filling factor defined as the ratio of the initial mask area to the area of the best fitting ellipse.
\item Columns [16] and [17], \texttt{SEMIAXISA} and \texttt{SEMIAXISB}: Minor and major axis of the ellipse that best fitted the extended mask.
\item Column [18], \texttt{ORIENTATION}: Orientation of the major axis counted anti-clockwise with respect the $b=0^{\circ}$ axis. 
\item Column [19], \texttt{TOTAL\_ROI}: Total sum of the input column density map over all the pixels of the extended mask of the filamentary region. 
\item Columns [20], \texttt{BACK\_ROI}: Total sum of the estimated background column density (model 2C1T, see Sect.~\ref{Sect:2C2Tmodel}) over all the pixels of the extended mask of the filamentary region.  
\item Column [21], \texttt{FIL\_ROI}: Total sum of the filament contribution to the measured column density over all the pixels of the extended mask of the filamentary region
\item Column [22], \texttt{AVERAGECD}: Average column density of the filament in the entire extended region.
\item Columns [23] and [24], \texttt{NH2MEANBRANCHES} and \texttt{NH2STDBRANCHES}: Mean and standard deviation of the measured column-density value along all pixels of the filament 1D branches.
\item Columns from [25] to [27], \texttt{CONTRAST\_BE12},  \texttt{CONTRAST\_BD} and \texttt{CONTRAST\_DE12}: Contrast value defined as the ratio between the average column density of the central branches, with respect to the average in the filament surroundings,  the central branches with respect to the filament region and the filament region with respect to its surroundings respectively (see Sect.~\ref{Sect:2C2Tmodel}).
\item Column [28], \texttt{RELEVANCE12}: Relevance of the candidate filament defined as ratio between the average column density on the central branches over the standard deviation of the measured column density in the filamentary region surroundings (see Sect.~\ref{Sect:2C2Tmodel}).
\item Column [29], \texttt{FLAGCAND}: Detection flag associated to the candidate region.
\item Columns from [30] to [33], \texttt{TMEANBRANCHES},  \texttt{TSTDBRANCHES}, \texttt{TMINBRANCHES} and \texttt{TMAXBRANCHEST}: Temperature mean, standard deviation and minimum and maximum respectively measured along the central branches of the filamentary region adopting the model 2C1T (see Sect.~\ref{Sect:2C2Tmodel})
\item Columns from [34] to [37], \texttt{TMEDIANROI},  \texttt{TSTDROI}, \texttt{TQ1ROI} and \texttt{TQ3ROI}: Median, first and third quartiles of the temperature distribution over the entire filamentary region.
\item Columns from [38] to [39], \texttt{NH2FILONLYMEANROI} and \texttt{NH2FILONLYSTDROI}: Average column density and standard deviation of the filament component measured in the entire filamentary extended mask assuming the model 2C2T (see Sect.~\ref{Sect:2C2Tmodel}).
\item Columns from [40] to [41], \texttt{TFILONLYMEDIANROI} and \texttt{TFILONLYSTDROI}: Average temperature and standard deviation of the filament component measured in the entire filamentary extended mask assuming the model 2C2T  (see Sect.~\ref{Sect:2C2Tmodel}).
\item Columns from [42] to [43], \texttt{NH2BCKONLYMEANROI} and \texttt{NH2BCKONLYSTDROI}:  Average column density and standard deviation of the background component measured in the entire filamentary extended mask assuming the model 2C2T (see Sect.~\ref{Sect:2C2Tmodel}).
\item Columns from [44] to [45], \texttt{TBCKONLYMEDIANROI} and \texttt{TBCKONLYSTDROI}: Average temperature and standard deviation of the background component measured in the entire filamentary extended mask assuming the model 2C2T  (see Sect.~\ref{Sect:2C2Tmodel}).
\item Columns from [46] to [48], \texttt{MED70}, \texttt{F1Q70} and \texttt{F3Q70}: Median, first and third quartiles of the flux measured at 70 $\mu$m within the region ascribed to the candidate.
\item Column [49], \texttt{NSOURCESASS}: Number of sources of the band-merged Hi-GAL catalogue spatially associated to the filament contour.
\item Column [50], \texttt{NROBASS}: Number of sources of the band-merged Hi-GAL catalogue sharing similar radial velocity (and same choice for the near/far distance ambiguity) and falling within the filament contour. This number differs from that in column [48] only when there are more than 3 sources spatially associated to the filament. 
\item Columns from [51] to [52], \texttt{RVROBASS} and \texttt{STDRVROBASS}: Mean and standard deviation of the radial velocities of the robust sources associated with the filament.
\item Columns from [53] to [54], \texttt{ROBDIST} and \texttt{STDROBDIST}: Mean and standard deviation of the distance of all the robust sources associated with the filaments. Distances are derived from the radial velocities, assuming the Galactic rotation curve of \citet{Russeil2017}.
\item Column [55], \texttt{EXTPC}:  Extension of the filament defined as the length of the major axis of the fitting ellipse to the filament extended mask region.
\item Column [56], \texttt{LENGTHPC}: Linear angular length measured along the filament main spine.
\item Column [57], \texttt{MASS1T}: Filament mass derived from the 2C1T model: the 2-component (filament and background) sharing the same temperature.
\item Column [58], \texttt{MASS2T}: Filament mass derived from the 2C2T model: the 2-component (filament and background) left free to have 2 different temperatures.
\item Column [59], \texttt{MLIN2T}: Filament linear mass derived as ratio between column [57] and [55], i.e. between the mass from the 2C2T and the angular length of the filament.
\item Column [60], \texttt{NIRDC}: Number of IRDCs whose contours eventually overlap with the filament. 
\item Column [61], \texttt{NATLASFIL}: Number of ATLASGAL filament associated to the Hi-GAL candidate filament.
\item Column [62], \texttt{SOURCESASS}: Source ID of the Full Hi-GAL extended catalogue spatially associated with the filamentary structure.
\item Columns from [63] to [64], \texttt{PROB10kms\_300pc} and \texttt{PROB10kms\_500pc}: Probability of association of the filament to the spiral arm model of \citet{Hou2009} assuming an arm full width of $W\,=\,600$\,pc and $W\,=\,1$\,kpc. 
\item Column [65], \texttt{FLAGDIST}: Flag on the reliability of the assigned distance. The flag tags the cases where $a)$  $R_{Gal}\,\leq\,5$\,kpc and $R_{Gal}\,\gtrsim\,22$\,kpc affected by large uncertainty, $b)$  the estimated RV exceeds the tangent point terminal velocity has been adopted, $c)$ distances of the majority of the associated compact sources are assigned by a different methods than the kinematic one.

\end{itemize}

The filament branches table includes both quantities measured along the branch segment (1-D branch) and in the portion of the filament spatially associated with the branch (2-D branch).
The column structure is described as: 
\begin{itemize}
\item Column [1], \texttt{IDBRANCH\_MOS}: Long integer that uniquely identifies the branch. 
\item Column [2], \texttt{IDMOS}: Long integer that identifies  the filament candidate to which the branch belongs. 
\item Column [3], \texttt{LENGTH}: Branch angular length in arc-seconds as the direct sum of all the positions along the 1-D segment.
\item Columns [4] and [5], \texttt{LIMIT1} and \texttt{LIMIT2}: IDs of the singular points tracing the extreme of the branch 1-D segment.
\item Column [6], \texttt{FLAGLIMITS}: String identifying the type of singular points located at the 1-D segment extremes (vertex, 'V', or node 'N')
\item Column [7], \texttt{DIRECTION}: Direction of the 1-D branch segment represented as the angle counted anti-clockwise with respect the $b=0^{\circ}$ axis.
\item Column [8], \texttt{FLAGSPINE}: Flag identifying if the 1-D branch segment is classified as belonging to the filament main spine, 'S', or not, 'B'.
\item Columns from [9] to [11], \texttt{MEANCD}, \texttt{STDCD} and \texttt{VARCD}: Mean, standard deviation and maximum variation (defined by the  difference between the maximum 
and minimum value) of the column density ascribed to the filament component along the 1-D segment assuming the 2C1T model.
\item Columns from [12] to [15], \texttt{MEANCDBACK}, \texttt{STDCDBACK} and \texttt{VARCDBACK}: Mean, standard deviation and maximum variation  of the column density ascribed to the background along the 1-D segment assuming the 2C1T model.
\item Column [16], \texttt{AREABRANCH}: Total area in arcsec$^{2}$  of the portion of the filamentary mask associated with the 2-D branch after the region segmentation.
\item Column [17], \texttt{TOTALCDROI}: Total sum of the measured  column density map over all the pixels associated with the 2-D portion of the filamentary mask associated with the 1-D branch.
\item Column [18], \texttt{TOTALBACKCDROI}: Total sum of the column density associated with the background component, assuming the 2C1T model over all the pixels associated with the 2-D portion of the filamentary mask associated with the 1-D branch.
\item Column [19], \texttt{BRANCHCDROI}: Total sum of the column density associated with the filament component, assuming the 2C1T model over all the pixels associated with the 2-D portion of the filamentary mask associated with the 1-D branch.
\item Column [20], \texttt{AVERBRANCHROI}: Mean column density associated with the filament component, assuming the 2C1T model over the 2-D portion of the filamentary mask associated with the 1-D branch.
\item Columns from [21] to [24], \texttt{TFULLMEANBR}, \texttt{TFULLSTDBR}, \texttt{TFULLMINBR}, \texttt{TFULLMAXBR}: Mean, standard deviation, minimum and maximum temperature measured along the 1-D branch segment assuming the 2C1T model.
\item Columns from [25] to [28], \texttt{NH2FILONLYMEANBR}, \texttt{NH2FILONLYSTDBR}, \texttt{NH2FILONLYMINBR}, \texttt{NH2FILONLYMAXBR}: Mean, standard deviation, minimum and maximum column density ascribed to the filament component measured along the 1-D branch segment assuming the 2C2T model.
\item Columns from [29] to [32], \texttt{NH2BACKONLYMEANBR}, \texttt{NH2BACKONLYSTDBR}, \texttt{NH2BACKONLYMINBR}, \texttt{NH2BACKONLYMAXBR}: Mean, standard deviation, minimum and maximum column density ascribed to the background component measured along the 1-D branch segment assuming the 2C2T model.
\item Columns from [33] to [36], \texttt{TFILONLYMEANBR}, \texttt{TFILONLYSTDBR}, \texttt{TFILONLYMINBR}, \texttt{TFILONLYMAXBR}: Mean, standard deviation, minimum and maximum temperature ascribed to the filament component measured along the 1-D branch segment assuming the 2C2T model.
\item Columns from [37] to [40], \texttt{TBACKONLYMEANBR}, \texttt{TBACKONLYSTDBR}, \texttt{TBACKONLYMINBR}, \texttt{TBACKONLYMAXBR}: Mean, standard deviation, minimum and maximum temperature ascribed to the background component measured along the 1-D branch segment assuming the 2C2T model.
\item Columns  [41] and [42], \texttt{TFULLMEANROI} and  \texttt{TFULLSTDROI}: Mean and standard deviation of the temperature measured assuming the 2C1T model over the 2-D portion of the filamentary mask associated with the 1-D branch.
\item Columns  [43] and [44], \texttt{NH2FILONLYMEANROI} and  \texttt{NH2FILONLYSTDROI}: Mean and standard deviation of the column density ascribed to the filament component,  assuming the 2C2T model over the 2-D portion of the filamentary mask associated  with the 1-D branch.
\item Columns  [45] and [46], \texttt{TFILONLYMEANROI} and  \texttt{TFILONLYSTDROI}: Mean and standard deviation of the temperature   ascribed to the filament  component  assuming the 2C2T model over the 2-D portion of the filamentary mask associated with the 1-D branch.
\item Columns  [47] and [48], \texttt{NH2BACKONLYMEANROI} and  \texttt{NH2BACKONLYSTDROI}: Mean and standard deviation of the column density ascribed to the background component  assuming the 2C2T model over the 2-D portion of the filamentary mask associated  with the 1-D branch.
\item Columns  [49] and [50], \texttt{TBACKONLYMEANROI} and  \texttt{TBACKONLYSTDROI}: Mean and standard deviation of the temperature   ascribed to the background  component  assuming the 2C2T model over the 2-D portion of the filamentary mask associated with the 1-D branch.
\end{itemize}

Finally, the table relative to the singular points has the following column structure:
\begin{itemize}
\item Column [1], \texttt{IDNODE}: Long integer that uniquely identifies the singular point. 
\item Column [2], \texttt{IDMOS}: Long integer that identifies  the filament candidate to which the singular point belongs. 
\item Column [3] and [4], \texttt{GLON} and \texttt{GLAT}: Position in Galactic longitude and latitude of the singular point. 
\item Column [5], \texttt{TYPE}: Flag identifying whether the singular point is a vertex or a node.
\item Column [6], \texttt{NCONNECTIONS}: Number of adjacent pixels belonging to the branches.
\end{itemize}

\bsp	
\label{lastpage}
\end{document}